\documentclass[%
 reprint,
 amsmath,amssymb,
 aps,
]{revtex4-2}

\usepackage{graphicx}
\usepackage[percent]{overpic}
\usepackage{dcolumn}
\usepackage{bm}
\usepackage{xcolor}
\usepackage{graphicx}
\usepackage{dcolumn}
\usepackage{bm}
\usepackage{amsmath,amssymb,bbm,bm,physics}
\usepackage{amsfonts,amssymb,txfonts}
\usepackage{epsfig,color}
\usepackage{graphicx}
\usepackage{hyperref}
\usepackage{float}
\usepackage{microtype}
\usepackage{placeins}
\usepackage{comment}
\usepackage{braket}
\usepackage[justification=raggedright,singlelinecheck=false]{caption}
\usepackage{booktabs}
\usepackage{tabularx}
\usepackage{array}
\newcolumntype{C}{>{\centering\arraybackslash}X}
\hypersetup{colorlinks=true,linkcolor=blue,anchorcolor=blue,citecolor=blue,filecolor=blue,urlcolor=blue,bookmarksnumbered=true,pdfview=FitB}

\usepackage{graphicx}
\usepackage{fancyhdr}
 \usepackage{cancel}
\usepackage{comment}

\renewcommand{\subsectionmark}[1]{}

\newcommand{\bk}{\mathbf k}
\newcommand{\bq}{\mathbf q}
\newcommand{\br}{\mathbf r}

\newcommand{\bP}{\mathbf P}

\newcommand{\ep}{\epsilon}

\newcommand{\ve}{\varepsilon}

\newcommand{\beq}{\begin{equation}}
\newcommand{\eeq}{\end{equation}}
\newcommand{\bea}{\begin{eqnarray}}
\newcommand{\eea}{\end{eqnarray}}

\newcommand{\kf}{k_\text{F}}

\usepackage{subcaption}
\newcommand{\kp}{{k_{\text{F}\perp}}}
\newcommand{\kl}{{k_{\text{F}\parallel}}}

\newcommand{\Tbgt}{{\tilde T_{\mathrm{BG}}}}
\newcommand{\Tbg}{{T_{\mathrm{BG}}}}
\newcommand{\Tbgp}{{T_{\mathrm{BG}\perp}}}
\newcommand{\Tbgl}{{T_{\mathrm{BG}\parallel}}}
\newcommand{\kb}{{k_{\text{B}}}}

\begin{document}

\preprint{APS/123-QED}

\title{Energy relaxation due to two-phonon scattering of electrons:\\ Breakdown of the energy diffusion model }

\author{Joshua Covey}
 \email{jcovey@ufl.edu}
\author{Dmitrii L. Maslov}%
\affiliation{%
 Department of Physics, University of Florida, Gainesville, Florida 32611 USA
}%

\date{\today}

\begin{abstract}
Recent THz spectroscopy of the quantum paraelectric SrTiO$_3$ (arXiv:2501.15771) and a high-$T_c$ cuprate (arXiv:2503.15646) has renewed interest in energy relaxation in correlated electron systems. We consider a situation in which single-phonon scattering is forbidden by symmetry or momentum conservation, while two-phonon scattering is allowed. Solving the Boltzmann equation, we show that above the Bloch-Gr\"uneisen temperature the energy relaxation rate from two soft transverse optical phonons exceeds the single-phonon one: while the latter scales as $1/T$, the former is linear in $T$. This dominance of two-phonon scattering invalidates the usual picture of energy diffusion due to frequent scattering by subthermal phonons; instead, energy relaxes via rare scattering events involving thermal phonons. Below the Bloch-Gr\"uneisen temperature, the energy relaxation rate scales as the single-particle rate, namely as $T^3$ for soft phonons. For anisotropic electron bands, an intermediate regime appears between two Bloch-Gr\"uneisen temperatures, in which both allowed single-phonon and two-phonon processes scale as $T^2$.
\end{abstract}

\maketitle

The rate at which electrons lose energy can vary significantly from the rate at which they lose momentum, thereby shedding unique light on the scattering mechanisms at play.
The nonequilibrium electron distribution can be decomposed into irreducible representations of the lattice symmetry group, one of which corresponds to a uniform shift of the distribution and hence to the current. Momentum relaxation describes decay of this projection, whereas energy relaxation describes return to the equilibrium distribution. The corresponding scattering times can differ parametrically~\cite{Mahan2000Many-ParticlePhysics}.

Motivated by recent nonlinear THz spectroscopy experiments on doped $\mathrm{SrTiO_3}$ (STO) thin films \cite{Kumar2025AbsenceSpectroscopy} and a high $T_c$ cuprate \cite{Chaudhuri2025PlanckianSuperconductors},
both showing momentum relaxation rates much larger than the energy relaxation rate, we study how hot electrons equilibrate with the lattice via two-phonon processes.

Two-phonon processes become important even at weak coupling when single-phonon scattering is forbidden by symmetry or kinematics. Doped quantum paraelectrics are prime examples \cite{Muller1979SrTiO3:K,Chandra2017ProspectsReview}: their proximity to a ferroelectric transition is associated with a soft transverse optical (TO) phonon whose $q=0$ frequency saturates at low temperature~\cite{Yamada1969NeutronSrTiO3}. Because of inversion symmetry, direct coupling to a single TO phonon is forbidden. Examples include perovskites such as STO, KTaO$_3$, and EuTiO$_3$, and rock salts such as PbTe.
Similar symmetry restrictions arise for flexural phonons in graphene membranes \cite{Mariani2008FlexuralGraphene} and for odd oxygen modes in high-$T_c$ cuprates \cite{Cuk2005AARPES,Johnston2010SystematicCuprates}. Another example is the antiferrodistortive transition of STO, where the TO branch softens near the Brillouin-zone boundary, such that for a small Fermi surface centered at $\Gamma$, the kinematic constraints can also favor two-phonon over one-phonon processes \cite{Scott1974Soft-modeTransitions}. Here we focus on 3D soft TO phonons with dispersion
\begin{flalign}
    \omega_\bq^2=\omega_0^2+s^2q^2, \label{ph dispersion}
\end{flalign}
for $q\ll q_\mathrm{BZ}$, where $q_\mathrm{BZ}$ is the boundary of the Brillouin zone \cite{Yamada1969NeutronSrTiO3,Courtens1993PhononRegime}.
Near the
ferroelectric
quantum critical point, $T\gg\omega_0$, one attains an acoustic form, $\omega_q\approx sq$.

For generic electron-phonon (eph) coupling of degenerate electrons to single phonons in the Bloch-Grueneisen (BG) regime---namely at
\begin{flalign}
T\ll \tilde T_\mathrm{BG}=\omega(q=2\kf) \qquad (\hbar,\kb=1),
\end{flalign}
where $\kf$ is the Fermi wavevector, 
each eph scattering event relaxes both energy and momentum. Hence, the ERR, $1/\tau_E$,
is of the same order as
as the single particle 
relaxation rate (SPRR), $1/\tau_{0}$
\cite{Kaganov1957RelaxationLattice,ZimanElectronsSolids,Allen1987TheoryMetals,Glorioso2022JouleMetals}.
The same 
will be shown to be true for two-phonon scattering. Note that $\tilde T_\mathrm{BG}=\sqrt{\omega_0^2+T_\mathrm{BG}^2}\sim \max\left\{\omega_0,T_\mathrm{BG}\right\}$ for the phonon dispersion in Eq.~\eqref{ph dispersion}, where $T_\mathrm{BG}\equiv 2\kf s$ is the BG temperature for an acoustic dispersion.

In the equipartition regime ($T\gg \tilde T_{BG}$), electrons lose energy much more slowly than momentum. The standard interpretation is the energy-diffusion model~\cite{Altshuler1982EffectsLocalisation,Gantmakher1987CarrierSemiconductors}: phonons are activated only up to $\omega_q\lesssim \tilde T_{BG}\ll T$, so each scattering event transfers only a small fraction of the electron thermal energy. Electrons therefore execute a random walk along the energy axis with diffusion coefficient $D_E\sim \tilde T_{BG}^2/\tau_0$.
The energy relaxation time is defined as
the time over which a random walk along the energy axis traverses the entire thermal window: 
\begin{flalign}
    \tau_E\sim \frac{T^2}{D_E}\sim \left(\frac{T}
    {\Tbgt}
    \right)^2\tau_{0}\gg\tau_{0}.\label{diffusive picture}
\end{flalign}
For single-phonon scattering in the equipartition regime with
$\tau_{0}\propto 1/T$, Eq.~\eqref{diffusive picture} 
yields $\tau_E\propto T$, in agreement with the actual calculation \cite{Ginzburg1955KineticEmission,Kaganov1957RelaxationLattice,Allen1987TheoryMetals}. 

\begin{figure*}[t]
\centering

\begin{subfigure}[c]{0.3\textwidth}
  \raggedright (a)\par\vspace{0.4em}
  \centering
  \includegraphics[width=1.4\linewidth]{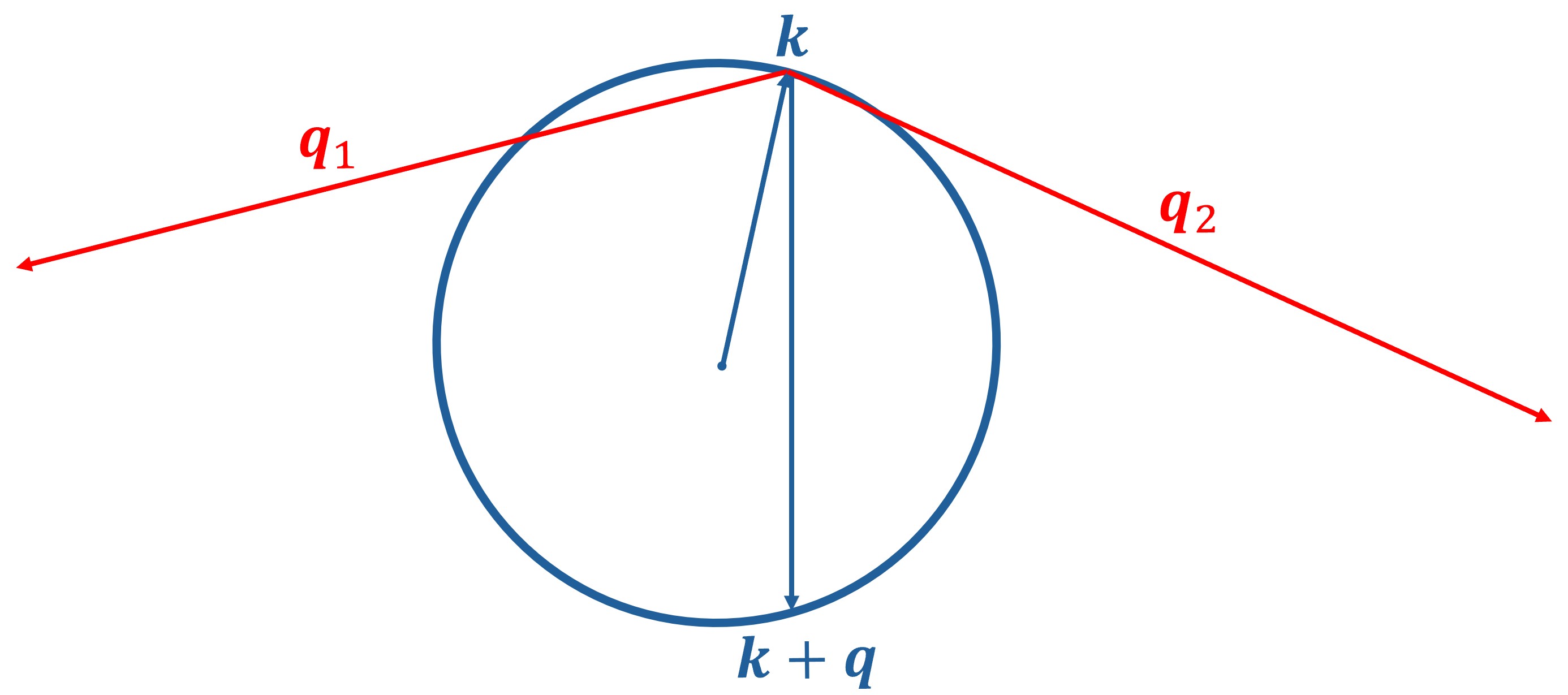}
  \caption{}
  \label{subfig:rare large}
\end{subfigure}
\hspace{1cm}
\begin{subfigure}[c]{0.3\textwidth}
  \raggedright (b)\par\vspace{0.4em}
  \centering
  \includegraphics[width=0.5\linewidth]{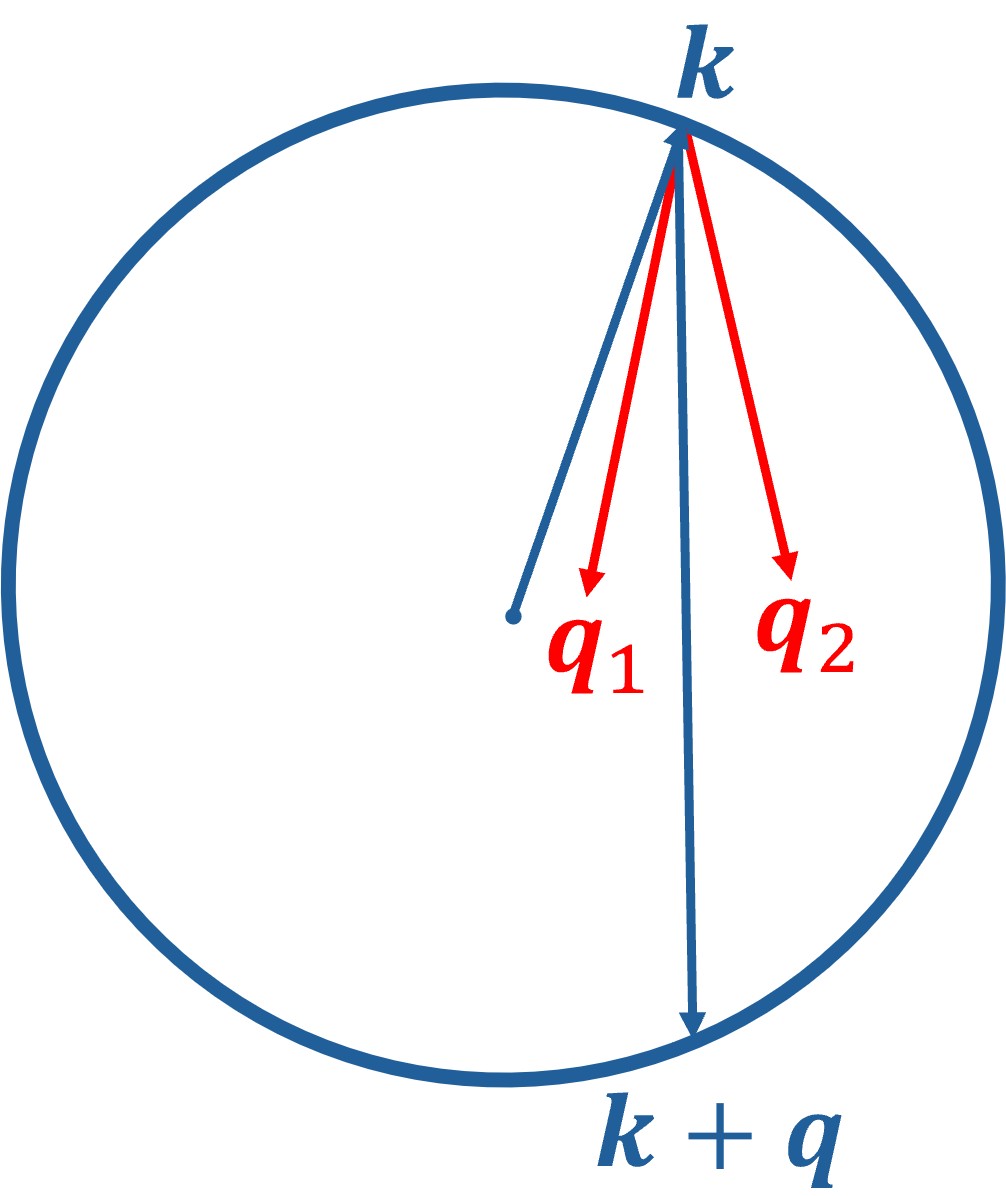}
  \caption{}
  \label{subfig:small transfer}
\end{subfigure}
\hspace{-1cm}
\begin{subfigure}[c]{0.3\textwidth}
  \raggedright (c)\par\vspace{0.4em}
  \centering
  \includegraphics[width=1.2\linewidth]{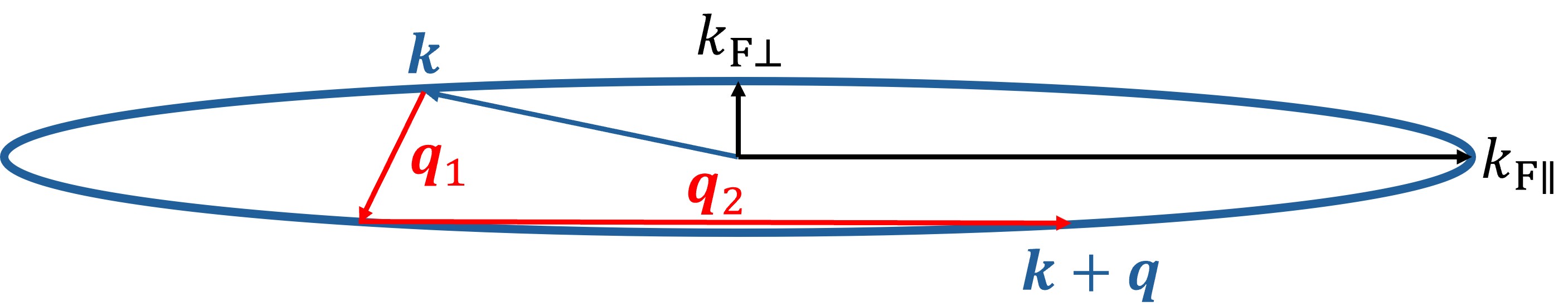}
  \caption{}
  \label{subfig:intermediate event}
\end{subfigure}

\caption{An initial electron state $\bk$ on the Fermi surface (FS) is scattered to a final state $\bk+\bq$, also on the FS. (a) Energy relaxation is controlled by thermal phonons with momenta $q_{1,2}\sim T/s$. At $T/\Tbg \rightarrow \infty$ the angle between $\bq_1$ and $\bq_2$ approaches $\pi$. (b) In contrast to (a), the single-particle and momentum-relaxation rates are controlled by phonons with momenta $q_{1,2}\sim \kf \ll T/s$. (c) An example of an intravalley two-phonon scattering event within a strongly ellipsoidal FS.}
\end{figure*}
In this Letter, we show that this picture fails for scattering by two soft TO (2TO) phonons in the equipartition regime, $T\gg \Tbgt$. Instead, energy relaxation is controlled by rare 2TO-phonon events with energy transfer of order $T$, rather than by frequent low-energy processes occurring at the SPRR for 2TO-phonon 
scattering, $1/\tau^\mathrm{2TO}_0$.These events are rare because the total momentum transfer $q=|\bq_1+\bq_2|$ is restricted by $2\kf$, while the individual phonon momenta are thermal, $q_{1,2}\sim T/s\gg \kf$, and hence nearly antiparallel; see Fig.~\ref{subfig:rare large}. By contrast, the SPRR and transport relaxation rate (TRR) are controlled by subthermal phonons with $q_{1,2}\sim \kf$ and $\omega_{q_{1,2}}\sim \Tbgt\ll T$; see Fig.~\ref{subfig:small transfer}. In that case, $1/\tau^\mathrm{2TO}_0\sim 1/\tau^\mathrm{2TO}_\mathrm{tr}\propto T^2$ \cite{Epifanov1981InteractionFerroelectrics,Kumar2021QuasiparticleParaelectrics},
simply following
the square of the phonon occupation number. Were the diffusion model applicable, Eq.~\eqref{diffusive picture} would predict $\tau^\mathrm{2TO}_E=\mathrm{const}$. Instead, we will show that
\begin{flalign}
    \tau_E^{\mathrm{2TO}}\sim \frac{T}{
    \Tbgt}\tau_{0}^{\mathrm{2TO}}
    \propto \frac{1}{T}.
   \label{2to not diff}
\end{flalign} 
While energy relaxation is still parametrically slow compared to momentum relaxation, it is parametrically faster than that in the diffusion picture.

We employ a semiclassical approach for degenerate electrons.
The electron energy exchange rate 
due to two-phonon coupling is determined by the weighted average of the Boltzmann collision integral over all states \cite{Gantmakher1987CarrierSemiconductors,ZimanElectronsSolids} (see Supplemental Material (SM) for details~\cite{5eq_energy_exchange})
\begin{widetext}
\begin{flalign}
\begin{split}
    \frac{\partial E^{\mathrm{2TO}}_
    \mathrm{e}}{\partial t} =2\sum_\bk\ve_\bk
    \left(\frac{\partial f_\bk}{\partial t}\right)^{\mathrm{
    2TO}}&=-\frac{4\pi}{ V^2}\sum_{\bk,\bq_1,\bq}|g_{\bq_1,\bq-\bq_1}|^2 
    \biggl\{
    \delta(\ve_\bk-\ve_{\bk-\bq}-\Omega_+)\Omega_+\biggl[(N_{\bq_1}+N_{\bq-\bq_1}+1)f_\bk(1-f_{\bk-\bq})+N_{\bq_1}N_{\bq-\bq_1}(f_\bk-f_{\bk-\bq})\biggr] \\
    &\hspace{3cm}+\delta(\ve_\bk-\ve_{\bk-\bq}-\Omega_-)\Omega_-\biggl[N_{\bq_1}N_{\bq-\bq_1}(f_\bk-f_{\bk-\bq})-N_{\bq_1}f_{\bk-\bq}(1-f_\bk)+N_{\bq-\bq_1}f_\bk(1-f_{\bk-\bq})\biggr]
    \biggr\}, \label{pow transfer}
    \end{split}
\end{flalign} 
\end{widetext}
where $\bq=\bq_1+\bq_2$ and $\Omega_\pm=\omega_{\bq_1}\pm\omega_{\bq-\bq_1}$ are the net momentum and energy transfers, respectively, and $f_\bk=f(\ve_\bk,T_\mathrm{e})$ and $N_\bq=N(\omega_\bq,T)$ are the Fermi and Bose distribution functions for electrons with dispersion $\ve_\bk$ at temperature $T_\mathrm{e}$ and phonons with dispersion $\omega_\bq$ and temperature $T$, respectively. The first line in Eq.~\eqref{pow transfer} corresponds to 
processes of simultaneous 
absorption and emission  of two phonons while the second 
one to absorption-emission
and emission-absorption
processes. In the equipartition limit, energy relaxation is controlled by double absorption and emission processes as the contribution from the latter two cancel to leading order in the expansion in $\Tbg/T$.

Equation \eqref{pow transfer} is valid in the two-temperature approximation
\cite{Ginzburg1955KineticEmission,Kaganov1957RelaxationLattice,Allen1987TheoryMetals}, 
which assumes that electrons are excited away from the lattice temperature $T$ and then
are quickly thermalized by electron-electron interaction to an equilibrium state with $T_\mathrm{e}>T$. After this stage, the energy flows from electrons to 
the lattice slowly, on the time scale $\tau_E$.  

At small overheating, when $\delta T\equiv T_\mathrm{e}-T\ll T$,
the ERR is related to the power transfer by
\begin{flalign}
\begin{split}
    \frac{1}{\tau_E}&=\frac{1}{\Delta E}\Bigg|\frac{\partial E_e}{\partial t}\Bigg|=\frac{1}{
    n_\mathrm{e}}\frac{2E_\mathrm{F}}{\pi^2k^2_\mathrm{B}T\delta T}\Bigg|\frac{\partial E_e}{\partial t}\Bigg|, \label{final rate}
\end{split}
\end{flalign}
where $\Delta E$ is the difference of thermal energies of electrons at temperatures $T_\mathrm{e}$ and $T$,
and $n_\mathrm{e}$ is the total number of conduction electrons. For an isotropic electron system, the most general form of the ERR due to two-phonon scattering is given by 
    \begin{flalign}
\begin{split}
   \frac{1}{\tau_E^{\mathrm{2ph}}} &= \frac{1}{T^3}\frac{E_\mathrm{F}m^2}{8\pi^7n}\int_0^\infty dq_1q_1^2 \int_{-1}^1 dt \int_0^{2\kf} dqq\abs{g_{\bq_1,\bq_-\bq_1}}^2  \\
    &\times\bigg\{
    \Omega_+^3 N(\omega_{\bq_1})  N(\omega_{\bq-\bq_1})(N(\Omega_+)+1)
    \\&
    \hspace{.5cm}+ 
    \Omega_-^3
    N(\omega_{\bq_1})(N(\omega_{\bq-\bq_1})+1)N(-\Omega_-)
    \biggr\},\label{ERR}
    \end{split} 
\end{flalign} 
where $m$ is the effective electron 
mass and $t=\cos\theta_\bq$.

We now focus on the electron-2TO phonon interaction described by
\cite{Epifanov1981InteractionFerroelectrics}
\begin{equation}
H_{\mathrm{int}}^{2\mathrm{TO}}
=
\frac{g_2}{2} \int d^3r\;n(\br)\,\bP(\br)\cdot\bP(\br),
\qquad
n(\br)=\psi^\dagger(\br)\psi(\br), \label{model}
\end{equation}
where the macroscopic polarization is  
\begin{flalign}
    \bP(\br)=\sum_{\bq}\frac{\boldsymbol{e}_\bq}{\sqrt{V}}A_\bq b_\bq e^{i\bq\cdot\br}+\mathrm{h.c.}, \qquad A_\bq=\sqrt{\frac{(\ep_0(\bq)-\ep_\infty)\omega_\bq}{4\pi}},
\end{flalign}
and $\omega_\bq$ is defined in Eq.~\eqref{ph dispersion}.
The squared scattering matrix element is
\begin{equation}
\abs{g_{\bq_1,\bq_2}}^2
=
\frac{g_2^2\Omega_0^4}{16\pi^2}\frac{1}{\omega_{\bq_1}\omega_{\bq_2}}\Big[1+(\hat\bq_1\cdot\hat\bq_2)^2\Big], \label{coupling elements}
\end{equation}
where $\Omega_0$ is related to the static permittivity and phonon dispersion via the Lyddane-Sachs-Teller relation \cite{Lyddane1941OnHalides,Kittel1987QuantumSolids},
$\Omega_0^2=\ep_0(\bq)\omega_\bq^2$.
 
The crucial difference between the ERR and the SPRR/TRR is the three powers of $\Omega_\pm$ in Eq.~\eqref{ERR}. One comes from Eq.~\eqref{pow transfer}, one from the integration over $\ve_\bk$, and one from expanding the Bose functions in $\delta T$. In the equipartition regime these factors shift the spectral weight to $q_1\approx |\bq-\bq_1|\sim T/s\gg \kf$, where the Bose factors are of order unity and $q\sim \kf$. Power counting then gives $1/\tau_E^\mathrm{2TO}\propto T$. In contrast, the corresponding integrals for the SPRR/TRR are controlled by the region $q_1\sim |\bq-\bq_1|\sim \kf$, where each Bose function attains its classical limit, $N\propto T$, so that $1/\tau^\mathrm{2TO}_{0}\sim 1/\tau^\mathrm{2TO}_\mathrm{tr}\propto T^2$. If $\omega_0\ll\Tbgt\approx\Tbg$, there is an intermediate BG regime, $\omega_0\ll T\ll \Tbg$. In this regime, not only the individual phonon momenta but also their sum are thermal, i.e., $q_1\sim |\bq-\bq_1|\sim q\sim T/s$. In this case, $1/\tau_E^\mathrm{2TO}\sim 1/\tau_0^\mathrm{2TO}\propto T^3$. Finally, all the scattering rates are suppressed exponentially for $T\ll\omega_0$.

The analytic form of the ERR due to soft 2TO phonon scattering is given by~\cite{11eq_2to_eer_isotropic}
\begin{flalign}
\begin{split}
   \frac{1}{\tau_E^{\mathrm{2TO}}} &= 
    \frac{T^3}{E_\mathrm{2TO}\Tbg}
    F_{\mathrm{2TO}}\left(
    \frac{T}{\Tbg}
    ,\frac{\omega_0}{\Tbg}\right), \label{isotropic 2to err}
    \end{split} 
\end{flalign}  
where $E_\mathrm{2TO}=64\pi^3s^4/mg_2^2\Omega_0^4$. At temperatures below the gap  ($T\ll\omega_0$) the ERR is exponentially suppressed regardless of the magnitude of $\Tbg$:
$1/\tau_E^\mathrm{2TO}\propto e^{-\omega_0/T}$. Outside this interval, the asymptotic limits of the function $F_\mathrm{2TO}$ are $F_\mathrm{2TO}\left(z\rightarrow0,\eta=0\right)=3\alpha/\pi^4$ with $\alpha\approx 79.1$ and 
$F_\mathrm{2TO}(z\rightarrow\infty,\eta\ll z)=4\beta/\pi^4z^2$ with $\beta\approx0.95$.

If $\omega_0\ll T_\mathrm{BG}$, there exists a BG temperature regime, $\omega_0\ll T\ll \Tbg$, where the ERR scales as $T^3$,\begin{flalign}
\begin{split}
   \frac{1}{\tau_E^{\mathrm{2TO}}} &=
    \frac{T^3}{E_\mathrm{2TO} \Tbg}\frac{3\alpha}{\pi^4} .\label{cube}
    \end{split} 
\end{flalign} 
In the equipartition regime which sets in for $T\gg\Tbg$,
the ERR scales linearly with temperature,
\begin{flalign}
     \frac{1}{\tau_E^\mathrm{2TO}}=T\frac{\Tbg}{E_\mathrm{2TO} }\frac{4\beta}{\pi^4}.\label{lin}
\end{flalign}
Since the SPRR scales quadratically with temperature, %
the relation in \eqref{2to not diff} is made evident. The 
picture of diffusive-like motion along the energy axis gives way to something resembling a damped chaotic pendulum with occasionally large swings away from the lattice temperature 
but eventually relaxing to it.

If $\omega_0\gg\Tbg$, the BG regime is absent, and the exponential regime for $T\ll\omega_0$ crosses over to the equipartition one with $1/\tau_E^\mathrm{2TO}$ in Eq.~\eqref{lin} already at $T\sim\omega_0$.

\begin{figure}[!]
\includegraphics[scale=0.45]{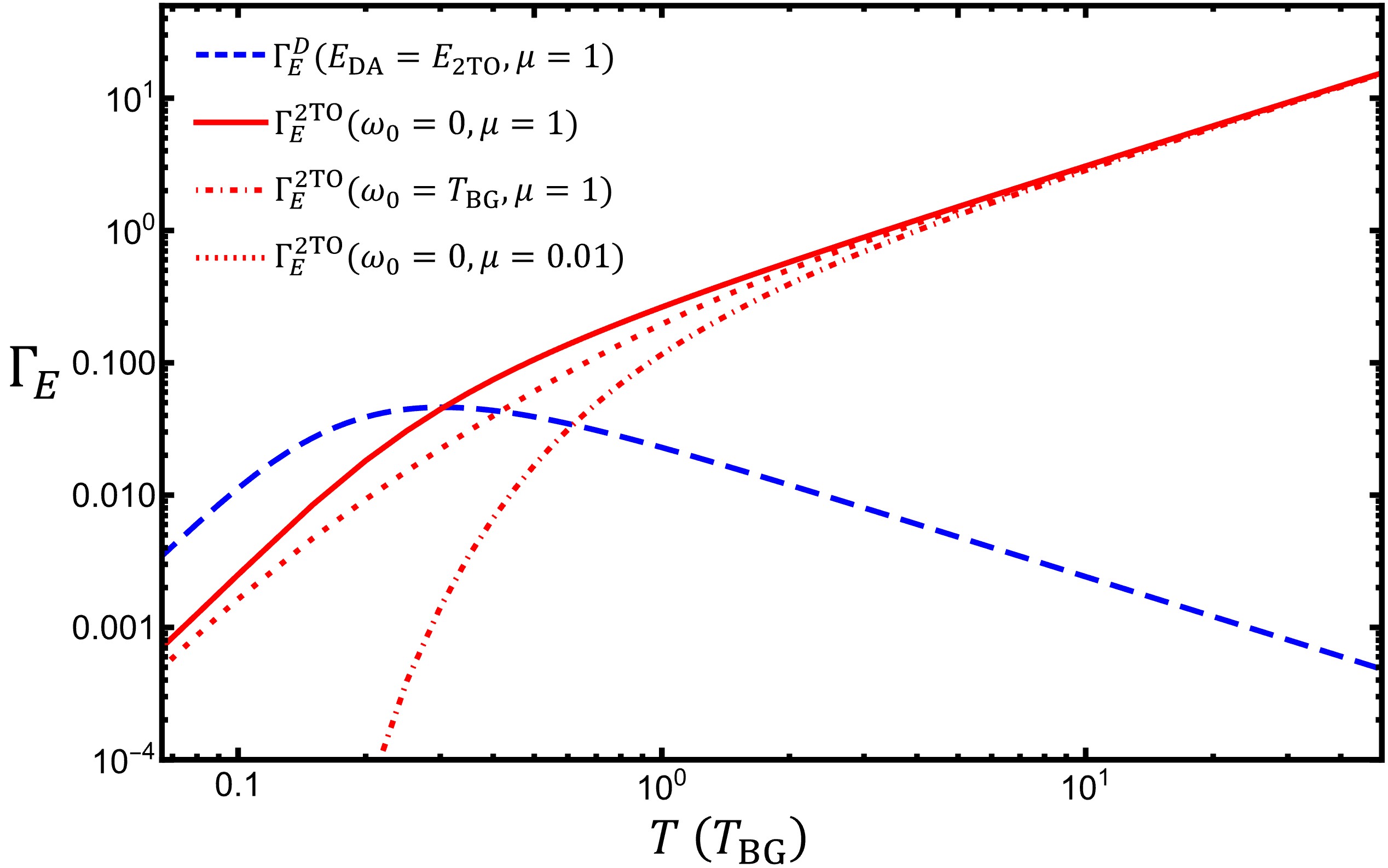}
  \caption{
  The energy relaxation rate, normalized to $1/\tau^\mathrm{ref}_E=T^2_\mathrm{BG}/E_\mathrm{2TO}$, as a function of temperature.  Solid line: scattering by two gapless TO phonons ($\eta=\omega_0/\Tbg=0$). Dashed line: scattering by single acoustic phonons. Dot-dashed line: scattering by two gapped TO phonons ($\eta=\omega_0/\Tbg=1$). Dotted line: scattering by two gapless TO phonons with a strongly anisotropic electron dispersion ($\mu=0.1$).}
  \label{fig:gamma plots}
\end{figure}

In Fig.~\ref{fig:gamma plots} we plot the dimensionless ERR $\Gamma^\mathrm{2TO}_E=\tau^\mathrm{ref}_E/\tau^{2TO}_E$, where $1/\tau^\mathrm{ref}_E\equiv T_\mathrm{BG}^2/E_\mathrm{2TO}$, over the entire range of $T$ for $\omega_0/\Tbg=0,1$. The parameter $\mu=1$ denotes the isotropic case, its definition to be discussed below. The solid line depicts the gapless case ($\omega_0=0$). Matching Eqs.~\eqref{cube} and \eqref{lin} gives a crossover scale $T_\mathrm{cr}=\sqrt{4\beta/3\alpha}\,\Tbg\approx 0.13\,\Tbg$. A numerical estimate based on the local exponent $n(T)\equiv d\ln\Gamma_E^\mathrm{2TO}/d\ln T$ and the condition $n(T_\mathrm{cr})=2$ gives $T_{cr}\approx 0.29\,\Tbg$~\cite{dBG_crossover}. In either estimate, the crossover occurs well below $\Tbg$. For the gapped case, the low-$T$ rate is exponentially suppressed, while the high-$T$ behavior is nearly unchanged.

In addition to 2TO phonon scattering, electrons also experience single-phonon scattering via the deformation potential coupling to the acoustic branch (DA scattering), and the corresponding ERRs need to be compared. The ERR due to DA-scattering was calculated in 
Refs.~\cite{Kaganov1957RelaxationLattice,Allen1987TheoryMetals}:
\begin{flalign}
   \frac{1}{\tau_E^\mathrm{DA}}=\frac{T^3}{E_\mathrm{DA}T_{\mathrm{BG}}}F_\mathrm{DA}\left(\frac{T}{\Tbg}\right), \label{isotropic acoustic err}
\end{flalign}
where $E_\mathrm{DA}=\rho s_\mathrm{a}^3/mD^2$, $\rho$ is the mass density per unit cell, $s_\mathrm{a}$ is the speed of sound for the acoustic mode, $D$ is the deformation potential energy, and the asymptotic limits of the scaling function are $F_\mathrm{DA}(z\to 0)=
360\zeta(5)/\pi^3 
$ and $F_\mathrm{DA}(z\gg 1)=
3/4\pi^3z^4
$.
To facilitate the  comparison,
we assume that $s_\mathrm{a}=s$, so that the
BG temperatures for the acoustic and TO phonons are the same.
The dimensionless ERR for DA-scattering, defined via the same reference rate as for 2TO scattering, 
$\Gamma^{\mathrm{DA}}_E=(E_\mathrm{2TO}/E_\mathrm{DA})
\left(1/\tau^\mathrm{DA}_E\right)$, 
is plotted by the 
dashed line in Fig.~\ref{fig:gamma plots} for $E_\mathrm{2TO}=E_\mathrm{DA}$. We see that  $\Gamma^{\mathrm{2TO}}_E$ exceeds $\Gamma^{\mathrm{DA}}_E$ by several orders of magnitude in the equipartition regime, whereas $\Gamma^{\mathrm{2TO}}_E$ is comparable to $\Gamma^{\mathrm{DA}}_E$ in the BG regime for the gapless 2TO case and significantly smaller than  $\Gamma^{\mathrm{DA}}_E$ in the exponential regime for the gapped case.

\begin{figure}[t]
\centering

\begin{subfigure}[c]{0.28\linewidth}
  \makebox[\linewidth][l]{(a)}
  \centering
  \includegraphics[width=\linewidth]{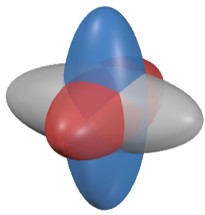}
  \caption{}
  \label{subfig:ellip_FS}
\end{subfigure}
\hspace{0.02\linewidth}
\begin{subfigure}[c]{0.65\linewidth}
  \makebox[\linewidth][l]{(b)}
  \centering
  \includegraphics[width=\linewidth]{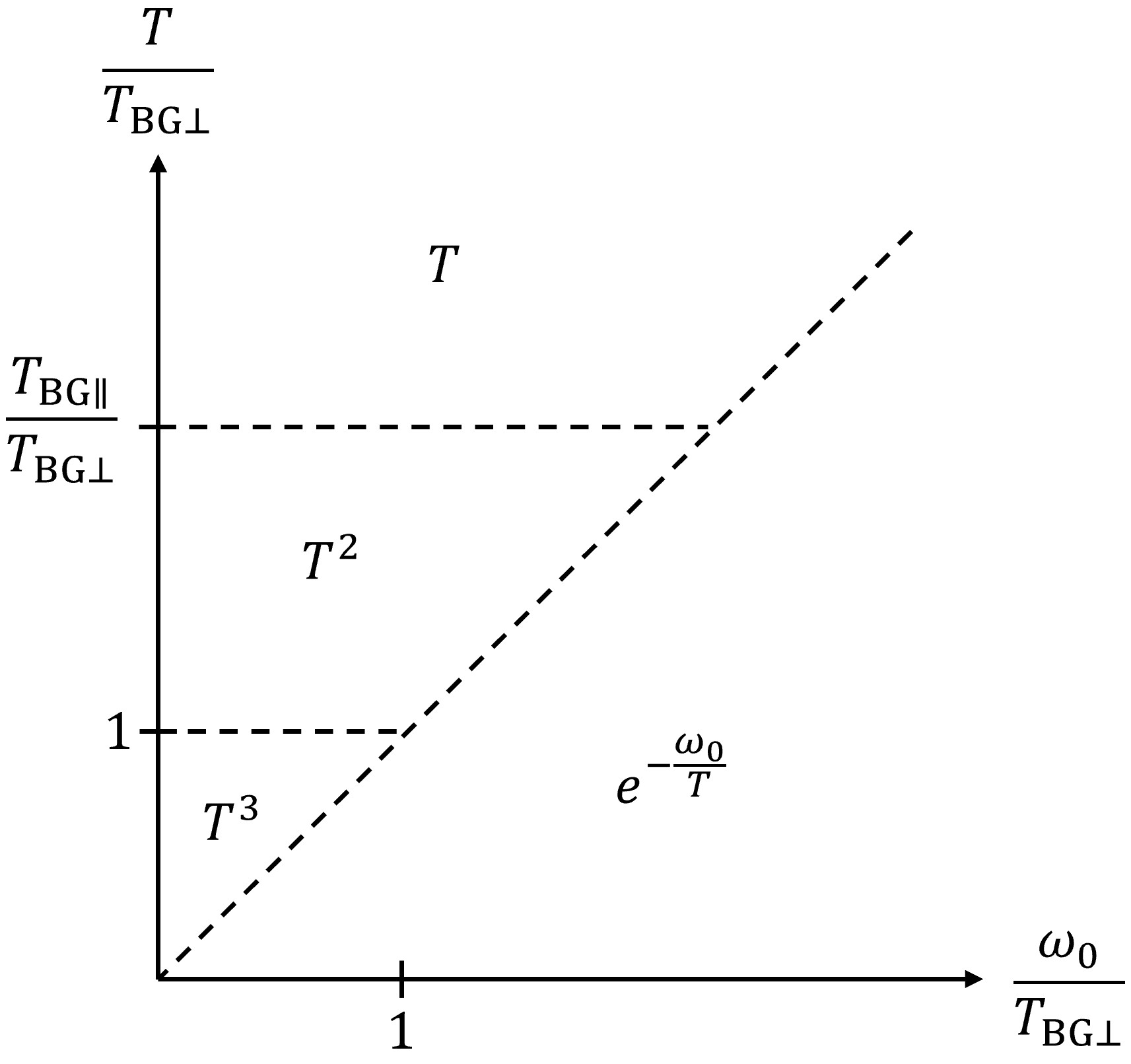}
  \caption{}
  \label{subfig:T_regimes}
\end{subfigure}

\caption{(a) A model STO Fermi surface consisting of the union of three mutually orthogonal ellipsoids. (b) Temperature scaling of the ERR in the various temperature regimes due to soft 2TO phonon scattering. In the isotropic case ($\Tbgp=\Tbgl=\Tbg$) the $T^2$ region vanishes.}
\label{fig:energy_transfer}
\end{figure}

The results presented above can be 
readily generalized to an anisotropic electron dispersion. 
As an example, we consider the band structure of STO. 
While its FS is nearly isotropic at the most dilute carrier densities, it becomes strongly anisotropic for $n\gtrsim
10^{19}\,\mathrm{cm}^{-3}$, when all three subbands derived from the $t_{2g}$ triplet are occupied 
\cite{vanderMarel2011CommonSrTiO3,Lin2015ScalableSurface}. 
Ignoring the tetragonal and spin-orbit splittings and assuming that 
$k_{\mathrm{F,max}}\ll q_\mathrm{BZ}$, the FS of STO can be modeled as the union of three mutually orthogonal ellipsoidal valleys, each defined as
\begin{flalign}
        \ve_\bk=\frac{k_\perp^2}{2m_\perp}+\frac{k_\parallel^2}{2m_\parallel},
\end{flalign}
with $k_{||}$ and $k_\perp$ 
along the major and minor axes, respectively---see Fig.~\ref{subfig:ellip_FS}.
An example of an intravalley two-phonon scattering event is shown in Fig.~\ref{subfig:intermediate event}. A distinct feature of the anisotropic case is the existence of
the intermediate temperature regime, $\Tbgp\ll T\ll\Tbgl$ \cite{Kukkonen1978TBismuth,Gantmakher1987CarrierSemiconductors,Kang2018Thermopower-conductivitySrTiO3}, where $T_\mathrm{BG\perp,\parallel}=2k_\mathrm{F\perp,\parallel}s$. 
In this regime, only momentum transfers orthogonal to the major axis are kinematically restricted 
to the range $\lesssim\kp$, while those parallel 
are large, on the order of $T/s\gg \kp$. 

In the anisotropic case, the ERRs for DA and 2TO mechanisms are given, respectively, by the following scaling forms~\cite{b16eq_da_eer_aniso,c17eq_da_eer_aniso}:
\begin{flalign}
    \frac{1}{\tau^\mathrm{DA}_E}
    &=
    \frac{T^3}{E_{\mathrm{DA},\perp}T_\mathrm{BG\perp}} \bar F_\mathrm{DA}\left(\frac{T}{\Tbgp},\mu\right),\label{ERR D generic}
    \\ 
  \frac{1}{\tau_E^{\mathrm{2TO}}} &= 
    \frac{T^3}{E_{\mathrm{2TO},\perp}T_\mathrm{BG\perp}}
   \bar F_\mathrm{2TO}\left(\frac{T}{\Tbgp},\frac{\omega_0}{\Tbgp},\mu\right) ,\label{ERR 2to generic}
\end{flalign}
where $E_{\mathrm{DA},\perp}=E_\mathrm{DA}(m\rightarrow m_\perp)$, $E_{\mathrm{2TO},\perp}=E_\mathrm{2TO}(m\rightarrow m_\perp)$, and $\mu=m_\perp/m_\parallel\leq 1$. 
As in the isotropic case, at temperatures below the gap ($T\ll\omega_0$) the ERR is exponentially suppressed regardless the magnitude of $\Tbgp$ or $\Tbgl$---hence, in each kinematic regime we consider $\omega_0\ll T$. 
The isotropic results are recovered at $\mu=1$.  
For $\mu\lesssim1$ ($\Tbgp\lesssim\Tbgl$) there are once again only two scaling regimes: 
the BG and equipartion ones.
In the BG temperature limit ($T\ll\Tbgp$) 
we find 
$\bar F_\mathrm{DA}(z\ll1,\mu) =360\zeta(5)\arccos(\mu)/\pi^3\sqrt{1-\mu}$
and $\bar F_\mathrm{2TO}(z\ll1,\eta\ll z,\mu)= 3\alpha\arccos(\mu)/\pi^4\sqrt{1-\mu}$ 
with $\alpha$ the same as in the isotropic case. 
In the equipartition regime $\bar F_\mathrm{DA}(z\gg1,\mu)=(2\mu+1)/4\pi^3z^4\mu^{3/2}$ 
and $F_\mathrm{2TO}(z\gg1,\eta\ll z,\mu) = 4\beta/\pi^4z^2\sqrt\mu$ with $\beta$ the same as in the isotropic case. The ERR for DA and 2TO scattering thereby scales as $T^3$ in the BG regime and $T^{-1}$ and $T$, respectively, in the equipartition regime. In the strongly anisotropic case,
$\sqrt\mu\ll 1$,  there is also 
an intermediate regime, $\Tbgp\ll T\ll\Tbgl$.
Here we have
$\bar F_\mathrm{DA}(z\gg 1,\mu\ll 1/z^2)=
4\pi/5z$
and 
$ \bar F_\mathrm{2TO}(z\gg 1,\eta\ll z,\mu\ll 1/z^2)=3\gamma/\pi^4 z$ with $\gamma\approx47.11$. Therefore, in this regime both the DA and 2TO ERRs scale quadratically with temperature,
\begin{flalign}
    \frac{1}{\tau^\mathrm{DA,2TO}_E}
    &\propto T^2. 
\end{flalign}
The intermediate temperature regime is akin to the BG 
one in that the quadratic scaling is the same for the DA and 2TO mechanisms.

To plot 
the 2TO EER for the anisotropic case
on equal footing with the isotropic results, we 
rewrite Eq.~\eqref{ERR 2to generic}  in  
the same form as 
Eq.~\eqref{isotropic 2to err},
\begin{flalign}
  \frac{1}{\tau_E^{\mathrm{2TO}}} &= 
    \frac{T^3}{E_{\mathrm{2TO}}T_\mathrm{BG}}
   \mu^{1/6}\bar F_\mathrm{2TO}\left(\frac{T}{\Tbg\ \mu^{1/6}},\frac{\omega_0}{\Tbg\ \mu^{1/6}},\mu\right),\label{ERR 2to generic rescaled}
\end{flalign}
and choose the volumes of an isotropic and an ellipsoidal Fermi surface to be the same, i.e., $\kf^3=k^2_\mathrm{F\perp}\kl$. Equation~\eqref{ERR 2to generic rescaled} is plotted by the dotted line  in Fig.~\ref{fig:gamma plots} 
for $\mu=0.01$ and for a gapless phonon dispersion ($\eta=0$).
One can see that the transition between the low and high $T$ scaling regimes is smeared in the case $\Tbgp\ll\Tbgl$ ($\sqrt\mu\ll 1$).

In conclusion, we have shown 
that in centrosymmetric crystals electron energy relaxation  is controlled by second order polar coupling in
the equipartition regime, a mechanism which invalidates the usual energy-diffusion picture.
For electrons with an ellipsoidal dispersion like those in STO, there is an intermediate temperature regime where coupling to either single acoustic or soft 2TO phonons at temperatures above the gap ($T\gg\omega_0$) leads to quadratic temperature scaling. Our analytic results are summarized in Fig.~\ref{subfig:T_regimes}.

Recent experiment~\cite{Kumar2025AbsenceSpectroscopy} reported an increase of the EER in STO over $T=10\text{--}50\,\mathrm{K}$, a temperature range where the resistivity goes as $T^2$~\cite{Lin2015ScalableSurface} such that two-phonon elastic scattering has been considered a possible culprit. The increasing dependence of the ERR was interpreted as being inconsistent with two-phonon elastic scattering and as evidence against the 2TO mechanism. Our results show that this interpretation is incorrect: in contrast to the single-phonon contribution, the two-phonon EER does increase with temperature. However, the experiment does not yet provide a definitive test of our theory, because its temperature range lies near the crossover between the BG and equipartition regimes. Using recent ARPES values $m_\perp=0.63m_e$ and $m_\parallel=8m_e$ \cite{Wakabayashi2025Orbital-resolvedARPES}, and the same numerical procedure as in the isotropic case~\cite{eBG_crossover2}, we obtain $T_{\mathrm{cr1}}\approx 18\,\mathrm{K}$ for the crossover from $T^3$ to $T^2$, and $T_{\mathrm{cr2}}\approx 63\,\mathrm{K}$ for the crossover from $T^2$ to $T$. Since the experimental window falls within this interval, measurements at higher temperatures are needed to probe the equipartition regime and test the 2TO scenario.

We are grateful to N. P. Armitage and S. A. Kivelson for stimulating discussions. This work was initiated at the 2025 Boulder Summer School for Condensed Matter and Materials Physics, supported by the National Science Foundation through the Materials Theory grant No.~NSF-2328793, and supported by the National Science Foundation under Grant No.~DMR-2224000.

\bibliography{jcovey,SM}

\end{document}


\title{Supplementary material: Energy relaxation due to \\two-phonon scattering of electrons}
\author{Joshua Covey and Dmitrii L. Maslov\\Department of Physics, University of Florida}
\date{\today}

\maketitle
\tableofcontents
\vspace{.35cm}
We calculate the energy relaxation rate (ERR) of degenerate electrons coupling either to the deformation potential via single acoustic phonons or to the macroscopic polarization field via simultaneous scattering off two soft transverse optical (2TO) phonons in the semiclassical Boltzmann picture under the assumptions of the two-temperature approximation \cite{Ginzburg1955KineticEmission,Kaganov1957RelaxationLattice,Allen1987TheoryMetals}---by soft we mean a phonon dispersion that can be modeled for $q\ll q_\mathrm{BZ}$ as $\omega_\bq^2=\omega_0^2+s^2q^2$ \cite{Yamada1969NeutronSrTiO3}. For the case of 2TO phonon scattering, we begin by restating the formulation of Kittel \cite{Kittel1987QuantumSolids} for the Froelich (polar) coupling mechanism and subsequently review the Lyddane-Sachs-Teller (LST) relation \cite{Lyddane1941OnHalides}. With the given Hamiltonian, we then determine the scattering matrix elements via Fermi's golden rule (FGR) under the Born approximation. With the collision integral as our starting point, we calculate the energy exchange rate (power transfer) and from there the ERR. For both cases, we present the results for an isotropic electron system below and above the Bloch-Grueneisen (BG) temperature, $\Tbg=2s\kf$, and then reconsider a $t_{2g}$ band structure analogous to that in $\mathrm{SrTiO_3}$ (STO) with its three mutually orthogonal Fermi surfaces (FS), and ignore tetragonal and spin orbit splitting. In this case there is an intermediate temperature regime, $\Tbgp\ll T\ll\Tbgl$, and for both single acoustic and 2TO phonon coupling we deduce the leading order temperature behavior.

\section{Energy relaxation via Boltzmann equation}
The generic in-out collision integral in the Boltzmann formalism is given by, 
\begin{flalign}
\frac{\partial f_\bk}{\partial t}=-\sum_{\bk'}\Big[
W_{\bk\to \bk'}\,f_{\bk}(1-f_{\bk'})-W_{\bk'\to \bk}\,f_{\bk'}(1-f_{\bk})
\Big],
\end{flalign}
where $f_\bk$ is the Fermi function for an electron with some dispersion $\ve_\bk$ at temperature $T_e$.
Under the adiabatic approximation, the scattering probabilities can be found via FGR in calculating the transition probability rates,
\begin{equation}
W_{i\to f}=2\pi\abs{M_{fi}}^2\,\delta(E_f-E_i), \qquad \hbar=1,\label{fgr}
\end{equation}
which will be done in the following sections for the specified interactions. As seen in \cite{Gantmakher1987CarrierSemiconductors}, the average energy exchange rate of electrons, i.e. the power transferred from conduction electrons to the lattice, is simply the weighted average of single particle collision integrals \cite{Gantmakher1987CarrierSemiconductors},
\begin{flalign}
    \frac{\partial E_e}{\partial t} =2\sum_\bk\ve_\bk\frac{\partial f_\bk}{\partial t}, \label{power}
\end{flalign}
where the factor of 2 accounts for the sum over spins. The ERR may then be defined as the energy exchange rate normalized by the difference in average thermal energy between hot and cold electrons, 
\begin{flalign}
\begin{split}
    \frac{1}{\tau_E}=\frac{1}{\Delta E}\abs{\frac{\partial E_e}{\partial t}}
    &=\frac{1}{N_e}\frac{2E_\mathrm{F}}{\pi^2T\delta T}\abs{\frac{\partial E_e}{\partial t}}, \qquad \kb=1,\label{generic rate}
\end{split}
\end{flalign}
where $\Delta E=\gamma_c(T_e^2-T^2)/2$, $\gamma_c=\pi^2 N_e/2E_\mathrm{F}$ is the specific heat for a free electron gas, and we expand to linear order in $\delta T=T_e-T$.

\section{Deformation potential interaction: A review of electrons coupled to microscopic lattice displacements}\label{sec:single la}
We restate the results for electrons coupling to the deformation potential via scattering off single acoustic phonons to set the stage for and subsequently compare with the following section on macroscopic polarization coupling at second order via 2TO phonon scattering. In an isotropic system, the interaction with the deformation potential is
\begin{equation}
H_{\mathrm{int}}^{D} = D\int d^3 r\;n(\br)\,\nabla\cdot\bu(\br).\label{isotropic D int}
\end{equation}
Its operator form is derived in a number of textbooks and we simply write the result,
\begin{equation}
H_{\mathrm{int}}^{D}
=\sum_{\bk,\bq} g_{\bq}\,c^\dagger_{\bk+\bq}c_{\bk}\left(b_{\bq}+b^\dagger_{-\bq}\right)\delta_{\bk',\bk+\bq}.
\end{equation}  
In this case electrons may only couple to longitudinal acoustic (LA) phonons whose matrix elements are
\begin{flalign}
    g_{\bq}^{D,\mathrm{LA}} = iD \sqrt{\frac{q}{2\rho V s}},
\end{flalign}
where $\rho=M/V_0$ is the mass density of the unit cell, $V=N_0V_0$ is the system volume, and $s$ is the speed of sound for LA phonons. The textbook collision integral, including both in (first line below) and out processes (second line) each with absorption (first terms of each line) and emission (second terms of each line), is
\begin{equation}
\begin{aligned}
\frac{\partial f_\bk}{\partial t}
&=
-\sum_{\bq}2\pi\abs{g_{\bq}}^2
\bigg\{
f_{\bk}\Big[
N_\bq\,(1-f_{\bk+\bq})\,\delta(\ve_{\bk+\bq}-\ve_{\bk}-\omega_{\bq})
+
(N_\bq+1)\,(1-f_{\bk-\bq})\,\delta(\ve_{\bk-\bq}-\ve_{\bk}+\omega_{\bq})
\Big]\\
&\hspace{3.4em}
-(1-f_{\bk})\Big[
N_\bq\,f_{\bk-\bq}\,\delta(\ve_{\bk}-\ve_{\bk-\bq}-\omega_{\bq})
+
(N_\bq+1)\,f_{\bk+\bq}\,\delta(\ve_{\bk}-\ve_{\bk+\bq}+\omega_{\bq})
\Big] 
\bigg\}. \label{coll la}
\end{aligned}
\end{equation}
Upon inserting Eq.~\eqref{coll la} into Eq.~\eqref{power}, followed by relabeling and combining terms, one arrives at the same result as Eq.~(6) in \cite{Allen1987TheoryMetals},
\begin{flalign}
    \begin{split}
        \frac{\partial E_e}{\partial t} = -4\pi \sum_{\bk,\bq}|g_\bq|^2 \omega_\bq\left[f_{\bk+\bq}\left(1-f_\bk\right)-N_\bq\left(f_\bk-f_{\bk+\bq}\right)\right]\delta(\ve_\bk-\ve_{\bk+\bq}+\omega_\bq). \label{exchange}
    \end{split}
\end{flalign}
This result is for generic matrix elements and dispersions and one may then insert that due to LA phonon coupling.

\subsection{Isotropic electron dispersion}
We first consider a simple isotropic electron system, namely one with a parabolic spectrum,
\begin{flalign}
    \ve_\bk= \frac{k^2}{2m}\label{parabolic}.
\end{flalign}
The sum in Eq.~\eqref{exchange} may be transformed to the following integral, $(1/V)\sum_\bk=\int d^3k/(2\pi)^3$. 
\paragraph{Kinematic constraint} 
The integral over solid angle is
\begin{flalign}
    2\pi\int d(\cos\theta)\,\delta\left(\frac{-1}{m}\left(kq\cos\theta+\frac{q^2}{2}\right)+ sq\right),
\end{flalign}
which gives a factor of $2\pi m/(kq)$ and enforces the following kinematic constraint on the magnitude of $q$,
\begin{flalign}
    \left|\frac{s}{\vf}-\frac{q}{2k}\right|<1,
\end{flalign}
where $\vf=\kf/m$ is the Fermi velocity. We consider the degenerate electron regime so that $\omega_\bq\lesssim T\ll E_\mathrm{F}$, and hence $s\ll\vf$ and $k\approx\kf$. Then q is restricted by
\begin{flalign}
    0<q<2\kf.
\end{flalign}
The integral over $k$ can be done by linearizing the spectrum about $\kf$ so that $dkk^2\approx \kf m\,d\ve_k$. The integral over electron energy is a standard result,
\begin{flalign}
    \int d\ve_k \left[f(\ve_k+\omega_q)(1-f(\ve_k))-N(\omega_q,T)\left(f(\ve_k)-f(\ve_k+\omega_q)\right)\right]=\omega_q\bigl[N(\omega_q,T_e)-N(\omega_q,T)\bigr]. \label{energy}
\end{flalign}
In the linear response regime we expand in the small temperature variation between the electronic and lattice subsystems, $\delta T=(T_e-T)\ll T$, which gives
\begin{flalign}
N(\omega_q,T_e)-N(\omega_q,T_L)= \delta T\frac{sq}{ T^2}\frac{e^{sq/ T}}{(e^{sq/ T}-1)^2}. \label{expand N}
\end{flalign}
With the use of Eq.~\eqref{generic rate}, we write the ERR due to electrons with parabolic dispersion coupling to the deformation potential via scattering off single LA phonons,
\begin{flalign}
    \frac{1}{\tau_E^\mathrm{DA}}=\frac{D^2s^2m^2E_\mathrm{F}}{2\pi^5\rho n}\frac{1}{T^3}
    \int_0^{2\kf } dqq^5\frac{e^{sq/ T}}{(e^{sq/ T}-1)^2}.\label{scat la}
\end{flalign}
We note the ERR is dependent on the particular coupling mechanism, where for scattering off single LA modes the squared matrix elements are  $|g^{\mathrm{LA}}_\bq|^2\propto q$. Unlike Froelich coupling which scales as $1/q$, $|g^{\mathrm{LA}}_\bq|^2$ suppresses coupling at small $q$. It is therefore to be emphasized that the following temperature dependence is not universal for single phonon scattering.

From here we can write the standard results in the low and high $T$ regimes. By introducing the dimensionless variable $x = sq/T$ and the Bloch-Grueneisen (BG) temperature $\Tbg=2\kf s$,
Eq.~\eqref{scat la} becomes
\begin{equation}
\frac{1}{\tau_E^\mathrm{DA}}
=\frac{D^2 m^2 E_F}{2\pi^5\rho ns^4}
T^3
\int_0^{T_{\mathrm{BG}}/T} dx\,
x^5 \frac{e^{x}}{(e^{x}-1)^2},
\label{eq:tauE_dimensionless}
\end{equation}
analogous to that found in Eq.~(7) of \cite{Kaganov1957RelaxationLattice}. Using $n=E_{\mathrm{F}}m\kf/(3\pi^2)$ and defining the energy constant 
\begin{equation}
    E_\mathrm{DA}=\frac{\rho s^3}{mD^2},
\end{equation}
we may rewrite Eq.~\eqref{eq:tauE_dimensionless} (see Eq. (14) in MT),
\begin{flalign}
\begin{split}
   \frac{1}{\tau_E^\mathrm{DA}}
&=\frac{T^3
}{E_\mathrm{DA}T_\mathrm{BG}}\frac{3}{\pi^3}
\int_0^{\Tbg/T} dx\,
x^5 \frac{e^{x}}{(e^{x}-1)^2}\\
&=\frac{T^3
}{E_\mathrm{DA}T_\mathrm{BG}}F_\mathrm{DA}\left(\frac{T}{\Tbg}\right),\label{isotropic da dimless variables}
\end{split}
\end{flalign}
where $F_\mathrm{DA}$ captures the integral part and constant $3/\pi^3$.

\subsubsection{Single particle picture of energy relaxation}\label{sec:energy relaxation intuition}
Here we emphasize the structure seen in the integrand of Eq.~\eqref{scat la}. There are two factors of $q$ coming from the measure of the integral, and then there is an additional factor $q^3\sim\omega_\bq^3$: one factor of $\omega_\bq$ comes from the expansion in $\delta T$, one from the initial factor of energy, and one from the integral over $\ve_\bk$. Therefore, the ERR in the linear response regime may be understood as the weighted average of the cubed energy transferred to or from a single electron. The same is true for 2TO phonon scattering as will be discussed in Sec.~\ref{sec:energy relaxation intuition 2TO}.

\subsubsection{Low temperature (Bloch--Grueneisen) limit: $T\ll T_{\mathrm{BG}}$ } 

In the Bloch--Grueneisen regime the upper limit may be extended to infinity:
\begin{equation}
F_\mathrm{DA}(z\rightarrow0)=\frac{3}{\pi^3}\int_0^{\infty} dx\, x^5 \frac{e^{x}}{(e^{x}-1)^2}
=\frac{3}{\pi^3}120\,\zeta(5).
\end{equation}
where $\zeta$ is the Riemann zeta function. Substituting this into Eq.~\eqref{isotropic da dimless variables} yields
\begin{equation}
\boxed{\frac{1}{\tau_E^\mathrm{DA}}
=\frac{T^3}{E_\mathrm{DA}\Tbg}\frac{360\,\zeta(5)}{\pi^3}},
\qquad
T\ll T_{\mathrm{BG}}. \label{isotropic low T da}
\end{equation}
Thus, in the Bloch--Grueneisen regime,
\begin{equation}
\frac{1}{\tau^\mathrm{DA}_E} \propto T^3,\qquad
T\ll T_{\mathrm{BG}}.
\end{equation}

\subsubsection{High temperature (equipartition) limit: $T\gg T_{\mathrm{BG}}$}

In the equipartition regime $x \ll 1$ over the entire integration range.
Upon expanding the Bose factor in small $x$,
the resulting integral is approximated as
\begin{equation}
F_\mathrm{DA}(z\gg1)\approx\frac{3}{\pi^3}\int_0^{1/z} dx\, x^3
=\frac{3}{4\pi^3}\frac{1}{z^4}. \label{high T scaling da isotropic}
\end{equation}
Substituting into Eq.~\eqref{isotropic da dimless variables} yields
\begin{equation}
\boxed{\frac{1}{\tau_E^\mathrm{DA}}
=\frac{T_\mathrm{BG}^3}{E_\mathrm{DA}T}\frac{3}{4\pi^3}},
\qquad
T\gg T_{\mathrm{BG}}. \label{high T err da isotropic}
\end{equation}
Hence, in the equipartition regime,
\begin{equation}
\frac{1}{\tau_E^\mathrm{DA}} \propto T^{-1},
\qquad
T\gg T_{\mathrm{BG}}.
\end{equation}

\subsection{Anisotropic electron dispersion: Intermediate temperature regime}\label{sec: aniso single}
We now consider a $t_{2g}$ electron band structure analogous to that of STO, having three mutually orthogonal ellipsoidal Fermi surface valleys centered at the $\Gamma$ point, with the one along the $z$-axis defined as
\begin{flalign}
    \ve_{z,\bk}=\frac{1}{2}\left(\frac{k_x^2+k_y^2}{m_\perp}+\frac{k_z^2}{m_\parallel}\right), \label{ellipz}
\end{flalign}
and the other two by permutation of $x$, $y$, and $z$. Each has minor axes length $2\kp$ and major axis length $2\kl$ corresponding to now two BG temperatures, $\Tbgp<\Tbgl$, with an intermediate temperature window in between. 

\subsubsection{Generalized Hamiltonian}
We must now generalize Eq.~\eqref{isotropic D int} for this new electron dispersion and its reduced symmetry. We let $\alpha,\beta\in\{x,y,z\}$ label the three Fermi surfaces and define
\[
n_{\alpha\beta}(\mathbf r)\equiv \psi_\alpha^\dagger(\mathbf r)\psi_\beta(\mathbf r),
\qquad
n_\alpha(\mathbf r)\equiv n_{\alpha\alpha}(\mathbf r),
\qquad
N_{\alpha\beta}(\mathbf r)\equiv n_{\alpha\beta}(\mathbf r)+n_{\beta\alpha}(\mathbf r)\;\;(\alpha\neq \beta).
\]
With displacement field $\mathbf u(\mathbf r)$, the symmetric strain tensor to which the electrons couple is
\[
\varepsilon_{ij}(\mathbf r)=\frac{1}{2}\left(\partial_i u_j(\mathbf r)+\partial_j u_i(\mathbf r)\right),\qquad i,j\in\{x,y,z\},
\]
which may be decomposed into hydrostatic and deviatoric (traceless) parts:
\begin{flalign}
\mathrm{Tr}\,\varepsilon(\mathbf r)=\varepsilon_{xx}+\varepsilon_{yy}+\varepsilon_{zz},
\qquad
\varepsilon^{\mathrm{dev}}_{ij}(\mathbf r)=\varepsilon_{ij}(\mathbf r)-\frac{1}{3}\delta_{ij}\,\mathrm{Tr}\,\varepsilon(\mathbf r),
\qquad
\sum_{i,j}\ve_{ij}^{\mathrm{dev}}=2\ve_{xy}+2\ve_{xz}+2\ve_{yz}.\label{hydro terms}
\end{flalign}

\subsubsection*{Intravalley (diagonal) deformation potential}
For the intravalley portion of the interaction potential, we need only consider a single Fermi surface (FS) valley at a time. For an ellipsoidal valley with azimuthal symmetry in the $xy$ plane we have
\begin{flalign}
\begin{split}
H_{\mathrm{intra},z}
&=
\int d^3r\,
n_z(\mathbf r)\left[\Lambda_\perp\left(\ve_{xx}(\br)+\ve_{yy}(\br)\right)+\Lambda_\parallel\ve_{zz}(\br)\right],\label{intra z}
\end{split}
\end{flalign}
where all off-diagonal terms drop out due to mirror symmetry. By rearranging terms using relations in Eq.~\eqref{hydro terms} and defining hydrostatic and uni-axial (deviatoric) constants, respectively 
\begin{flalign}
\Xi_d=\frac{1}{3}\Lambda_\parallel+\frac{2}{3}\Lambda_\perp
\qquad
\mathrm{and}
\qquad
\Xi_u=\Lambda_\parallel-\Lambda_\perp,
\end{flalign}
Eq.~\eqref{intra z} becomes 
\begin{flalign}
\begin{split}
H_{\mathrm{intra},z}
&=
\int d^3r\,
n_z(\mathbf r)\left[
\Xi_d\Tr\,\ve(\br)+\Xi_u\left(\ve_{zz}(\br)-\frac{1}{3}\Tr\ve(\br)\right)
\right].\label{intra z}
\end{split}
\end{flalign}
Notice, for an isotropic system $\Xi_u=0$ and the result in Eq.~\eqref{isotropic D int} is recovered.

\subsubsection*{Intervalley (off-diagonal) shear deformation potential}
On the other hand, given the $t_{2g}$ symmetric triplet at $\Gamma$ in STO, mirror symmetry invariance in the energy forbids diagonal-strain components but allows shear-strain, yielding for the total intervalley interaction
\begin{flalign}
H_{\mathrm{inter}}
=
\int d^3r\;2\Xi_s\left[
\varepsilon_{xy}(\mathbf r)\,N_{xy}(\mathbf r)
+\varepsilon_{yz}(\mathbf r)\,N_{yz}(\mathbf r)
+\varepsilon_{zx}(\mathbf r)\,N_{zx}(\mathbf r)
\right].
\end{flalign}
The total interaction is 
\[
H_{\mathrm{eph}}=H_{\mathrm{intra}}+H_{\mathrm{inter}},
\]
where $H_{\mathrm{intra}}$ includes contributions from all three FS valleys.

\subsubsection*{Fourier transforms and phonon normal modes}
The wavefunctions now include surface index $\alpha$,
\[
\psi_\alpha(\mathbf r)=\frac{1}{\sqrt V}\sum_{\mathbf k} c_{\alpha\mathbf k}\,e^{i\mathbf k\cdot\mathbf r},
\qquad
\psi_\alpha^\dagger(\mathbf r)=\frac{1}{\sqrt V}\sum_{\mathbf k} c_{\alpha\mathbf k}^\dagger\,e^{-i\mathbf k\cdot\mathbf r}.
\]
The displacement field is quantized in terms of normal modes with dispersion $\omega_{\mathbf q\lambda}$, polarization $\mathbf e_{\mathbf q\lambda}$, and mass density $\rho$:
\[
\mathbf u(\mathbf r)=\frac{1}{\sqrt V}\sum_{\mathbf q,\lambda}\mathbf u_{\mathbf q\lambda}\,e^{i\mathbf q\cdot\mathbf r},
\qquad
\mathbf u_{\mathbf q\lambda}
=
\sqrt{\frac{1}{2\rho\,\omega_{\mathbf q\lambda}}}\;
\mathbf e_{\mathbf q\lambda}\,\left(b_{\mathbf q\lambda}+b_{-\mathbf q\lambda}^\dagger\right).
\]
Then the symmetric strain tensor is
\begin{flalign}
\varepsilon_{ij}(\mathbf r)=\frac{1}{\sqrt V}\sum_{\mathbf q,\lambda}\varepsilon_{ij,\bq\lambda}\,e^{i\mathbf q\cdot\mathbf r},
\qquad
\varepsilon_{ij,\bq\lambda}=\frac{i}{2}\left(q_i u_{j,\mathbf q\lambda}+q_j u_{i,\mathbf q\lambda}\right), \label{ve fourier}
\end{flalign}
where $u_{i,\mathbf q\lambda}$ is the $i$th component of $\mathbf u_{\mathbf q\lambda}$.
In particular,
\begin{flalign}
\mathrm{Tr}\,\varepsilon_{\bq\lambda}=i\,\mathbf q\cdot\mathbf u_{\mathbf q\lambda}
=
i\sqrt{\frac{1}{2\rho\,\omega_{\mathbf q\lambda}}}\;
(\mathbf q\cdot\mathbf e_{\mathbf q\lambda})\left(b_{\mathbf q\lambda}+b_{-\mathbf q\lambda}^\dagger\right). \label{ve op}
\end{flalign}

\subsubsection*{Polarizations and acoustic dispersions}
No longer can we only consider the longitudinal branch. We assume isotropic acoustic modes with one longitudinal (L) and two degenerate transverse (T) branches:
\[
\omega_{\bq \mathrm{L}}=s_\mathrm{L} q,\qquad \omega_{\bq \mathrm{T}}=s_\mathrm{T} q,
\qquad q\equiv |\bq|.
\]
The polarization vectors may then be defined as
\[
\be_{\bq \mathrm{L}}=\hat{\bq},\qquad \be_{\bq \mathrm{T}_{1,2}}\cdot \hat{\bq}=0,\qquad
\sum_{\lambda=\mathrm{L},\mathrm{T}_1,\mathrm{T}_2} e_{i,\bq\lambda}e_{j,\bq\lambda}=\delta_{ij}.
\]
Summing the transverse components gives the projection operator,
\[
\sum_{t=\mathrm{T}_1,\mathrm{T}_2} e_{i,\bq t}e_{j,\bq t}=P_{ij}(\hat{\bq})\equiv \delta_{ij}-\hat q_i\hat q_j.
\]

\subsubsection*{Momentum-space Hamiltonian and vertices}
Upon inserting the Fourier expansions and performing the real space integral, one obtains the general eph Hamiltonian,
\[
H_{\mathrm{eph}}
=
\sum_{\mathbf k,\mathbf q,\lambda}\sum_{\alpha,\beta}
g^{\alpha\beta}_{\bq,\lambda}\;
c_{\beta,\mathbf k+\mathbf q}^\dagger\,c_{\alpha,\mathbf k}\;
\left(b_{\mathbf q\lambda}+b_{-\mathbf q\lambda}^\dagger\right),
\]
with intravalley and intervalley vertices defined next.

\subsubsection{Matrix elements}
\subsubsection*{Intravalley vertex $g^{\alpha\alpha}_{\bq\lambda}$ and squared matrix elements (Longitudinal and Transverse)}
Inserting Eq.~\eqref{ve fourier} and Eq.~\eqref{ve op} in the definition of $H_{\mathrm{intra}}$ in Eq.~\eqref{intra z} yields for a given valley $\alpha$,
\begin{flalign}
\begin{split}
g^{\alpha\alpha}_{\bq\lambda}
&=
i\sqrt{\frac{1}{2\rho\,\omega_{\bq\lambda}}}\;
\Big[
\Xi_D\,(\bq\cdot\be_{\bq\lambda})
+
\Xi_u\, q_\alpha\, e_{\alpha,\bq\lambda}
\Big],
\qquad
\Xi_D\equiv \Xi_d-\frac13\Xi_u,
\end{split}
\end{flalign}
where 
\[
q_\alpha\equiv \bq\cdot\hat{\alpha}=q\cos\theta_\alpha
\qquad \mathrm{and} \qquad
e_{\alpha,\bq\lambda}\equiv \be_{\bq\lambda}\cdot\hat{\alpha}
\]
are the projections of the momentum transfer and displacement unit vector onto the major axis (associated with dispersion $\ve_{\alpha,\bk}$).

\paragraph{Longitudinal (L)}
For $\be_{\bq \mathrm{L}}=\hat{\bq}$,
\[
\bq\cdot\be_{\bq \mathrm{L}}=q,\qquad e_{\alpha,\bq \mathrm{L}}=\hat{\bq}\cdot\hat{\alpha}=\cos\theta_\alpha,
\]
hence
\begin{flalign}
\big|g^{\alpha\alpha}_{\bq \mathrm{L}}\big|^2
=
\frac{1}{2\rho\,\omega_{\bq \mathrm{L}}}
\Big(q\Xi_D+\Xi_u\,q\cos^2\theta_\alpha\Big)^2
=
\frac{q}{2\rho\,s_\mathrm{L}}\Big(\Xi_D+\Xi_u\cos^2\theta_\alpha\Big)^2.\label{sq long matx}
\end{flalign}

\paragraph{Transverse (T).}
For any transverse mode $t$, $\bq\cdot\be_{\bq t}=0$, so the hydrostatic part drops out and
\begin{flalign}
g^{\alpha\alpha}_{\bq\mathrm{T}_{1,2}}
=
i\,\Xi_u \sqrt{\frac{1}{2\rho\,\omega_{\bq \mathrm{T}_{1,2}}}}\;q_{\alpha}\, e_{\alpha,\bq \mathrm{T}_{1,2}}.
\end{flalign}
Using the projector, we sum over the two degenerate transverse vectors:
\[
\sum_{t=\mathrm{T}_1,\mathrm{T}_2} (e_{\alpha,\bq t})^2
=
\sum_{t=\mathrm{T}_1,\mathrm{T}_2} (\hat{\alpha}\cdot\be_{\bq t})^2
=
\hat{\alpha}_i P_{ij}(\hat{\bq})\hat{\alpha}_j
=
1-(\hat{\alpha}\cdot\hat{\bq})^2
=
\sin^2\theta_\alpha.
\]
Therefore, the two-mode transverse \emph{sum} is
\begin{flalign}
\sum_{t=\mathrm{T}_1,\mathrm{T}_2} \big|g^{\alpha\alpha}_{\bq t}\big|^2
=
\frac{\Xi_u^2}{2\rho\,\omega_{\bq \mathrm{T}}}\;q_{\alpha}^2\,\sin^2\theta_\alpha
=
\frac{q}{2\rho\,s_\mathrm{T}}\;\Xi_u^2\,\cos^2\theta_\alpha\sin^2\theta_\alpha.\label{sq tran matx}
\end{flalign}

\subsubsection*{Intervalley shear vertex $g^{\alpha\beta}_{\bq\lambda}$, $\alpha\neq \beta$, $ (\alpha\beta)\in\{(xy),(yz),(zx)\}$ and squared matrix elements (L and T)}
For the allowed pairs we have
\begin{flalign}
g^{\alpha\beta}_{\bq\lambda}
=
i\,\Xi_s\sqrt{\frac{1}{2\rho\,\omega_{\mathbf q\lambda}}}\;
\left(q_\alpha e_{\beta,\mathbf q\lambda}+q_\beta e_{\alpha,\mathbf q\lambda}\right),
\qquad \alpha\neq \beta,\;\;(\alpha\beta)\in\{xy,yz,zx\},
\end{flalign}
where it is evident $g^{\alpha\beta}=g^{\beta\alpha}$. The squared matrix elements are
\begin{flalign}
\begin{split}
\left|g^{\alpha\beta}_{\bq\lambda}\right|^2
&=
\frac{\Xi_s^2}{2\rho\,\omega_{\mathbf q\lambda}}
\left(q_\alpha e_{\beta,\mathbf q\lambda}
+q_\beta e_{\alpha,\mathbf q\lambda}\right)^2. 
\end{split}
\end{flalign}

\paragraph{Longitudinal (L).}
With $\be_{\bq \mathrm{L}}=\hat{\bq}$, $e_{i,\bq \mathrm{L}}=q_i/q$, one can write
\[
q_\alpha e_{\beta,\bq \mathrm{L}}+q_\beta e_{\alpha,\bq \mathrm{L}}
=
q_\alpha\frac{q_\beta}{q}+q_\beta\frac{q_\alpha}{q}
=
\frac{2q_\alpha q_\beta}{q},
\]
hence
\begin{flalign}
\left|g^{\alpha\beta}_{\bq \mathrm{L}}\right|^2
=
\frac{\Xi_s^2}{2\rho\,\omega_{\bq \mathrm{L}}}\left(\frac{2q_\alpha q_\beta}{q}\right)^2
=
\frac{q}{\rho\,s_\mathrm{L}}2\Xi_s^2\cos^2\theta_\alpha\cos^2\theta_\beta.\label{sq long matx inter}
\end{flalign}

\paragraph{Transverse (T): two-mode sum.}
Summing over the two degenerate transverse polarizations using $P_{ij}(\hat{\bq})$ gives:
\[
\sum_{t=\mathrm{T_1},\mathrm{T_2}} \left(q_\alpha e_{\beta,\bq t}+q_\beta e_{\alpha,\bq t}\right)^2
=
q_\alpha^2 P_{\beta\beta}+q_\beta^2 P_{\alpha\alpha}+2q_\alpha q_\beta P_{\alpha\beta}.
\]
For $\alpha\neq\beta$,
\[
P_{\alpha\alpha}=1-\frac{q_\alpha^2}{q^2},\qquad
P_{\beta\beta}=1-\frac{q_\beta^2}{q^2},\qquad
P_{\alpha\beta}=-\frac{q_\alpha q_\beta}{q^2},
\]
so
\[
\sum_{t=\mathrm{T}_1,\mathrm{T}_2} \left(q_\alpha e_{\beta,\bq t}+q_\beta e_{\alpha,\bq t}\right)^2
=
q_\alpha^2+q_\beta^2-\frac{4q_\alpha^2 q_\beta^2}{q^2}.
\]
Therefore the two-mode transverse \emph{sum} is
\begin{flalign}
\sum_{t=\mathrm{T}_1,\mathrm{T}_2} \left|g^{\alpha\beta}_{\bq t}\right|^2
=
\frac{\Xi_s^2}{2\rho\,\omega_{\bq \mathrm{T}}}
\left(q_\alpha^2+q_\beta^2-\frac{4q_\alpha^2 q_\beta^2}{q^2}\right)
=
\frac{q}{2\rho\,s_\mathrm{T}}\;\Xi_s^2\left(\cos^2\theta_\alpha+\cos^2\theta_\beta-4\cos^2\theta_\alpha\cos^2\theta_\beta\right).\label{sq tran matx inter}
\end{flalign}

\subsubsection{Collision integral and power transfer (most general multivalley form)}

\paragraph{Collision integral } 
\paragraph{Scattering matrix elements (Born approximation)}
An initial state is denoted by $|\alpha,\mathbf k; \{N_{\mathbf q\lambda}\}\rangle$ and final state by
$|\beta,\bk'; \{N'_{\mathbf q\lambda}\}\rangle$,
where $\{N_{\mathbf q\lambda}\}$ is the set of all Bose occupation numbers,
\[
N_{\mathbf q\lambda}=\frac{1}{e^{\omega_{\mathbf q\lambda}/k_BT}-1}.
\]
On applying FGR, the eph collision integral that obeys fermion statistics and includes both in and out processes is
\begin{flalign}
\begin{split}
\frac{\partial f_{\alpha,\bk}}{\partial t}
&=2\pi\sum_{\beta,\mathbf q,\lambda}
\left|g^{\alpha\beta}_{\bq,\lambda}\right|^2
\Bigg\{
N_{\mathbf q\lambda}\,
\Big[
f_{\beta,\bk-\bq}\big(1-f_{\alpha,\bk}\big)\,
\delta\big(\varepsilon_{\alpha,\mathbf k}-\varepsilon_{\beta,\mathbf k-\mathbf q}-\omega_{\mathbf q\lambda}\big)
\\
&\hspace{3.2cm}
-
f_{\alpha,\bk}\big(1-f_{\beta,\bk+\bq}\big)\,
\delta\big(\varepsilon_{\beta,\mathbf k+\mathbf q}-\varepsilon_{\alpha,\mathbf k}-\omega_{\mathbf q\lambda}\big)
\Big]
\\
&\hspace{1.6cm}
+
(N_{\mathbf q\lambda}+1)\,
\Big[
f_{\beta,\bk-\bq}\big(1-f_{\alpha,\bk}\big)\,
\delta\big(\varepsilon_{\alpha,\mathbf k}-\varepsilon_{\beta,\mathbf k-\mathbf q}+\omega_{\mathbf q\lambda}\big)
\\
&\hspace{3.2cm}
-
f_{\alpha,\bk}\big(1-f_{\beta,\bk+\bq}\big)\,
\delta\big(\varepsilon_{\beta,\mathbf k+\mathbf q}-\varepsilon_{\alpha,\mathbf k}+\omega_{\mathbf q\lambda}\big)
\Big]
\Bigg\}.
\end{split}
\end{flalign}
where $\varepsilon_{\alpha,\mathbf k}$ is the electron dispersion on valley $\alpha$, $f_{\alpha,\bk}$ is the electronic distribution function on valley $\alpha$, and the first and second lines correspond to out processes while the third and fourth lines correspond to in processes. Notice, a common factor $|g^{\alpha\beta}_{\bq\lambda}|^2$ has been pulled out for both intravalley and intervalley coupling types given their definitions above along with the relations $\omega_{\bq\lambda}=\omega_{-\bq\lambda}$ and $\be_{\bq\lambda}=\be_{-\bq\lambda}$. 
Note, this is the most general collision integral for single-phonon deformation potential scattering,
valid for a generic electron dispersion $\varepsilon_{a,\mathbf k}$.
The intravalley and intervalley structure enters entirely through the vertices $g^{\alpha\beta}_{\bq\lambda}$ given above.

\paragraph{Energy exchange rate} 
Upon inserting the collision integral into Eq.~\eqref{power}, we now have a generalized power transfer
\begin{flalign}
\begin{split}
\frac{\partial E_e}{\partial t}
&=4\pi\sum_{\alpha,\bk,\beta,\mathbf q,\lambda}
\ve_{\alpha,\bk}\left|g^{\alpha\beta}_{\bq,\lambda}\right|^2
\Bigg\{
N_{\mathbf q\lambda}\,
\Big[
f_{\beta,\bk-\bq}\big(1-f_{\alpha,\bk}\big)\,
\delta\big(\varepsilon_{\alpha,\mathbf k}-\varepsilon_{\beta,\mathbf k-\mathbf q}-\omega_{\mathbf q\lambda}\big)
\\
&\hspace{3.2cm}
-
f_{\alpha,\bk}\big(1-f_{\beta,\bk+\bq}\big)\,
\delta\big(\varepsilon_{\beta,\mathbf k+\mathbf q}-\varepsilon_{\alpha,\mathbf k}-\omega_{\mathbf q\lambda}\big)
\Big]
\\
&\hspace{1.6cm}
+
(N_{\mathbf q\lambda}+1)\,
\Big[
f_{\beta,\bk-\bq}\big(1-f_{\alpha,\bk}\big)\,
\delta\big(\varepsilon_{\alpha,\mathbf k}-\varepsilon_{\beta,\mathbf k-\mathbf q}+\omega_{\mathbf q\lambda}\big)
\\
&\hspace{3.2cm}
-
f_{\alpha,\bk}\big(1-f_{\beta,\bk+\bq}\big)\,
\delta\big(\varepsilon_{\beta,\mathbf k+\mathbf q}-\varepsilon_{\alpha,\mathbf k}+\omega_{\mathbf q\lambda}\big)
\Big]
\Bigg\}.
\end{split}
\end{flalign}
This result includes 6 different scattering processes ($x\leftrightarrow y,x\leftrightarrow z,y\leftrightarrow z,x\rightarrow x,y\rightarrow y,z\rightarrow z$) where the double arrow denotes scattering both to and from ellipsoids with major axis $\alpha$ and $\beta$. Each combination can be reduced to the same form as that in Eq.~\eqref{exchange},
\begin{flalign}
    \begin{split}
        \frac{\partial E_e^{\mathrm{\alpha\beta}}}{\partial t} = -4\pi \sum_{\bk,\bq,\lambda}\left|g^{\alpha\beta}_{\bq\lambda}\right|^2 \omega_{\bq\lambda}
        \left[f_{\beta,\bk+\bq}\left(1-f_{\alpha,\bk}\right)-N_{\bq\lambda}\left(f_{\alpha,\bk}-f_{\beta,\bk+\bq}\right)\right]
        \delta(\ve_{\alpha,\bk}-\ve_{\beta,\bk+\bq}+\omega_{\bq\lambda}). \label{exchange general}
    \end{split}
\end{flalign}

\subsubsection{Energy relaxation rate} \label{sec: intravalley contribution}
Finally, we calculate the ERR. Here we mention that all squared matrix elements, shown in Eqs.~\eqref{sq long matx}, \eqref{sq tran matx}, \eqref{sq long matx inter}, and \eqref{sq tran matx inter}, have the same $q$ dependence, differing only in their angular dependence and prefactors. Out of convenience, we let $\Xi_d\gg\Xi_{u,s}$ so that we may neglect all but the longitudinal intravalley scattering events. We thereby set $\Xi_D=D$ and $s_\mathrm{L}=s_\mathrm{a}$ to specify an acoustic mode in contrast to the value $s$ defined in the context of 2TO scattering. In Sec.~\ref{subsec:intervalley da} we show that the temperature scaling is indeed the same for intervalley events.
For scattering within a single valley, Eq.~\eqref{ellipz} may be written
\begin{flalign}
    \ve_\bk=\frac{1}{2}\left(\frac{k_\perp^2}{m_\perp}+\frac{k_\parallel^2}{m_\parallel}\right), \label{ellip par}
\end{flalign}
and we can rescale the major axis so that the spectrum is isotropic in $\kt$,
\begin{flalign}
    \ve_\bk=\frac{\kt^2}{2m_\perp}=\ve_{\kt}.\label{kt}
\end{flalign}
The sum over $\bk$ in Eq.~\eqref{exchange general} may be transformed to the following integral, $(1/V)\sum_\bk=\int d^3k/(2\pi)^3=\sqrt{m_\parallel/m_\perp}\int d^3\kt/(2\pi)^3$, where $\kt_{\text{F}}=\kp$. The integral over solid angle of the rescaled variable $\kt$, whose polar angle is $\tilde\theta$, is then given by
\begin{flalign}
    2\pi\int d(\cos\tilde{\theta})\,\delta\left(\frac{-1}{m_\perp}\left(\kt\qt\cos\tilde{\theta}+\frac{\qt^2}{2}\right)+ s_\mathrm{a} q\right),
\end{flalign}
which gives a factor of $2\pi m_\perp/\kt\qt$ in front and enforces the following kinematic constraint on the magnitude of $q$,
\begin{flalign}
    \left|\frac{s_\mathrm{a}}{\vfp}\frac{q}{\qt}-\frac{\qt}{2\kt}\right|<1,
\end{flalign}
where $\vfp=\kp/m_\perp$ is the Fermi velocity along the minor axis and in the degenerate electron regime, $\omega_\bq\lesssim T\ll E_\mathrm{F}$, i.e. $s_\mathrm{a}\ll\varv_{\mathrm{F}\perp,\parallel}$.
There are now two limiting cases depending on the angle of q: (i) since $q^{\text{max}}_\perp=\qt^{\text{max}}$, it is clear that the kinematic constraint along the minor axis is given by
\begin{flalign}
    \qp<2\kp,
\end{flalign}
corresponding to the lesser of the two Bloch-Grueneisen temperatures, $\Tbgp=2 s_\mathrm{a}\kp$, and (ii) when $q=q_\parallel=\sqrt{m_\parallel/m_\perp}\qt$, the relation $\sqrt{m_\parallel/m_\perp}\kp=\kl$ gives the kinematic constraint along the major axis,
\begin{flalign}
    q_\parallel<2\kl,
\end{flalign}
where the larger Bloch-Grueneisen temperature is then $\Tbgl=2s_\mathrm{a}\kl$. Thus, in the case of a highly elongated ellipsoid where $\Tbgl\gg\Tbgp$, there may be an intermediate temperature range of considerable extent one need consider, namely $\Tbgp\ll T\ll\Tbgl$. In general, one can determine the bounds on $q$ as a function of its polar angle $\theta$, 
\begin{flalign}
    0<q<2\kp\left(\frac{1}{\sqrt{1-(1-\mu)\cos^2\theta}}+\frac{s_\mathrm{a}/\vfp}{1-(1-\mu)\cos^2\theta}\right)\approx\frac{2\kp}{\sqrt{1-(1-\mu)\cos^2\theta}},\quad \mu=\frac{m_\perp}{m_\parallel},
\end{flalign}
where the rescaled momentum transfer is $\qt=q\sqrt{\sin^2\theta+\mu\cos^2\theta}$, and we assume $s_\mathrm{a}/\vfp\ll1$. The integral over $\kt$ can be done just as in the usual isotropic case where one linearizes the spectrum about $\kt_{\text{F}}=\kp$ so that $d\kt\kt^2\approx \kp m_\perp d\ve_{\kt}$. The integral over electron energy is performed just as in Eq.~\eqref{energy},
\begin{flalign}
    \int d\ve_{\kt} \left[f(\ve_{\kt}+\omega_q)(1-f(\ve_{\kt}))-N(\omega_q,T)\left(f(\ve_{\kt})-f(\ve_{\kt}+\omega_q)\right)\right]=\omega_q\bigl[N(\omega_q,T_e)-N(\omega_q,T)\bigr].
\end{flalign}  
Upon inserting the matrix elements from Eq.~\eqref{sq long matx}, relabeling $\cos\theta=t$, expanding as in Eq.~\eqref{expand N}, and using Eq.~\eqref{generic rate} we write the ERR due to electrons coupling to single acoustic phonons within a single Fermi surface valley,
\begin{flalign}
    \frac{1}{\tau^{\mathrm{DA}}_E}=\frac{D^2s_\mathrm{a}^2m_\perp\sqrt{m_\perp m_\parallel}E_\mathrm{F}}{4\pi^5\rho n}\frac{1}{T^3}
    \int_{-1}^1dtA(\mu,t)\int_0^{2\kp A(\mu,t)} dqq^5\frac{e^{sq/ T}}{(e^{sq/ T}-1)^2},\label{scat la anisotropic}
\end{flalign}
where 
\begin{flalign}
   A(\mu,t)=\left(1-(1-\mu) t^2\right)^{-1/2}. \label{anisotropic function}
\end{flalign}
It is clearly evident how Eq.~\eqref{scat la anisotropic} reduces to the isotropic form in Eq.~\eqref{scat la} when $\mu=1$. We rewrite Eq.~\eqref{scat la anisotropic} in terms of a rescaled integration variable $x=sq/T$ as follows (see Eq. (16) in MT),
\begin{flalign}
\begin{split}
    \frac{1}{\tau^{\mathrm{DA}}_E}&=
    \frac{m_\perp D^2}{\rho s_\mathrm{a}^3 \Tbgp}T^3
    \frac{3}{2\pi^3}\int_{-1}^1dt A(\mu,t)
    \int_0^{A(\mu,t)\Tbgp/T} dxx^5
    \frac{e^{x}}{(e^{x}-1)^2}
    \\
    &=\frac{T^3}{E_{\mathrm{DA},\perp}T_\mathrm{BG\perp}} \bar F_\mathrm{DA}\left(\frac{T}{\Tbgp},\mu\right), \label{anisotropic da dimless variables}
    \end{split}
\end{flalign}
where $\bar F_\mathrm{DA}$ denotes anisotropy as compared to $F_\mathrm{DA}$, and we used $n=E_{\mathrm{F}}\sqrt{m_\perp m_\parallel}\kp/(3\pi^2)$ to define $E_\mathrm{DA,\perp}=\rho s_\mathrm{a}^2/m_\perp D^2$. 

To properly compare Eq.~\eqref{anisotropic da dimless variables} with Eq.~\eqref{isotropic da dimless variables} and later on Eq.~\eqref{isotropic 2to dimless variables}, we use the definitions of dispersion in the isotropic and anisotropic cases (Eq.~\eqref{parabolic} and Eq.~\eqref{ellip par}, respectively) and the subsequent relation under the assumption that the volume of the Fermi sea is preserved under deformation,
\begin{flalign}
    \kf^3=k_\mathrm{F\perp}^2\kl.
\end{flalign}
Combining this with the relation $\kl/\kp=\sqrt{m_\parallel/m_\perp}$ we have 
\begin{flalign}
    \frac{\kf}{\kp}=\frac{1}{\mu^{1/6}}.
\end{flalign}
With $T_\mathrm{BG\perp}^2/E_{\mathrm{DA},\perp}=(k_\mathrm{F\perp}^2m_\perp/\kf^2m)(T_\mathrm{BG}^2/E_\mathrm{DA})=\mu^{2/3}(T_\mathrm{BG}^2/E_\mathrm{DA})$ (under the assumption $D$ is the same in both isotropic and anisotropic cases), we can write Eq.~\eqref{anisotropic da dimless variables} with the same variables and in the same units as Eq.~\eqref{isotropic da dimless variables},
\begin{flalign}
\begin{split}
   \frac{1}{\tau_E^{\mathrm{DA}}} &= 
    \frac{T^3}{E_{\mathrm{DA}}\Tbg}
   \mu^{1/6}\bar F_\mathrm{DA}\left(\frac{z}{\mu^{1/6}},\mu\right) .\label{anisotropic da dimless variables}
    \end{split} 
\end{flalign}

\subsubsection{Asymptotic limits: intravalley contribution}
Here we define the asymptotic limits of the scaling function, 
\begin{flalign}
\begin{split}
    \bar F_\mathrm{DA}(z,\mu)&=
    \frac{3}{2\pi^3}\int_{-1}^1 \frac{dt}{\sqrt{1-(1-\mu)t^2}}\int_0^{1/\left(z\sqrt{1-(1-\mu)t^2}\right)} dxx^5 \frac{e^x}{(e^x-1)^2},\label{scaling da}
    \end{split}
\end{flalign}
and subsequent ERR, where $\mu\leq1$. It will be useful to let
\[
w=\sqrt{1-(1-\mu)t^2},
\]
and given the integrand is even with respect to $t$, we have
\begin{flalign}
\begin{split}
    \bar F_\mathrm{DA}(z,\mu)&=
    \frac{3}{\pi^3}\frac{1}{\sqrt{1-\mu}}\int_{\sqrt\mu}^1 \frac{dw }{\sqrt{1-w^2}}J\left(\frac{1}{z w}\right),\label{scaling da}
    \end{split}
\end{flalign}
where 
\[
J(a)\equiv\int_0^{a} dxx^5 \frac{e^x}{(e^x-1)^2}.
\]

\subsubsection*{Low temperature limit: $T\ll\Tbgp,\Tbgl$ } \label{subsec:low T da}
In the BG regime, where $T\ll\Tbgp$ ($T$ is below the lesser of the two BG temperatures) the upper limit in the integral over $x$ can be taken to infinity for all $w$ and the asymptotic limit of the scaling function is 
\begin{flalign}
\begin{split}
    \bar F_\mathrm{DA}(z\ll1,\mu)&=
    \frac{3}{\pi^3}\frac{1}{\sqrt{1-\mu}}\int_{\sqrt\mu}^1 \frac{dw}{\sqrt{1-w^2}}\int_0^\infty dx x^5\frac{e^x}{(e^x-1)^2}\\
    &
    =\frac{360\zeta(5)}{\pi^3}M_\mathrm{L}(\mu),\qquad 
    M_\mathrm{L}(\mu)=\frac{\arccos(\sqrt\mu)}{\sqrt{1-\mu}}.
    \end{split}
\end{flalign}
The ERR is therefore
\begin{flalign}
    \boxed{\frac{1}{\tau^{\mathrm{DA}}_E}
    =\frac{T^3}{E_{\mathrm{DA},\perp}\Tbgp}\frac{360\zeta(5)}{\pi^3}M_\mathrm{L}(\mu)},\qquad T\ll\Tbgp ,
\end{flalign}
which clearly yields the isotropic result in Eq.~\eqref{isotropic low T da} when $\mu=1$ ($M_\mathrm{L}(1)=1$).  
In summary, in the BG regime we have
\begin{flalign}
    \frac{1}{\tau^{\mathrm{DA}}_E}\propto T^3,\qquad T\ll\Tbgp.
\end{flalign}

\subsubsection*{High temperature limit: $T\gg\Tbgp,\Tbgl$}
In the equipartition regime, $\Tbgp/Tw\ll1$ for all $w$---at the upper limit, $w=1$, the argument of $J$ is $ \Tbgp/T\ll1$ and at the lower limit, $w=\sqrt{\mu}=\Tbgp/\Tbgl$, the argument of $J$ is $\Tbgl/T\ll1$. Then the Bose functions may be expanded everywhere so $J(1/zw\ll1)\approx 1/4z^4w^4$. The scaling function becomes
\begin{flalign}
\begin{split}
    \bar F_\mathrm{DA}(z\gg1,\mu\gg1/z^2)&=
    \frac{3}{4\pi^3z^4}\frac{1}{\sqrt{1-\mu}}\int_{\sqrt\mu}^1 \frac{dw}{w^4\sqrt{1-w^2}}\\
    &=\frac{3}{4\pi^3z^4}M_\mathrm{H}(\mu),
    \end{split}
\end{flalign}
where
\begin{flalign}
    M_\mathrm{H}(\mu)=\frac{2+\frac{1}{\mu}}{3\sqrt{\mu}}.
\end{flalign}
The ERR is 
\begin{flalign}
\begin{split}
\boxed{\frac{1}{\tau_E^\mathrm{DA}}
=\frac{T_\mathrm{BG\perp}^3}{E_\mathrm{DA,\perp}T}\frac{3}{4\pi^3} M_\mathrm{H}(\mu)},\qquad
T\gg \Tbgp,\Tbgl.
\end{split}
\end{flalign}
In the isotropic case, $\mu=1$ and $M_\mathrm{H}(1)=1$ and the result in 
Eq.~\eqref{high T err da isotropic} is recovered. In the strongly anisotropic limit, $M_\mathrm{H}(\mu\rightarrow0)$ diverges and so too would the energy relaxation rate. However, in reality even at large $T/\Tbgp\gg1$, if $T/\Tbgp$ remains finite and $\mu$ is sufficiently decreased, eventually $\sqrt{\mu}T/\Tbgp=T/\Tbgl$ will no longer be $\gg1$ and one enters the intermediate regime to be discussed below. In summary, we have
\begin{equation}
\frac{1}{\tau_E^\mathrm{DA}} \propto T^{-1},\qquad
T\gg \Tbgp,\Tbgl.
\end{equation}

\subsubsection*{Intermediate temperature limit: $\Tbgp\ll T\ll\Tbgl$} \label{subsec:intermed T da}
Now at the lower limit, $w=\sqrt\mu=\Tbgp/\Tbgl\ll \Tbgp/T\ll1$ so that the $w$ interval now includes a region $w\lesssim\Tbgp/T\ll1$ where the argument of $J$, and hence $J$ itself, is $O(1)$. We may expand in small $w$ and $\mu$ so that $\sqrt{1-w^2}\approx1$ and $\sqrt{1-\mu}\approx1$. We let $r=z_\perp w$ and then have for the scaling function,
\begin{flalign}
\begin{split}
    \bar F_\mathrm{DA}(z\gg 1,\mu\ll1/z^2)&\approx
    \frac{3}{\pi^3z}
    \int_{\sqrt\mu z}^{z} drJ(1/r)\\
    &\approx\frac{3}{\pi^3 z}
    \int_0^\infty dr \int_0^{1/r}dxx^5\frac{e^x}{(e^x-1)^2}.
    \end{split}
\end{flalign} 
By swapping the order of integration, the integral becomes
\begin{flalign}
     \int_0^{\infty} dxx^5\frac{e^x}{(e^x-1)^2}\int_0^{1/x} dw=\int_0^\infty dx x^4 \frac{e^x}{(e^x-1)^2}=\frac{4\pi^4}{15}.
\end{flalign}
Then we have for the scaling function,
\begin{flalign}
\begin{split}
    \bar F_\mathrm{DA}(z\gg 1,\mu\ll1/z^2)&=
    \frac{4\pi}{5z}.
    \end{split}
\end{flalign} 
The ERR is
\begin{flalign}
\begin{split}
   \boxed{\frac{1}{\tau_E^{\mathrm{DA}}} =\frac{T^2}{E_{\mathrm{DA,\perp}}}
   \frac{4\pi}{5}}, \qquad (\Tbgp\ll T\ll\Tbgl). \label{intermediate DA asymptotic}
    \end{split} 
\end{flalign}
In the next section we show how this result is produced by taking the cylindrical limit, $\mu=0$.
In summary, we have quadratic temperature scaling in the intermediate regime,
\begin{flalign}
    \frac{1}{\tau_E^\mathrm{DA}}\propto T^2, \qquad (\Tbgp\ll T\ll\Tbgl).
\end{flalign}

\subsubsection{Cylindrical limit: $\mu=0$} 
At $\mu=0$, the scaling function is
\begin{flalign}
\begin{split}
    \bar F_\mathrm{DA}(z,\mu=0)&=
    \frac{3}{2\pi^3}\int_{-1}^1 \frac{dt}{\sqrt{1-t^2}}\int_0^{1/\left(z\sqrt{1-t^2}\right)} dxx^5 \frac{e^x}{(e^x-1)^2},
    \end{split}
\end{flalign}
which is divergent. On transforming to cylindrical coordinates we have,
\begin{flalign}
    A(\mu=0,\theta)=\frac{1}{\sqrt{1-t^2}}=\frac{1}{\sin\theta}=\frac{x}{\xp}.\label{A def}
\end{flalign}
where the Fermi surface is transformed to a cylindrical one cut at $\pm\kl$. 
Then 
\begin{flalign}
\begin{split}
   &\int_{-1}^1 \frac{dt}{\sqrt{1-t^2}}\int_0^{1/\left(z\sqrt{1-t^2}\right)} dxx^5 \frac{e^x}{(e^x-1)^2}
   \quad \rightarrow\quad
   \int_{0}^{1/z}dx_\perp\int_{-1/z_\parallel}^{1/z_\parallel}dx_\parallel x^4\frac{e^x}{(e^x-1)^2}. \label{cylindrical limit}
   \end{split}
\end{flalign}
where $z_\parallel=T/\Tbgl$ and $x=\sqrt{x_\perp^2+x_\parallel^2}$. On taking the limit $\Tbgl/T\rightarrow\infty$ and noting $x_\perp\leq \Tbgp/T\ll1$, we may expand $x\approx x_\parallel$ which yields
\begin{flalign}
\begin{split}
    \bar F_\mathrm{DA}(z,\mu=0)&=
    \frac{3}{\pi^3}\frac{1}{z}
    \int_0^\infty dx_\parallel x_\parallel^4\frac{e^{x_\parallel}}{(e^{x_\parallel}-1)^2}\\
    &=
    \frac{4\pi}{5z}.
    \end{split}
\end{flalign}
Equation~\eqref{intermediate DA asymptotic} is clearly recovered.
Notice, in the low temperature limit, temperature dependence in the scaling function drops out as one may take both $1/z,1/z_\parallel \rightarrow\infty$ in Eq.~\eqref{cylindrical limit} and the $T^3$ results ensues. Similarly, in the equipartition regime, the Bose factors may be expanded since both $x_{\perp,\parallel}\ll1$ in the entire interval. Then the $1/T$ result ensues up to a numerical prefactor.

\subsubsection{Intervalley contribution} \label{subsec:intervalley da}
\begin{figure}[t]
\centering
  \includegraphics[scale=0.1]{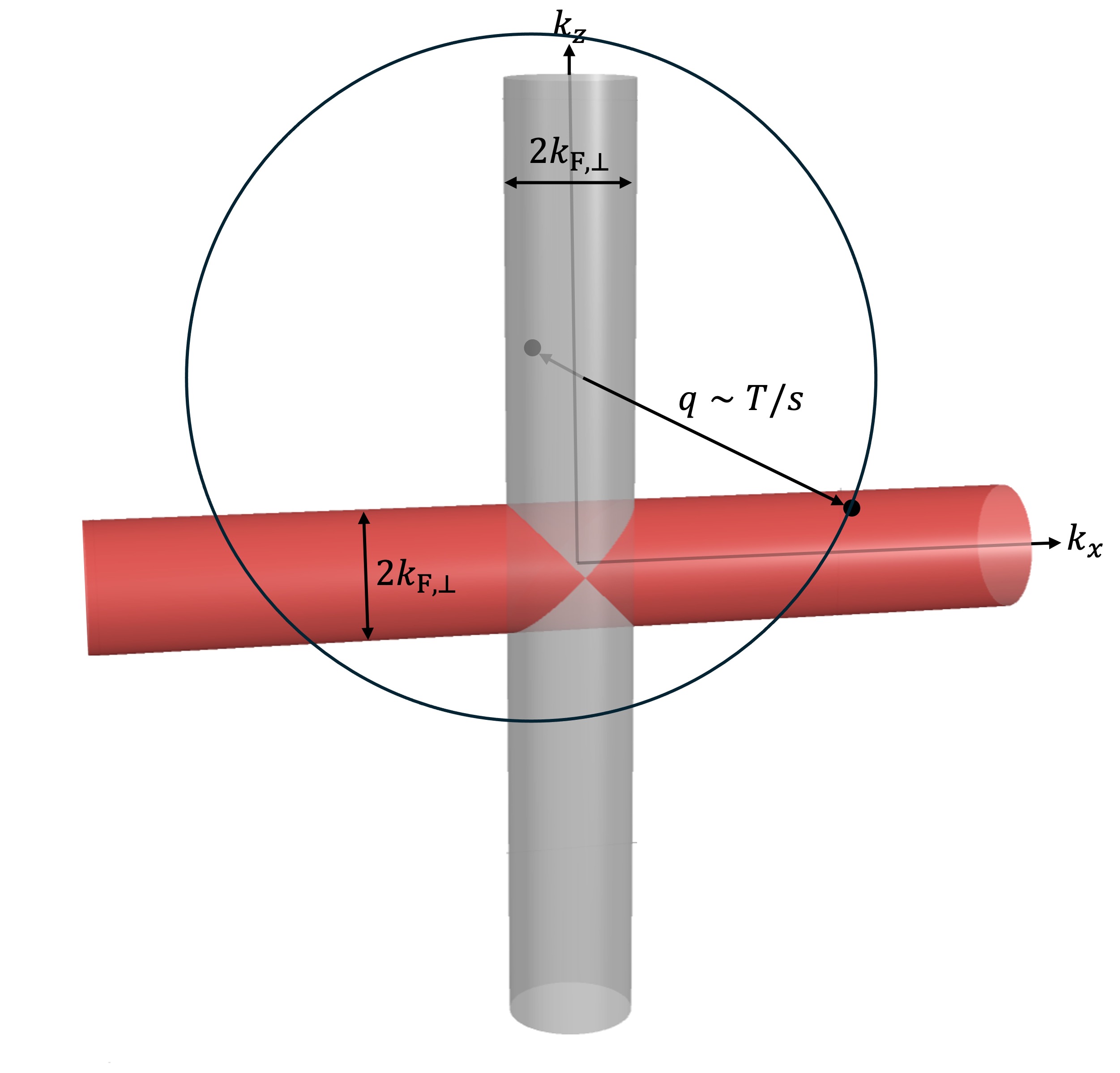}
  \caption{Intervalley scattering $\ve_{z,\bk}\to\ve_{x,\bk+\bq}$: Given an initial $k_z$, for a particular $q_y$, which is restricted kinematically to be within $(0,2\kp)$, $q_z$ is fixed to a single value. If $q_y$ and $q_z$ allow for a final state $\ve_{x,\bk+\bq}$, $q_x$ is kinematically restricted only by the length of the cylinder $2\kl$. }  
  \label{fig: intervalley}
\end{figure}
Here we consider scattering between two ellipsoids, say from one with major axis along the $z$ direction to one with major axis along the $x$ direction,
\begin{flalign}
    \ve_{z,\bk}=\frac{1}{2}\left(\frac{k_x^2+k_y^2}{m_\perp}+\frac{k_z^2}{m_\parallel}\right),\\
    \ve_{x,\bk}=\frac{1}{2}\left(\frac{k_x^2}{m_\parallel}+\frac{k_y^2+k_z^2}{m_\perp}\right),\\
\end{flalign}
each with minor axes length $\kp$ and major axis length $\kl$; see Fig.~\ref{fig: intervalley}. One cannot simultaneously rescale the major axes of the two valleys so out of convenience we again consider the case $\mu=0$ so that the ellipsoids become cylinders,
\begin{flalign}
    \ve_{z,\bk}=\frac{1}{2m_\perp}\left(k_x^2+k_y^2\right), \\
    \ve_{x,\bk}=\frac{1}{2m_\perp}\left(k_y^2+k_z^2\right). \label{cyl xy} 
\end{flalign}
As seen in the previous section, this choice is valid to determine the leading order temperature dependence. Once again these are cut at $\pm\kl$. Using the delta function in Eq.~\eqref{exchange general},
\[
\delta(\ve_{z,\bk}-\ve_{x,\bk+\bq}+\omega_{\bq\lambda})=m_\perp\delta\left(k_x^2-(k_z+q_z)^2+(2k_yq_y+q_y^2)+2m_\perp s_\lambda q\right)
\]
to instead integrate over $k_z=k_\parallel$, we have for the remaining integral over $k_\perp=\sqrt{k_x^2+k_y^2}$ and $\phi_\bk$,
\begin{flalign}
    \begin{split}
        &\int dk_\perp k_\perp
        \left[N_q\big(f(\ve_{k_\perp})-f(\ve_{k_\perp}+\omega_q)\big)-f(\ve_{k_\perp}+\omega_q)\big(1-f(\ve_{k_\perp})\big)\right] \\
        &\times \int d\phi_\bk 
        \frac{m_\perp}{\sqrt{k_\perp^2\cos^2\phi_\bk-2k_\perp \sin\phi_\bk\,q_y-q_y^2+2m_\perp s_\lambda q}} \label{intervalley first integral}
    \end{split}
\end{flalign}
where we note that in the cylindrical limit $\ve_{z,\bk}$ is independent of $k_z$ and upon transforming to polar coordinates in the $xy$-plane we rewrote,
\[
\ve_{z,\bk}\rightarrow\ve_{k_\perp}=\frac{k_\perp^2}{2m_\perp}\qquad \mathrm{and} \qquad\ve_{x,\bk+\bq}=\ve_{k_\perp}+\omega_{q\lambda}.
\]
We then linearize about the FS and set all factors of $k_\perp=\kp$. The integral over $k_\perp$ is then given by the usual result seen in Eq.~\eqref{energy}.

\subsubsection*{Kinematic constraints}
\paragraph{Constraint on $q_z$:} For a final state $\ve_{x,\bk+\bq}$, it is required that
\[
|q_z|\in(|k_z-\kp|,|k_z+\kp|).
\]
However, once a specific $q_y$ has been chosen, $q_z$ is pinned to a single value within this interval, thereby cutting the integral over $q_z$.

\paragraph{Constraint on $q_y$:} The upper and lower kinematic constraints on $q_y$ are determined by its roots under the square root in Eq.~\eqref{intervalley first integral},
\[
q_y=\kp\left(\pm1-\sin\phi_\bk\right),
\]
where we have assumed $2m_\perp s_\lambda q/k^2_{\mathrm{F}\perp}\ll1$---since it may be that $q\sim2\kl$ this requires $4s_\lambda/\vfp\ll\kp/\kl$, however, keeping the full form will not affect leading order temperature behavior. The kinematic constraint on $q_y$ varies according to $k_y=\kp\sin\phi_\bk$, the $y$ component of the initial state. For example, if $\phi_\bk=\pm\pi/2$, it must be that $|q_y|<2\kp$ while for $\phi_\bk=0$ or $\pi$, $|q_y|<\kp$. While $q_{y,z}$ are tightly restricted, $q_x$ is kinematically free up to $\mathcal{O}(\kl)$. 

Expanding in $\delta T$ yields a similar result as in the previous section for the ERR except for the presence of a complicated integral over the azimuthal angle of $\bk$ and corresponding bounds on $q_{y,z}$. We now integrate over $\bq$ in Cartesian coordinates and insert the squared intervalley matrix elements from Eq.~\eqref{sq long matx inter} and Eq.~\eqref{sq tran matx inter}. We keep only the term that will dominate upon taking $\Tbgp\ll T\ll\Tbgl$, namely the transverse elements can be shown to reduce to the following here in this limit:
\[
\frac{\Xi_s^2}{2\rho\,s_\mathrm{T}}\;q\left(\frac{q_x^2+q_z^2}{q^2}-4\frac{q_x^2 q_z^2}{q^4}\right)\rightarrow\frac{\Xi_s^2}{2\rho\,s_\mathrm{T}}\; q_x.
\]
Then we have for the $z\to x$ intervalley contribution to the ERR,
\begin{flalign}
\begin{split}
    \frac{1}{\tau_E^{\mathrm{DA,inter,(z,x)}}}
    &=
    \frac{s^2m_\perp^2\Xi_s^2E_{\mathrm{F}}}{128\pi^{10}\rho n}\frac{1}{T^3}
    \int_0^{2\kl} dq_x 
    q_x^4
    \frac{e^{sq_x/\kb T}}{(e^{sq_x/\kb T}-1)^2} \\ &\times
    \int d\phi_\bk
    \int_{-\kp\,g_-(\phi_\bk)}^{\kp\, g_+(\phi_\bk)} dq_y
    \int_{|-\kp+\mathcal{F}(q_z,q_y,\phi_\bk)|}^{|\kp+\mathcal{F}(q_z,q_y,\phi_\bk)|}dq_z
    \frac{1}{\sqrt{\kp^2\cos^2\phi_\bk-2\kp \sin\phi_\bk\,q_y-q_y^2}}. \label{tau e inter}
\end{split}
\end{flalign}
where $\mathcal{F}$ denotes the functional relation between $k_z$ and $q_z$, $q_y$, and $\phi_\bk$, $g_\pm$ denotes the functional dependence of the upper and lower limits of $q_y$ on $\phi_\bk$, and we have let all remaining factors of $q\sim q_x\sim\kl\gg \kp$ in the intermediate temperature limit (we must also shift all $q_z$ to be within $(-\kp,\kp)$ so that $|q_z|\ll|q_x|$) .

\subsubsection*{Intermediate temperature limit: $\Tbgp\ll T\ll\Tbgl$} \label{subsec:intermed T da}
We take the limit $\Tbgl/T\rightarrow\infty$ and shift all $q_z$ to be within $(-\kp,\kp)$ so that $q\sim q_x$, and on rescaling $q_x$ we have
\begin{flalign}
\begin{split}
    \frac{1}{\tau_E^{\mathrm{DA,inter,(z,x)}}}
    &=
    \frac{m_\perp^2\Xi_s^2E_{\mathrm{F}}}{128\pi^{10}\rho ns^3}T^2
    \int_0^\infty dx
    x^4
    \frac{e^{x}}{(e^{x}-1)^2} \\ &\times
    \int d\phi_\bk
    \int_{-\kp\,g_-(\phi_\bk)}^{\kp\, g_+(\phi_\bk)} dq_y
    \frac{1}{\sqrt{\kp^2\cos^2\phi_\bk-2\kp \sin\phi_\bk\,q_y-q_y^2}}. \label{tau e inter}
\end{split}
\end{flalign}
As mentioned previously, for any particular $q_y$, $q_z$ is pinned to a specific value within its available range, thereby cutting this integral. Given the kinematic bounds on $q_y$ for fixed $\phi_\bk$, the integral over $q_y$ contains only integrable endpoint singularities and is finite for all $\phi_\bk$. Hence, it does not provide any additional temperature dependence and the resulting ERR due to intervalley scattering therefore scales quadratically with temperature, 
\begin{flalign}
\begin{split}
    \frac{1}{\tau_E^{\mathrm{DA,inter}}}\propto T^2 .
\end{split}
\end{flalign}
So in either case of intravalley or intervalley scattering, energy relaxation scales quadratically with temperature, differing only by a prefactor. In the former case, relaxation is controlled by momentum transfers along the major axis of a single valley while in the latter it is controlled by momentum transfers that bring the electron far along the major axis of the valley orthogonal to the original one.

\section{Review: Froelich (polar) coupling and the Lyddane-Sachs-Teller relation}
\subsection{Polarization (macroscopic) field }
The field felt by an electron in a material under an external field $\bE(\br,t)$ is given by,
\begin{flalign}
    \bD=\ep\bE=\bE+4\pi\bP.
\end{flalign}
The linear response relation in the frequency domain is
\begin{flalign}
    \bP_\bq(\Omega)=[\chi_{\bq,i}(\Omega)+\chi_{\bq,e}(\Omega)]\bE_\bq(\Omega),
\end{flalign}
where $\chi_{i,e}$ capture the material response due to the lattice and conduction electrons, respectively, and we restrict to the long wavelength regime $q\ll q_\mathrm{BZ}$, where $q_\mathrm{BZ}$ denotes the Brillouin zone boundary. For small external frequencies compared to typical phonon or electron energies, $\Omega\ll\omega_\bq,\ve_\bk$, we can take for the limiting form of the dielectric constant
\begin{flalign}
    \ep_{\bq,\Omega\rightarrow0}=\ep_{\bq,0}=1+4\pi[\chi_{\bq,i}(0)+\chi_{\bq,e}(0)], \label{static1}
\end{flalign}
where $\chi_{i,e}(0)$ are the static dielectric responses of the lattice ions and electrons, respectively.
For a degenerate electron system, $\omega_\bq\lesssim T\ll\ve_\bk\sim E_\mathrm{F}$, so the high external frequency regime may be defined to include those still much less than any relevant electron energy, $\omega_\bq\ll\Omega\ll\ve_\bk$. In this regime the slow lattice response may be neglected,
\begin{flalign}
    \ep_{\Omega\rightarrow\infty}=\ep_\infty=1+4\pi\chi_e(0)
\end{flalign}
Then one can define the displacement field for all $\Omega\ll E_\mathrm{F}$,
\begin{flalign}
\begin{split}
    \bD_\bq(\Omega)&=\ep_{\bq,\infty}\bE_\bq(\Omega)+4\pi\bP_\bq(\Omega),
    \quad
    \bP_\bq(\Omega)=\chi_{\bq,i}(\Omega)\bE_\bq(\Omega),\label{disp field}
    \end{split}
\end{flalign}
where the polarization field $\bP$ is redefined to be solely that of the lattice response.
Within the Born effective-charge description the polarization may be written as the following sum over charged displacements for each sub-lattice $\kappa$,
\begin{flalign}
    \bP_\bq(\Omega)=\sum_\kappa \bZ^*_\kappa \bu_{\bq,\kappa}(\Omega),
\end{flalign}
where $\bZ^*_\kappa$ are Born effective-charge tensors (charge per displacement, with units $e/V_0$, and $V_0$ being the volume of a unit cell). Each sub-lattice displacement vector $\bu_{\bq,\kappa}$ is a sum over normal coordinates,
\begin{equation}
\bu_{\bq,\kappa}=\sum_{\lambda}
\sqrt{\frac{1}{2M_\kappa\omega_{\bq\lambda}}}\,
\be_{\bq\kappa\lambda}\left(b_{\bq\lambda}+b^\dagger_{-\bq\lambda}\right).
\end{equation}
Then we have for the polarization operator,
\begin{flalign}
    \begin{split}
\bP_\bq&=\sum_{\kappa,\lambda}
\bZ^*_\kappa\sqrt{\frac{1}{2M_\kappa\omega_{\bq\lambda}}}\,\be_{\bq\kappa\lambda}\,
\left(b_{\bq\lambda}+b^\dagger_{-\bq\lambda}\right)=\sum_\lambda Z^*_{\lambda}\,Q_{\bq\lambda}\,\be_{\bq\lambda}, \label{pol op first}
\end{split}
\end{flalign}
where the effective charge vector and normal coordinate for each mode are, respectively,
\begin{flalign}
Z_{\lambda}^*\be_{\bq\lambda}=\sum_\kappa\sqrt{\frac{M}{M_\kappa}}\bZ_\kappa^*\be_{\bq\kappa\lambda},
\quad
    Q_{\bq\lambda}=\sqrt{\frac{1}{2M\omega_{\bq\lambda}}}\,
\left(b_{\bq\lambda}+b^\dagger_{-\bq\lambda}\right), \label{normal def}
\end{flalign}
where $M=\sum_\kappa M_\kappa$, and for a TO mode, $\bq\cdot\be_{\bq\mathrm{TO}}=0$. Note that $Z^*_\lambda$ vanishes except for IR-active optical modes; in centrosymmetric crystals, these correspond to the odd-parity modes.

In the harmonic approximation the normal coordinate for a TO mode, $Q_{\bq\mathrm{TO}}$, with associated effective mass $M$, obeys the following equation of motion,
\begin{equation}
M\ddot Q_{\bq\mathrm{TO}} =-M\omega_{\bq\mathrm{TO}}^2 Q_{\bq\mathrm{TO}} +Z^*_{\mathrm{TO}}\,E_\bq(t),
\end{equation}
where $E_\bq(t)=E_\bq e^{i \Omega t}$ is the component of the external field parallel to $\be_{\bq\mathrm{TO}}$. Assuming harmonic time dependence we obtain in frequency space
\begin{equation}
M(\omega_{\bq\mathrm{TO}}^2-\Omega^2)Q_{\bq\mathrm{TO}}=Z^*_{\mathrm{TO}}E_\bq
\quad\Rightarrow\quad
P_{\bq\mathrm{TO}}=Z^*_{\mathrm{TO}}Q_{\bq\mathrm{TO}}=\frac{(Z^*_{\mathrm{TO}})^2}{M(\omega_{\bq\mathrm{TO}}^2-\Omega^2)}E_\bq.
\end{equation}
Combining with Eq.~\eqref{disp field}, we have for the dielectric function,
\begin{flalign}
    \ep_\bq(\Omega)=\ep_{\bq,\infty}+4\pi\frac{(Z^*_{\mathrm{TO}})^2}{M(\omega_{\bq\mathrm{TO}}^2-\Omega^2)}, \label{ep function}
\end{flalign}
which in the static limit is
\begin{flalign}
    \ep_{\bq,0}=\ep_{\bq,\infty}+4\pi\frac{(Z^*_{\mathrm{TO}})^2}{M\omega_{\bq\mathrm{TO}}^2}. \label{static2}
\end{flalign}
Using Eqs.~\eqref{pol op first} and \eqref{static2}, the polarization operator may now be written,
\begin{equation}
\bP_\bq=\sum_\mathrm{TO}\be_{\bq\mathrm{TO}}\,A_{\bq\mathrm{TO}} X_{\bq\mathrm{TO}},
 \label{pol op}
\end{equation}
where
\begin{equation}
A_{\bq\mathrm{TO}}=\sqrt{\frac{\omega_{\bq\mathrm{TO}}}{2}}\,
\sqrt{\frac{\ep_{\bq,0}-\ep_{\bq,\infty}}{4\pi}}\qquad \mathrm{and} \qquad
\quad
X_{\bq\mathrm{TO}}\equiv b_{\bq\mathrm{TO}}+b^\dagger_{-\bq\mathrm{TO}}. \label{pol op defs}
\end{equation}

\subsection{Lyddane Sachs Teller (LST) Relation}
With the use of Maxwell's equation, 
\begin{flalign}
\begin{split}
    \nabla\cdot\bE&=4\pi\rho=4\pi(\rho_f+\rho_b),
    \end{split}
\end{flalign}
where $f$ and $b$ correspond to free and bound charges, respectively, one may consider a charge-neutral bulk with no free carriers, namely 
\begin{flalign}
    \nabla\cdot\bE=4\pi\rho_b
    \quad
    \mathrm{and}
    \quad
    \nabla\cdot\bD=0=4\pi\rho_f, \label{long comp vanish}
\end{flalign}
where the longitudinal component of the displacement field vanishes, $\bq\cdot \bD_\bq=0$. The longitudinal optical (LO) eigenmode corresponds to a nontrivial solution of $\nabla\cdot\bD=0$, namely $\ep_\bq(\omega_{\bq\mathrm{LO}})=0$.
Imposing $\ep_\bq(\omega_{\bq LO})=0$ in Eq.~\eqref{ep function}, we obtain
\begin{equation}
\omega_{\bq\mathrm{LO}}^2=\omega_{\bq\mathrm{TO}}^2+\frac{4\pi (Z^*_{\mathrm{TO}})^2}{M\ep_{\bq,\infty}}. \label{lo to}
\end{equation}
Using this, along with the static result in Eq.~\eqref{static2}, yields the standard LST relation 
\begin{flalign}
\frac{\omega_{\bq\mathrm{LO}}^2}{\omega_{\bq\mathrm{TO}}^2}=\frac{\ep_{\bq,0}}{\ep_{\bq,\infty}},
\end{flalign}
where $\bq$ dependence holds only for $q\ll q_\mathrm{BZ}$.
As in \cite{Kumar2021QuasiparticleParaelectrics}, one often considers $\ep_0\gg\ep_\infty$ so that $\ep_\infty$ may be absorbed to form a reference frequency $\Omega_0^2=\ep_\infty\omega_{\mathrm{LO}}^2$.
Then we have for the TO mode,
\begin{equation}
A_{\bq\mathrm{TO}}=\frac{\Omega_0}{\sqrt{8\pi\omega_{\bq\mathrm{TO}}}}.
\end{equation}
To match the normalization convention of \cite{Kumar2021QuasiparticleParaelectrics}, we absorb the factor $1/\sqrt2$ in $Q_{\bq\lambda}$ (defined in Eq.~\eqref{normal def}) within $Z_\lambda^*$, such that
\begin{flalign}
Z_{\lambda}^*/\sqrt2\rightarrow Z_{\lambda}^*.
\end{flalign}
Finally, we have
\begin{equation}
A_{\bq\mathrm{TO}}=\frac{\Omega_0}{\sqrt{4\pi\omega_{\bq\mathrm{TO}}}}.
\end{equation}

\section{2TO interaction and scattering matrix elements}\label{sec:two-phonon}

\subsection{Real-space interaction Hamiltonian}
The 2TO density interaction term for an isotropic phonon spectrum is
\begin{equation}
H_{\mathrm{int}}^{2\mathrm{TO}}
=
\frac{g_2}{2} \int d^3r\;n(\br)\,\bP(\br)\cdot\bP(\br),
\qquad
n(\br)=\psi^\dagger(\br)\psi(\br).
\end{equation}
Expanding the density in its operator form gives
\begin{equation}
\psi(\br)=\frac{1}{\sqrt{V}}\sum_{\bk_1} c_{\bk_1}\,e^{i\bk_1\cdot\br}
\quad\Rightarrow\quad
n(\br)=\frac{1}{V}\sum_{\bk_1,\bk_2} c^\dagger_{\bk_2}c_{\bk_1}\,e^{i(\bk_1-\bk_2)\cdot\br},
\end{equation}
The second order polarization operator is then given by
\begin{flalign}
\bP(\br)\cdot\bP(\br)=\frac{1}{V}\sum_{\bq_1,\lambda_1}\sum_{\bq_2,\lambda_2}
(\be_{\bq_1\lambda_1}\cdot\be_{\bq_2\lambda_2})A_{\bq_1\lambda_1}A_{\bq_2\lambda_2}
X_{\bq_1\lambda_1}X_{\bq_2\lambda_2}\,e^{i(\bq_1+\bq_2)\cdot\br},
\end{flalign}
where we have generalized Eq.~\eqref{pol op} to the case of two independent modes ($\lambda_i=1,2$) with the following relations for TO modes specifically: $\be_{\bq\lambda}\cdot\hat\bq=0$ and $\be_{\bq\lambda}\cdot \be_{\bq\lambda'}=\delta_{\lambda\lambda'}$. Now one can insert $n(\br)$ and $\bP(\br)$ into $H_{\mathrm{int}}^{2\mathrm{TO}}$, 
and upon integrating $\int d^3r\,e^{i(\bk_1-\bk_2+\bq_1+\bq_2)\cdot\br}=V\,\delta_{\bk_1-\bk_2+\bq_1+\bq_2,0}$,
one obtains
\begin{equation}
H_{\mathrm{int}}^{2\mathrm{TO}}
=
\frac{1}{2V}\sum_{\substack{\bk_1,\bk_2,\bq_1,\bq_2,\\\lambda_1,\lambda_2}}
g_{\lambda_1\lambda_2}(\bq_1,\bq_2)\;
c^\dagger_{\bk_2}c_{\bk_1}\;
X_{\bq_1\lambda_1}X_{\bq_2\lambda_2}\;
\delta_{\bk_1-\bk_2+\bq_1+\bq_2,0}, \label{2to int}
\end{equation}
with the (operator-stripped) two-phonon kernel,
\begin{equation}
g_{\lambda_1\lambda_2}(\bq_1,\bq_2)
\equiv
g_2\,
(\be_{\bq_1\lambda_1}\cdot\be_{\bq_2\lambda_2})\,
A_{\bq_1\lambda_1}A_{\bq_2\lambda_2}. \label{kernel}
\end{equation}
For both in and out processes, there are now four types of two-phonon interactions one must account for, namely double absorption, absorption-emission, emission-absorption, and double emission defined by the following bilinear pairs,
\begin{equation}
X_{\bq_1\lambda_1}X_{\bq_2\lambda_2}
=
\underbrace{b_{\bq_1\lambda_1}b_{\bq_2\lambda_2}}_{\text{abs,abs}}
+\underbrace{b_{\bq_1\lambda_1}b^\dagger_{-\bq_2\lambda_2}}_{\text{abs,em}}
+\underbrace{b^\dagger_{-\bq_1\lambda_1}b_{\bq_2\lambda_2}}_{\text{em,abs}}
+\underbrace{b^\dagger_{-\bq_1\lambda_1}b^\dagger_{-\bq_2\lambda_2}}_{\text{em,em}}. \label{four processes}
\end{equation}

\subsection{Scattering matrix elements: e.g. double emission}
The phonon scattering matrix elements corresponding to a double emission out process ($\bk\rightarrow\bk'=\bk-\bq_1-\bq_2$) is now derived, where the seven other processes are produced in a similar manner.
Once again FGR introduces a sum over all final many-body states, which are, in the unperturbed basis, a direct product of an electron momentum eigenstate and set of phonon occupation states:
\begin{equation}
\ket{i}=\ket{\bk;\{N_{\bq\lambda}\}},\qquad
\ket{f}=\ket{\bk';\{N_{\bq\lambda}'\}},
\end{equation}
where $\{N\}\equiv\{N_{\bq\lambda}\}$ specifies the occupation number for every mode $(\bq,\lambda)$.
The total transition rate out of $\ket{i}$ is
\begin{equation}
W_{i\to \mathrm{all}}
=
2\pi\sum_{f}
\abs{\mel{f}{H_{\mathrm{int}}}{i}}^2\,
\delta(E_f-E_i),
\qquad
\sum_f \equiv \sum_{\bk'}\sum_{\{N_{\bq\lambda}'\}}.
\end{equation}
Ignoring the energy delta for now, and using completeness of the full many-body basis:
\begin{equation}
\sum_f \ket{f}\bra{f}=\mathbbm{1},
\end{equation}
then
\begin{align}
\sum_f \abs{\mel{f}{H_{\mathrm{int}}}{i}}^2
&=
\sum_f \mel{i}{H_{\mathrm{int}}^\dagger}{f}\mel{f}{H_{\mathrm{int}}}{i}
=
\mel{i}{H_{\mathrm{int}}^\dagger H_{\mathrm{int}}}{i}. \label{closure}
\end{align}

\subsubsection{Application to the 2TO ($P^2$) interaction and survival of only direct and exchange terms}

\paragraph{Example channel: emission--emission.}
Upon applying momentum conservation in Eq.~\eqref{2to int} via the Kronecker delta, the interaction term associated with double emission is,
\begin{equation}
H^{\mathrm{em,em}}
=
\frac{1}{2V}\sum_{\bk_1}
\sum_{\substack{\bq_1,\lambda_1\\\bq_2,\lambda_2}}
g_{\lambda_1\lambda_2}(-\bq_1,-\bq_2)\;
c^\dagger_{\bk_1-\bq_1-\bq_2}c_{\bk_1}\;
b^\dagger_{\bq_1\lambda_1}b^\dagger_{\bq_2\lambda_2}.
\end{equation}
while its Hermitian conjugate is
\begin{equation}
(H^{\mathrm{em,em}})^\dagger
=
\frac{1}{2V}\sum_{\bk_1}
\sum_{\substack{\bq_1,\lambda_1\\\bq_2,\lambda_2}}\textbf{}
g^*_{\lambda_1\lambda_2}(-\bq_1,-\bq_2)\;
c^\dagger_{\bk_1}c_{\bk_1-\bq_1-\bq_2}\;
b_{\bq_1\lambda_1}b_{\bq_2\lambda_2},
\end{equation}
where we relabeled $\bq_1,\bq_2\rightarrow-\bq_1,-\bq_2$ in both. Using the closure relation to sum over final states as in Eq.~\eqref{closure}, we consider diagonal matrix elements of
\begin{align}
\left(H^{\mathrm{em,em}}\right)^\dagger H^{\mathrm{em,em}}
&=
\frac{1}{4V^2}\sum_{\bk_1}
\sum_{\substack{\bq_1,\lambda_1\\\bq_2,\lambda_2}}
\sum_{\substack{\bq_3,\lambda_3\\\bq_4,\lambda_4}}
g^{*}_{\lambda_1\lambda_2}(-\bq_1,-\bq_2)\;
g_{\lambda_3\lambda_4}(-\bq_3,-\bq_4)\;
\Big[c^\dagger_{\bk_1}c_{\bk_1-\bq_1-\bq_2}\,c^\dagger_{\bk_1-\bq_3-\bq_4}c_{\bk_1}\Big]\;
\Big[b_{\bq_2\lambda_2}b_{\bq_1\lambda_1}\,b^\dagger_{\bq_3\lambda_3}b^\dagger_{\bq_4\lambda_4}\Big],
\end{align}
between states $\ket{i}=\ket{\bk;\{N_{\bq\lambda}\}}$.

\paragraph{Electron selection: net momentum transfer must match.}
For a single occupied electron momentum state $\bk$ in $\ket{i}$, the electron operator string yields a nonzero result only when
\[
\bq_1+\bq_2=\bq_3+\bq_4.
\]

\paragraph{Phonon selection: only direct and exchange contractions survive.}
Because $\ket{i}$ is a product of number states, 
the expectation
\[
\expval{b_{\bq_2\lambda_2}b_{\bq_1\lambda_1}\,b^\dagger_{\bq_3\lambda_3}b^\dagger_{\bq_4\lambda_4}}_{\{N_{\bq\lambda}\}}
\]
vanishes unless the annihilated modes exactly coincide with the created modes, and hence the only nonzero contributions are:
\begin{align}
\textbf{Direct:}\quad &
(\bq_1,\lambda_1)=(\bq_3,\lambda_3)\ \text{and}\ (\bq_2,\lambda_2)=(\bq_4,\lambda_4),\\
\textbf{Exchange:}\quad &
(\bq_1,\lambda_1)=(\bq_4,\lambda_4)\ \text{and}\ (\bq_2,\lambda_2)=(\bq_3,\lambda_3).
\end{align}
Both direct and exchange contributions yield the same Bose factor:
\begin{equation}
\expval{b_{\bq_2\lambda_2}b_{\bq_1\lambda_1}\,b^\dagger_{\bq_1\lambda_1}b^\dagger_{\bq_2\lambda_2}}_{\{N\}}
=
(N_{\bq_1\lambda_1}+1)(N_{\bq_2\lambda_2}+1).
\end{equation}

\paragraph{Resulting weight (direct + exchange).}
The surviving contribution after summing over final phonon states has the structure
\begin{equation}
\sum_f \abs{\mel{f}{H^{\mathrm{em,em}}}{i}}^2
\;\propto\;
\frac{1}{4V^2}\sum_{\substack{\bq_1,\lambda_1\\\bq_2,\lambda_2}}
\Big[
\abs{g_{\lambda_1\lambda_2}(-\bq_1,-\bq_2)}^2
+
g_{\lambda_1\lambda_2}(-\bq_1,-\bq_2)\,g^{*}_{\lambda_2\lambda_1}(-\bq_2,-\bq_1)
\Big]\,
(N_{\bq_1\lambda_1}+1)(N_{\bq_2\lambda_2}+1),
\end{equation}
with the appropriate electron energy and momentum constraints understood, namely
\[
\ve_{\bk'}=\ve_{\bk-\bq_1-\bq_2}=\ve_\bk-\omega_{\bq_1}-\omega_{\bq_2}
\]
for double emission. Given the definitions in Eq.~\eqref{pol op defs} and Eq.~\eqref{kernel}, we note that the kernel is symmetric under inversion and exchange,
\begin{equation}
g_{\lambda_1\lambda_2}(\bq_1,\bq_2)=g_{\lambda_1\lambda_2}(-\bq_1,-\bq_2)=g_{\lambda_2\lambda_1}(\bq_2,\bq_1),
\end{equation}
and therefore the bracket simplifies to
\begin{equation}
\abs{g_{\lambda_1\lambda_2}(-\bq_1,-\bq_2)}^2
+
g_{\lambda_1\lambda_2}(-\bq_1,-\bq_2)\,g^{*}_{\lambda_2\lambda_1}(-\bq_2,-\bq_1)
=
2\,\abs{g_{\lambda_1\lambda_2}(\bq_1,\bq_2)}^2.
\end{equation}

\subsubsection{Squared matrix elements and summation over phonon branches}
We have the following sum over the two TO branches:
\begin{equation}
\sum_{t_1=1,2}\sum_{t_2=1,2}\abs{g_{t_1t_2}(\bq_1,\bq_2)}^2\propto\sum_{t_1=1,2}\sum_{t_2=1,2}(\be_{\bq_1t_1}\cdot\be_{\bq_2t_2})^2
\end{equation}
where we assume the branches share the same dispersion $\omega_{\bq,1}=\omega_{\bq,2}$, and hence $A_{\bq t}$ is branch-independent ($A_{\bq t}\equiv A_{\bq}$). 
Once again we use the transverse projection,
\[
\sum_{t=\mathrm{T}_1,\mathrm{T}_2} e_{i,\bq t}e_{j,\bq t}=P_{ij}(\hat{\bq})\equiv \delta_{ij}-\hat q_i\hat q_j.
\]
where the subsequent identity follows,
\begin{equation}
\sum_{t_1,t_2}(\be_{\bq_1t_1}\cdot \be_{\bq_2t_2})^2
=\mathrm{Tr}\!\left[P_\mathrm{T}(\hat\bq_1)P_\mathrm{T}(\hat\bq_2)\right]
=1+(\hat\bq_1\cdot \hat\bq_2)^2,
\end{equation}
and we have as a result,
\begin{equation}
\abs{g_{\bq_1,\bq_2}}^2=\sum_{t_1,t_2}\abs{g_{t_1t_2}(\bq_1,\bq_2)}^2
=
\frac{g_2^2\Omega_0^4}{16\pi^2}\frac{1}{\omega_{\bq_1}\omega_{\bq_2}}\Big[1+(\hat\bq_1\cdot\hat\bq_2)^2\Big]. \label{coupling elements}
\end{equation}
matching that found in \cite{Kumar2021QuasiparticleParaelectrics}.

\section{Two-phonon Energy Relaxation Rate}
\subsection*{Collision integral}
The collision integral for 2TO phonon scattering is
\begin{flalign}
\begin{split}
    \frac{\partial f_\bk}{\partial t}^{\mathrm{2ph}}
    &=-\frac{4\cdot2\pi}{4V^2}\sum_{\bq_1,\bq_2}
\abs{g_{\bq_1,\bq_2}}^2\Bigg\{
f_{\bk}\Bigg[
\,(N_{\bq_1}+1)(N_{\bq_2}+1)\,(1-f_{\bk-\bq_1-\bq_2})\,
\delta\Big(\ve_{\bk-\bq_1-\bq_2}-\ve_{\bk}+\omega_{\bq_1}+\omega_{\bq_2}\Big)
\\
&\hspace{5.2em}
+\,N_{\bq_1}N_{\bq_2}\,(1-f_{\bk+\bq_1+\bq_2})\,
\delta\Big(\ve_{\bk+\bq_1+\bq_2}-\ve_{\bk}-\omega_{\bq_1}-\omega_{\bq_2}\Big)
\\
&\hspace{5.2em}
+\,(N_{\bq_1}+1)N_{\bq_2}\,(1-f_{\bk-\bq_1+\bq_2})\,
\delta\Big(\ve_{\bk-\bq_1+\bq_2}-\ve_{\bk}+\omega_{\bq_1}-\omega_{\bq_2}\Big)
\\
&\hspace{5.2em}
+\,N_{\bq_1}(N_{\bq_2}+1)\,(1-f_{\bk+\bq_1-\bq_2})\,
\delta\Big(\ve_{\bk+\bq_1-\bq_2}-\ve_{\bk}-\omega_{\bq_1}+\omega_{\bq_2}\Big)
\Bigg]
\\[0.5em]
&\hspace{2.6em}
- 
(1-f_{\bk})\Bigg[
\,(N_{\bq_1}+1)(N_{\bq_2}+1)\,f_{\bk+\bq_1+\bq_2}\,
\delta\Big(\ve_{\bk}-\ve_{\bk+\bq_1+\bq_2}+\omega_{\bq_1}+\omega_{\bq_2}\Big)
\\
&\hspace{5.2em}
+\,N_{\bq_1}N_{\bq_2}\,f_{\bk-\bq_1-\bq_2}\,
\delta\Big(\ve_{\bk}-\ve_{\bk-\bq_1-\bq_2}-\omega_{\bq_1}-\omega_{\bq_2}\Big)
\\
&\hspace{5.2em}
+\,(N_{\bq_1}+1)N_{\bq_2}\,f_{\bk+\bq_1-\bq_2}\,
\delta\Big(\ve_{\bk}-\ve_{\bk+\bq_1-\bq_2}+\omega_{\bq_1}-\omega_{\bq_2}\Big)
\\
&\hspace{5.2em}
+\,N_{\bq_1}(N_{\bq_2}+1)\,f_{\bk-\bq_1+\bq_2}\,
\delta\Big(\ve_{\bk}-\ve_{\bk-\bq_1+\bq_2}-\omega_{\bq_1}+\omega_{\bq_2}\Big)
\Bigg]
\Bigg\}.\label{coll 2to}
    \end{split} 
\end{flalign}
where the factor of 4 in the numerator accounts for both the sum over spins and the direct and exchange contractions for identical bosons when summing over all ordered pairs $(\bq_1,\lambda_1),(\bq_2,\lambda_2)$. The factor $2\pi$ comes from FGR and the factor of 4 in the denominator comes from the definition of the Hamiltonian with $g_2/2$. In addition we have assumed a lattice symmetry and phonon dispersion that allows one to pull out the common factor, $\abs{g_{\bq_1,\bq_2}}^2$.  Note, while the matrix elements have been specified for the particular case of 2TO phonon (polar) coupling, the form of Eq.~\eqref{coll 2to} is general for two-phonon scattering with the same inversion and exchange symmetry. The first four terms correspond to out processes and the last four correspond to in processes, with each group of four in the order of double emission, double absorption, emission-absorption, and absorption-emission. 

\subsection*{Power transfer}
Upon inserting Eq.~\eqref{coll 2to} into Eq.~\eqref{power}, followed by combining double absorption with double emission processes and emission-absorption with absorption-emission processes, and shifting $\bk\rightarrow\bk\pm\bq_1\pm\bq_2$ as needed, we have the following simplified relation for the power transfer,
\begin{flalign}
\begin{split}
    \frac{\partial E_e^{\mathrm{2ph}}}{\partial t} 
    &=-\frac{4\pi}{ V^2}\sum_{\bk,\bq_1,\bq_2}|g_{\bq_1,\bq_2}|^2 
    \\ &\times
    \biggl\{
    \delta(\ve_\bk-\ve_{\bk-\bq_1-\bq_2}-\omega_{\bq_1}-\omega_{\bq_2})(\omega_{\bq_1}+\omega_{\bq_2})\biggl[(N_{\bq_1}+N_{\bq_2}+1)f_\bk(1-f_{\bk-\bq_1-\bq_2})+N_{\bq_1}N_{\bq_2}(f_\bk-f_{\bk-\bq_1-\bq_2})\biggr] \\
    &+\delta(\ve_\bk-\ve_{\bk-\bq_1+\bq_2}-\omega_{\bq_1}+\omega_{\bq_2})(\omega_{\bq_1}-\omega_{\bq_2})\biggl[N_{\bq_1}N_{\bq_2}(f_\bk-f_{\bk-\bq_1+\bq_2})-N_{\bq_1}f_{\bk-\bq_1+\bq_2}(1-f_\bk)+N_{\bq_2}f_\bk(1-f_{\bk-\bq_1+\bq_2})\biggr]
    \biggr\}. 
    \end{split}
\end{flalign} 
We then relabel $\bq_2\rightarrow-\bq_2$ in the second term and let $\bq=\bq_1+\bq_2$, yielding (Eq. (5) in the Main Text (MT))
\begin{flalign}
\begin{split}
    \frac{\partial E_e^{\mathrm{2ph}}}{\partial t}
    &=-\frac{4\pi}{ V^2}\sum_{\bk,\bq_1,\bq}|g_{\bq_1,\bq-\bq_1}|^2 
    \\ &\times
    \biggl\{
    \delta(\ve_\bk-\ve_{\bk-\bq}-\Omega_+)\Omega_+\biggl[(N_{\bq_1}+N_{\bq-\bq_1}+1)f_\bk(1-f_{\bk-\bq})+N_{\bq_1}N_{\bq-\bq_1}(f_\bk-f_{\bk-\bq})\biggr] \\
    &+\delta(\ve_\bk-\ve_{\bk-\bq}-\Omega_-)\Omega_-\biggl[N_{\bq_1}N_{\bq-\bq_1}(f_\bk-f_{\bk-\bq})-N_{\bq_1}f_{\bk-\bq}(1-f_\bk)+N_{\bq-\bq_1}f_\bk(1-f_{\bk-\bq})\biggr]
    \biggr\}, \label{pow transfer to}
    \end{split}
\end{flalign} 
where $\Omega_\pm=\omega_{\bq_1}\pm\omega_{\bq-\bq_1}$. Just as in Eq.~\eqref{exchange}, it is straightforward to show that the power transfer, and hence the ERR, vanish when $T=T_e$. 

\subsection{Isotropic electron dispersion}
\subsection*{Kinematic constraints}
Each $\delta$ function in Eq.~\eqref{pow transfer to} may be written 
\begin{flalign}
    \delta(\ve_k-\ve_{\bk-\bq}-\Omega_\pm)=\frac{\delta(\cos\theta_{\bk}-\frac{q}{2k}-\frac{m\Omega\pm}{kq})}{\frac{kq}{m}}. \label{delta spherical}
\end{flalign}
Once again, the $\delta$ function is the only term dependent on $\cos\theta_{\bk}$, and upon integration over the solid angle produces the following constraints on $q$ which must be satisfied simultaneously,
\begin{flalign}
    q^2\lessgtr \pm2\kp q-2m\Omega_\pm.
\end{flalign}
Both $\Omega_\pm\lessgtr0$ must be considered and $q$ is thereby restricted between the following bounds,
\begin{flalign}
    \kf\Bigg(1-\sqrt{1-\frac{\Omega_\pm}{E_\mathrm{F}}}\Bigg) < \ &q<\kf\Bigg(1+\sqrt{1-\frac{\Omega_\pm}{E_\mathrm{F}}}\Bigg), \ \  \Omega_\pm>0, \label{O+}\\ 
    \kf\Bigg(-1+\sqrt{1+\frac{|\Omega_\pm|}{E_\mathrm{F}}}\Bigg) < \ &q<\kf\Bigg(1+\sqrt{1+\frac{|\Omega_\pm|}{E_\mathrm{F}}}\Bigg), \ \ \Omega_\pm<0. \label{-O}
\end{flalign}
For a degenerate Fermi system, $\Omega_\pm\lesssim T\ll E_\mathrm{F}$, so that to leading order one may restrict $0<q<2\kf$ in all cases---we mention that this amounts to the condition $T\gg ms^2$, and while the transport rate has a different scaling regime for temperatures $T\ll ms^2$ (see SM in \cite{Kumar2021QuasiparticleParaelectrics}), the ERR scales with $T$ the same in the BG regime whether $T$ is above or below $ms^2$.
Having integrated over all three azimuthal angles, Eq.~\eqref{pow transfer to} becomes
\begin{flalign}
\begin{split}
    \frac{\partial E_e^{\mathrm{2ph}}}{\partial t} &= -\frac{8\pi V}{(2\pi)^6}m\int_0^\infty dq_1q_1^2 \int_{-1}^1 dt \int_0^{2\kf} dqq\abs{g_{\bq_1,\bq_-\bq_1}}^2  \\
    &\times\int dkk\,\biggl\{\Omega_+\biggl[\bigg(N_{\bq_1}+N_{\bq-\bq_1}+1\bigg)f(\ve_k)\bigg(1-f(\ve_k-\Omega_+)\bigg)+N_{\bq_1}N_{\bq-\bq_1}\bigg(f(\ve_k)-f(\ve_k-\Omega_+)\bigg)\biggr] \\ 
    &\quad\quad\quad+\Omega_-\bigg[N_{\bq_1}N_{\bq-\bq_1}\bigg(f(\ve_k)-f(\ve_k-\Omega_-)\bigg)-N_{\bq_1}f(\ve_k-\Omega_-)\bigg(1-f(\ve_k)\bigg)+N_{\bq-\bq_1}f(\ve_k)\bigg(1-f(\ve_k-\Omega_-)\bigg)\bigg]
    \biggr\},
    \end{split} 
\end{flalign} 
where $t=\cos\theta_\bq$.
Per usual, we may linearize the electron dispersion near the FS such that $\ve_k\approx\vf (k-\kf)$, where $\vf=\kf/m$, and set all instances of $k=\kf$. The integrals over $\ve_k$ may be performed just as in Eq.~\eqref{energy}, yielding
\begin{flalign}
\begin{split}
    \frac{\partial E_e^{\mathrm{2ph}} }{\partial t} &= -\frac{4 V}{(2\pi)^5}m^2\int_0^\infty dq_1q_1^2 \int_{-1}^1 dt \int_0^{2\kf} dqq\abs{g_{\bq_1,\bq_-\bq_1}}^2  \\
    &\times\biggl\{\Omega_+^2\biggl[\bigg(N_{\bq_1}+N_{\bq-\bq_1}+1\bigg)\tilde N(\Omega_+)-N_{\bq_1}N_{\bq-\bq_1}\biggr] +\Omega_-^2\bigg[-N_{\bq_1}N_{\bq-\bq_1}+N_{\bq_1}\tilde N(-\Omega_-)+N_{\bq-\bq_1}\tilde N(\Omega_-)\bigg)\bigg]
    \biggr\},
    \end{split} 
\end{flalign} 
where $\tilde{N}$ specifies a Bose function at the electron temperature $T_e$. The Bose functions may be expanded in $\delta T/T$ and we have
\begin{flalign}
\begin{split}
    \frac{\partial E_e^{\mathrm{2ph}} }{\partial t}&= -\frac{\delta T}{T^2}\frac{4 V}{(2\pi)^5}m^2\int_0^\infty dq_1q_1^2 \int_{-1}^1 dt \int_0^{2\kf} dqq\abs{g_{\bq_1,\bq_-\bq_1}}^2  \\
    &\times\bigg\{\Omega_+^3  \frac{e^{\Omega_+/T}}{(e^{\omega_{\bq_1}/T}-1)(e^{\omega_{\bq-\bq_1}/T}-1)(e^{\Omega_+/T}-1)}+ \Omega_-^3
          \frac{e^{\omega_{\bq_1}/T}-e^{\omega_{\bq-\bq_1}/T}}{(e^{\omega_{\bq_1}/T}-1)(e^{\omega_{\bq-\bq_1}/T}-1)(e^{\Omega_-/T}-1)(e^{-\Omega_-/T}-1)}\bigg\}
    \biggr\}.
    \end{split} 
\end{flalign} 
where it is now readily apparent that the power transfer vanishes at $T_e=T$. 

\subsection*{Two-phonon ERR} 
Finally, we use the relation in Eq.~\eqref{generic rate} to write the ERR,
\begin{flalign}
\begin{split}
   \frac{1}{\tau_E^{\mathrm{2ph}}} &= \frac{1}{T^3}\frac{E_\mathrm{F}m^2}{8\pi^7n}\int_0^\infty dq_1q_1^2 \int_{-1}^1 dt \int_0^{2\kf} dqq\abs{g_{\bq_1,\bq_-\bq_1}}^2  \\
    &\times\bigg\{\Omega_+^3  \frac{e^{\Omega_+/T}}{(e^{\omega_{\bq_1}/T}-1)(e^{\omega_{\bq-\bq_1}/T}-1)(e^{\Omega_+/T}-1)}+ \Omega_-^3
          \frac{e^{\omega_{\bq_1}/T}-e^{\omega_{\bq-\bq_1}/T}}{(e^{\omega_{\bq_1}/T}-1)(e^{\omega_{\bq-\bq_1}/T}-1)(e^{\Omega_-/T}-1)(e^{-\Omega_-/T}-1)}
    \biggr\}.\label{eer}
    \end{split} 
\end{flalign} 
Here we again emphasize that the ERR's temperature scaling is indeed dependent on the specific form of the coupling matrix elements---i.e. there is no universal scaling law. We now consider the specific case of soft 2TO polar coupling and simultaneously assume proximity to the quantum critical point. We can thereby insert the coupling matrix elements defined in Eq.~\eqref{coupling elements} with a phonon dispersion modeled as \cite{Yamada1969NeutronSrTiO3,Courtens1993PhononRegime}
\begin{flalign}
    \omega_\bq=\sqrt{\omega_0^2+s^2 q^2},\qquad q\ll q_\mathrm{BZ}. \label{ph dispersion}
\end{flalign}
We may then integrate over dimensionless variables, $\bx=s\bq/T$ and $\by=s\bq_1/T$, and writing $u=\omega_0/T$, Eq.~\eqref{eer} becomes
\begin{flalign}
\begin{split}
   \frac{1}{\tau_E^{\mathrm{2TO}}} &= T^3\frac{m g_2^2\Omega_0^4}{64\pi^3 s^4\Tbg}
    \frac{3}{\pi^4}
    \int_0^{\Tbg/T} dxx
   \int_{-1}^1 dt_{\bx,\by}\int_0^\infty dyy^2 
   \frac{1+t_{\bx,\by}^2}{\sqrt{u^2+y^2}\sqrt{u^2+|\bx-\by|^2}}\\
    &\times\left\{\left(\sqrt{u^2+y^2}+\sqrt{u^2+|\bx-\by|^2}\right)^3  
    \frac{e^{\sqrt{u^2+y^2}+\sqrt{u^2+|\bx-\by|^2}}}{\left(e^{\sqrt{u^2+y^2}}-1\right)\left(e^{\sqrt{u^2+|\bx-\by|^2}}-1\right)\left(e^{\sqrt{u^2+y^2}+\sqrt{u^2+|\bx-\by|^2}}-1\right)}\right.
    \\
    &\left.+\left(\sqrt{u^2+y^2}-\sqrt{u^2+|\bx-\by|^2}\right)^3
          \frac{e^{\sqrt{u^2+y^2}}-e^{\sqrt{u^2+|\bx-\by|^2}}}{\left(e^{\sqrt{u^2+y^2}}-1\right)\left(e^{\sqrt{u^2+|\bx-\by|^2}}-1\right)\left(e^{\sqrt{u^2+y^2}-\sqrt{u^2+|\bx-\by|^2}}-1\right)\left(e^{-\sqrt{u^2+y^2}+\sqrt{u^2+|\bx-\by|^2}}-1\right)}
    \right\},
    \\
    &=T^3\frac{m g_2^2\Omega_0^4}{64\pi^3 s^4 \Tbg}F_\mathrm{2TO}\left(\frac{T}{\Tbg},\frac{\omega_0}{\Tbg}\right),\label{isotropic eer to}
    \end{split} 
\end{flalign}  
where $F_\mathrm{2TO}$ captures the integral part and constant $3/\pi^4$, $t_{\bx,\by}=\cos\theta_{\bx,\by}$ and $|\bx-\by|=\sqrt{x^2+y^2-2xyt_{\bx,\by}}$, and we used $n=E_\mathrm{F}m\kf/(3\pi^2)$. Matching the upper energy scale defined in \cite{Kumar2021QuasiparticleParaelectrics}, we have 
\begin{equation}
E_\mathrm{2TO}=\frac{64\pi^3s^4}{m g_2^2\Omega_0^4}, \label{E0}
\end{equation}
and subsequently (see Eq. (11) in MT)
\begin{flalign}
\begin{split}
   \frac{1}{\tau_E^{\mathrm{2TO}}} &= 
    \frac{T^3}{E_\mathrm{2TO}T_\mathrm{BG}}
   F_\mathrm{2TO}\left(\frac{T}{\Tbg},\frac{\omega_0}{\Tbg}\right) \label{isotropic 2to dimless variables}
    \end{split} 
\end{flalign}

\subsubsection{Single particle picture of energy relaxation}\label{sec:energy relaxation intuition 2TO}
Just as in Sec.~\ref{sec:energy relaxation intuition}, we emphasize the structure of the integrand in Eq.~\eqref{eer}. There is a factor $\Omega_\pm^3$: one factor of $\Omega_\pm$ comes from the expansion in $\delta T$, one from the initial factor of energy in the power transfer, and one from the integral over $\ve_\bk$. Therefore, the ERR in the linear response regime may be understood as the weighted average of the cubed energy transferred to or from a single electron.

\subsection{Asymptotic Limits}
We now present the limiting forms of Eq.~\eqref{isotropic 2to dimless variables} for the two regimes associated with the relation between the gap, $\omega_0$, and the kinematic energy scale, $\Tbg$.

\subsubsection{Small gap: $\omega_0\ll \Tbg$  }

\subsubsection*{Low temperature (Bloch--Grueneisen) limit: $ T\ll T_{\mathrm{BG}}$ }
In the Bloch--Grueneisen regime, the Bose factors suppress any $x,y\gg1$ ($q,q_1\gg T/s$) so that the upper limit of $x$ may be extended to infinity. We then have the following two cases.

\paragraph{1) Gapped phonons $T\ll\omega_0$ } 
In Eq.~\eqref{isotropic 2to dimless variables} we may let $A=\sqrt{u^2+y^2}\geq u$ and $B=\sqrt{u^2+r^2}\geq u$, where $r=|\bx-\by|$. With $u\gg1$ we have $1/(e^{A,B}-1)\approx e^{-A,-B}$. The sum in the curly brackets is then
\begin{flalign}
    (A+B)^3e^{-A-B}-(A-B)^2e^{-B}.
    \end{flalign}
Therefore the ERR is exponentially suppressed in this regime,
\begin{equation}
\boxed{\frac{1}{\tau_E^{\mathrm{2TO}}} \propto e^{-\frac{\omega_0}{T}}}.
\end{equation}

\paragraph{2) Near gapless phonons in the BG regime $\omega_0\ll T\ll \Tbg$ } 
Here we have cubic scaling,
\begin{flalign}
\begin{split}
   \boxed{\frac{1}{\tau_E^{\mathrm{2TO}}} =
    \frac{T^3}{E_\mathrm{2TO} \Tbg}\frac{3\alpha}{\pi^4}}, \qquad \omega_0\ll T\ll \Tbg \label{isotropic low T 2to} ,
    \end{split} 
\end{flalign} 
where 
\[
F_\mathrm{2TO}\left(z\ll1,\eta\ll1\right)=\frac{3\alpha}{\pi^4},\qquad\alpha\approx 79.1.
\]
The asymptotic limit is therefore
\begin{equation}
\frac{1}{\tau_E^{\mathrm{2TO}}} \propto T^3,\qquad\omega_0\ll T\ll \Tbg.
\end{equation}
In the low temperature regime all scattering events relax both energy and momentum, hence why the single particle relaxation rate and ERR---regardless of the specific phonon coupling mechanism---scale the same with temperature.

\subsubsection*{High temperature (equipartition) limit: $T\gg T_{\mathrm{BG}},\omega_0$}
In the equipartition regime, we see an interesting distinction between momentum and energy relaxation such that the well known diffusion model breaks down, namely the relation between energy relaxation and single particle times \cite{Altshuler1982EffectsLocalisation,Gantmakher1987CarrierSemiconductors},
\[
\tau_E\propto\frac{T^2}{\Tbg^2}\tau_{0},
\]
does \textit{not} hold.
We can see this by making the usual a priori assumption that $x$ and $y$ are small over the entire integration interval, which is justified by the resulting integrand in the case of single phonon scattering or when solving the transport relaxation rate for 2TO phonon scattering (see SM in \cite{Kumar2021QuasiparticleParaelectrics}). However, this is not the case for energy relaxation: if we expand the Bose factors for small $x,y\ll1$,
\[
\frac{1}{e^x-1}\approx 1/x,
\] 
the terms inside the curly brackets in Eq.~\eqref{isotropic 2to dimless variables} simply reduce to a factor of 4, and the resulting integral,
\begin{flalign}
     F_\mathrm{2TO}\left(z\rightarrow\infty,\eta\ll z\right)&=
   4\int_0^{1/z} dxx
   \int_{-1}^1 dt_{\bx,\by}\int_0^\infty dyy 
   (1+t_{\bx,\by}^2)\frac{y}{|\bx-\by|},
\end{flalign}
is clearly divergent. However, the original integral in Eq.~\eqref{isotropic 2to dimless variables} is convergent, and since $x$ is small everywhere, we may expand in $x$. To leading order we have
\begin{flalign}
\begin{split}
    F_\mathrm{2TO}(z\rightarrow\infty,\eta\ll z) &= 
    \frac{8}{\pi^4}
    \int_0^{1/z} dxx
   \int_0^\infty dyy^3
   \frac{e^{2y}}{(e^{y}-1)^2(e^{2y}-1)},
   \\
   &=
   \frac{4\beta}{\pi^4z^2}  \label{high T 2to scaling}
    \end{split}
\end{flalign}
where 
\[
\int_{-1}^1dt_{\bx,\by}(1+t_{\bx,\by}^2)=8/3,\quad \mathrm{and}\quad\beta = \int_0^\infty dyy^3
   \frac{e^{2y}}{(e^{y}-1)^2(e^{2y}-1)}\approx0.95.
\]
Thus, the ERR in the equipartition regime in the near gapless limit is 
\begin{flalign}
    \boxed{\frac{1}{\tau_E^{\mathrm{2TO}}}=T\frac{\Tbg}{E_\mathrm{2TO}}\frac{4\beta}{\pi^4}}, \qquad T\gg T_{\mathrm{BG}},\omega_0, \label{high T err 2to isotropic}
\end{flalign}
 which scales linearly with temperature,
\begin{flalign}
\frac{1}{\tau_E^{\mathrm{2TO}}}\propto T, \qquad T\gg T_{\mathrm{BG}},\omega_0.
\end{flalign}
We then have the following relation between single particle and energy relaxation times,
\begin{flalign}
    \tau_E^{\mathrm{2TO}}\sim \frac{T}{\Tbg}\tau_{0}^{\mathrm{2TO}}.
\end{flalign}
Just like the diffusion scenario, energy relaxation is slow on the single particle time scale, $\tau_E^{\mathrm{2TO}}\gg\tau_{0}^{\mathrm{2TO}}$, but faster than diffusion by a factor of $T/\Tbg$. It is therefore evident that energy relaxation does not follow a diffusive picture. This can be explained qualitatively as follows: rather than energy relaxation being a random walk process with many small steps along the energy axis, it can instead be viewed as a process controlled by rare but large energy transfers\footnote{The only term in $F_\mathrm{2TO}$ that survives expansion in small $x$ is the one associated with $\Omega_+=\omega_{\bq_1}+\omega_{\bq_2}$, namely double absorption and emission processes dominate.}. While the resulting momentum transfer must be small, namely $q=|\bq_1+\bq_2|\ll T/s$, there is no such restriction on individual momentum transfers, namely $q_{1,2}\sim T/s$.

\subsubsection{Crossover temperature} \label{crossover isotropic}
\begin{figure}[t]
\centering
  \includegraphics[scale=0.7]{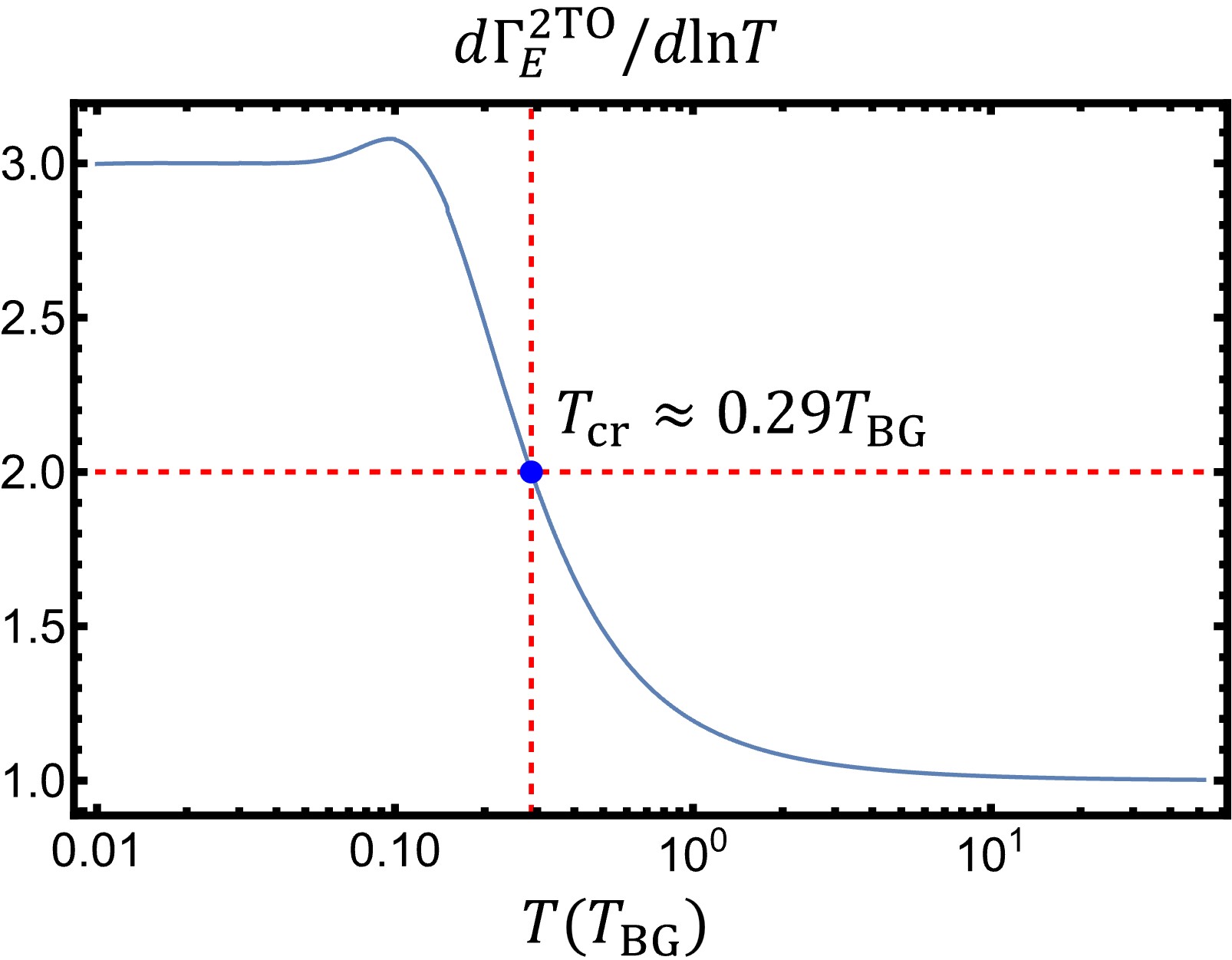}
  \caption{The effective exponent $n(T)$ is plotted. We define the crossover temperature by the relation $n(T_\mathrm{cr})=2$, which gives $T_\mathrm{cr}\approx0.29\Tbg$. }  
  \label{fig:crossover isotropic}
\end{figure}

The crossover temperature between the two scaling regimes can be determined by matching the asymptotic limits in Eqs.~\eqref{isotropic low T 2to} and \eqref{high T err 2to isotropic}. We find a theoretical crossover between 
the BG and equipartition regimes to occur at $T_\mathrm{cr}=\sqrt{4\beta/3\alpha}\ \Tbg\approx 0.13\ \Tbg$. 
We also determined $T_\mathrm{cr}$ numerically by calculating the temperature at which the scaling exponent, 
\[
n(T)\equiv d \ln \Gamma^\mathrm{2TO}_E/d\ln T,
\] 
equals the average of the low and high $T$ limits, namely, when 
$n(T_\mathrm{cr})=2$---the dimensionless energy relaxation rate, $\Gamma_E^\mathrm{2TO}=\tau_E^\mathrm{ref}/\tau_E^\mathrm{2TO}$, is the ERR shown in Eqn.~\eqref{isotropic 2to dimless variables} normalized by the reference rate $1/\tau_E^\mathrm{ref}=T^2_\mathrm{BG}/E_\mathrm{2TO}$. This methods yields $T_\mathrm{cr} \approx0.29\Tbg$, as shown in Fig.~\ref{fig:crossover isotropic}. In either case, $T_\mathrm{cr}$ is significantly lower than $\Tbg$.

\subsubsection{Large gap: $\omega_0\gg \Tbg$}
\subsubsection*{Low temperature (Bloch--Grueneisen) limit: $ T\ll T_{\mathrm{BG}}$ }
The cubic scaling regime is fully replaced by a region of exponential suppression.

\subsubsection*{High temperature (equipartition) limit: $T\gg T_{\mathrm{BG}},\omega_0$ }
Regardless of the relation between $\omega_0$ and $\Tbg$, as long as $T$ is much larger than both, we are back to the regime of linear temperature scaling.

\subsection{Anisotropic electron dispersion}

\subsubsection*{Real-space interaction Hamiltonian (t$_{2g}$ symmetry)}
As in Sec.~\ref{sec: aniso single}, we consider the $t_{2g}$ symmetry of STO and its corresponding mutually orthogonal, ellipsoidal Fermi surface valleys. We repeat various definitions here,
\[
n_{\alpha\beta}(\mathbf r)\equiv \psi_\alpha^\dagger(\mathbf r)\psi_\beta(\mathbf r),\qquad
n_a(\mathbf r)\equiv n_{\alpha\alpha}(\mathbf r),\qquad
N_{\alpha\beta}(\mathbf r)\equiv n_{\alpha\beta}(\mathbf r)+n_{\beta\alpha}(\mathbf r)\ (\alpha\neq\beta).
\]
The most general local 2TO coupling consistent with the $t_{2g}$ triplet once again splits into hydrostatic and deviatoric (intravalley) and shear (intervalley) parts:
\begin{align}
H^{2\mathrm{TO}}_{\mathrm{int}}
&=H_h^{2\mathrm{TO}}+H_u^{2\mathrm{TO}}+H_s^{2\mathrm{TO}},\\[2pt]
H_h^{2\mathrm{TO}}
&=\Xi_h^{(2)}\int d^3r\;\Big(\sum_\alpha n_\alpha(\mathbf r)\Big)\; \bP(\mathbf r)\!\cdot\!\bP(\mathbf r),\label{H2_h}\\[2pt]
H_u^{2\mathrm{TO}}
&=\Xi_u^{(2)}\int d^3r\;\sum_\alpha n_\alpha(\mathbf r)\left[P_{\alpha}^2(\mathbf r)-\frac{1}{3}\bP(\mathbf r)\!\cdot\!\bP(\mathbf r)\right],\label{H2_u}\\[2pt]
H_{s,\alpha\beta}^{2\mathrm{TO}}
&=\Xi_s^{(2)}\int d^3r\;N_{\alpha\beta}(\br)P_\alpha(\br)P_\beta(\br)\qquad (\alpha,\beta)\in\{(x,y),(x,z),(y,z)\}.\label{H2_s}
\end{align}
where once again $\alpha$ is associated with the major axis of an ellipsoidal Fermi surface, e.g. $\alpha=z$ denotes
\begin{flalign}
    \ve_{z,\bk}=\frac{1}{2}\left(\frac{k_x^2+k_y^2}{m_\perp}+\frac{k_z^2}{m_\parallel}\right), 
\end{flalign}
and $\Xi_h^{(2)},\Xi_u^{(2)},\Xi_s^{(2)}$ are the (long-wavelength) 2TO coupling constants.

\subsubsection{Fourier transforms and TO normal modes}
The electron fields on valley $\alpha$ are repeated here
\[
\psi_\alpha(\mathbf r)=\frac{1}{\sqrt V}\sum_{\bk}c_{\alpha\bk}\,e^{i\bk\cdot\mathbf r},\qquad
\psi_\alpha^\dagger(\mathbf r)=\frac{1}{\sqrt V}\sum_{\bk}c_{\alpha\bk}^\dagger\,e^{-i\bk\cdot\mathbf r}.
\]
The polarization field is quantized in TO normal coordinates:
\begin{align}
P_\alpha(\mathbf r)
&=\frac{1}{\sqrt V}\sum_{\bq,t} P_{\alpha,\bq t} \,e^{i\bq\cdot\mathbf r},\\
P_{\alpha,\bq \mathrm{TO}}    
&=A_{\bq t}\,e_{\alpha,\bq \mathrm{TO}}\,\Big(b_{\bq \mathrm{TO}}+b^\dagger_{-\bq \mathrm{TO}}\Big),\qquad A_{\bq\mathrm{TO}}=\frac{\Omega_0}{\sqrt{4\pi\omega_{\bq\mathrm{TO}}}}.
\end{align}
We keep fully isotropic TO dispersions and polarizations: $\omega_{\bq \mathrm{TO}}$ and $\be_{\bq \mathrm{TO}}$
do not depend on the direction of $\bq$, and the two transverse branches are labeled $t=1,2$.

\subsubsection{Momentum-space Hamiltonian and ordered two-phonon vertices}
Inserting the Fourier expansions and performing $\int d^3r$ yields the generic two-phonon form
\begin{equation}
H^{2\mathrm{TO}}_{\mathrm{int}}
=
\frac{1}{2V}\sum_{\substack{\alpha,\beta \\ \bk_1,\bq_1,t_1\\ \bq_2,t_2}}
\Gamma^{\alpha\beta}_{t_1t_2}(\bq_1,\bq_2)\;
c^\dagger_{\alpha,\bk_1-\bq_1-\bq_2}\,c_{\beta,\bk_1}\;
X_{\bq_1t_1}X_{\bq_2t_2},
\label{H2_k}
\end{equation}
where $X_{\bq\lambda}\equiv b_{\bq\lambda}+b^\dagger_{-\bq\lambda}$ and the Kronecker delta enforcing
$\bk_1-\bk_1'+\bq_1+\bq_2=0$ has been used to eliminate $\bk'$.
The two-phonon vertices for the three symmetry channels are:

\paragraph{(i) Intraband hydrostatic ($h$), $\alpha=\beta$:}
\begin{equation}
\Gamma^{\alpha\alpha,(h)}_{t_1t_2}(\bq_1,\bq_2)
=
\Xi_h^{(2)}\,A_{\bq_1t_1}A_{\bq_2t_2}\;
\big(\be_{\bq_1t_1}\cdot\be_{\bq_2t_2}\big).
\label{Gamma_h}
\end{equation}

\paragraph{(ii) Intraband deviatoric ($u$), $\alpha=\beta$:}
\begin{equation}
\Gamma^{\alpha\alpha,(u)}_{t_1t_2}(\bq_1,\bq_2)
=
\Xi_u^{(2)}\,A_{\bq_1t_1}A_{\bq_2t_2}\;
\left(e^\alpha_{\bq_1t_1}\,e^\alpha_{\bq_2t_2}-\frac{1}{3}\be_{\bq_1t_1}\cdot\be_{\bq_2t_2}\right).
\label{Gamma_u}
\end{equation}

\paragraph{(iii) Interband shear ($s$), $\alpha\neq\beta$ (allowed cyclic pairs only, $(\alpha,\beta)\in\{(x,y),(x,z),(y,z)\}$):}
\begin{equation}
\Gamma^{\alpha\beta,(s)}_{t_1t_2}(\bq_1,\bq_2)
=
\Xi_s^{(2)}\,A_{\bq_1t_1}A_{\bq_2t_2}\;
e^{\alpha}_{\bq_1t_1}\,e^{\beta}_{\bq_2t_2},
\qquad \alpha\neq\beta.
\label{Gamma_s}
\end{equation}

\subsubsection{Vertex chosen for calculation} \label{sec: final vertex}
For intraband scattering $\alpha=\beta$, the total vertex is $\Gamma^{\alpha\alpha}=\Gamma^{\alpha\alpha,(h)}+\Gamma^{\alpha\alpha,(u)}$. For interband (shear) scattering $\alpha\neq\beta$ (only allowed cyclic pairs), $\Gamma^{\alpha\beta}=\Gamma^{\alpha\beta,(s)}$.
Notice, given the definitions of the deviatoric and shear vertices, there is no explicit $q$ dependence in either \textit{beyond} that of the hydrostatic term, and therefore, out of convenience, we consider only the hydrostatic part to deduce leading order temperature behavior in the intermediate regime. Then we may simply work with the same initial form of the power transfer given in Eq.~\eqref{pow transfer to} for the isotropic case, only now the electron dispersion is ellipsoidal. Note, by considering only the hydrostatic part, we neglect intervalley scattering entirely. However, as was shown for the single phonon case, for leading order temperature behavior one need only consider the case of intravalley scattering, so we relabel
\[
\Xi_h^{(2)}\rightarrow g_2
\]
and consider only a single valley at a time.

\subsubsection{Anisotropic kinematic constraints}
Provided the discussion in the preceding section, we need only specify the dispersion associated with a single ellipsoidal Fermi surface as defined in Eq.~\eqref{ellip par}, and rescaling this axis as in Eq.~\eqref{kt}, each $\delta$ function in Eq.~\eqref{pow transfer to} may be written 
\begin{flalign}
    \delta(\ve_{\kt}-\ve_{\bkt-\bqt}-\Omega_\pm)=\frac{\delta\left(\cos\theta_{\bkt}-\frac{\qt}{2\kt}-\frac{m_\perp\Omega\pm}{\kt\qt}\right)}{\frac{\kt\qt}{m_\perp}}, \label{delta spherical}
\end{flalign}
Once again we rewrite the sum $(1/V)\sum_\bk=\int d^3k/(2\pi)^3=\sqrt{m_\parallel/m_\perp}\int d^3\kt/(2\pi)^3$, and $\kt_{\text{F}}=\kp$. We present a similar analysis to Sec.~\ref{sec: intravalley contribution} now in the context of 2TO scattering. The $\delta$ function is the only term dependent on $\cos\theta_{\bkt}$, and upon integration produces the following constraints on $\qt$ which must be satisfied simultaneously,
\begin{flalign}
    \qt^2\lessgtr \pm2\kp \qt-2m_\perp\Omega_\pm.
\end{flalign}
Both $\Omega_\pm\lessgtr0$ must be considered and $\qt$ is thereby restricted between the following bounds,
\begin{flalign}
    \kp\Bigg(1-\sqrt{1-\frac{\Omega_\pm}{E_\mathrm{F}}}\Bigg) < \ &\qt<\kp\Bigg(1+\sqrt{1-\frac{\Omega_\pm}{E_\mathrm{F}}}\Bigg), \ \  \Omega_\pm>0, \label{O+}\\ 
    \kp\Bigg(-1+\sqrt{1+\frac{|\Omega_\pm|}{E_\mathrm{F}}}\Bigg) < \ &\qt<\kp\Bigg(1+\sqrt{1+\frac{|\Omega_\pm|}{E_\mathrm{F}}}\Bigg), \ \ \Omega_\pm<0. \label{-O}
\end{flalign}
For a degenerate Fermi system, $\Omega_\pm\lesssim T\ll E_\mathrm{F}$, and to leading order one may restrict $0<\qt<2\kp$ in all cases. Then the bound on $q$ is a function of its polar angle $\theta$,
\[
0<q<\frac{2\kp}{\sqrt{1-(1-\mu) t^2}}=2\kp A(\mu,t),
\]
and we repeat some definitions here:
\[
t=\cos\theta,\qquad A(\mu,t)=\left(1-\left(1-\mu\right) t^2\right)^{-1/2},\qquad \mu=\frac{m_\perp}{m_\parallel} .
\] 
Having integrated over all three azimuthal angles and since we will consider particular coupling matrix elements with appropriate symmetry, we use the isotropic form of the power transfer in Eq.~\eqref{pow transfer to} which can now be rewritten
\begin{flalign}
\begin{split}
    \frac{\partial E_e^{\mathrm{2ph}}}{\partial t} &= -\frac{4\pi V}{(2\pi)^6}\sqrt{m_\perp m_\parallel }\int_{-1}^1 dt_1\int_0^\infty dq_1q_1^2 \int_{-1}^1 dtA(\mu,t) \int_0^{2\kp A(\mu,t)} dqq\abs{g_{\bq_1,\bq_-\bq_1}}^2  \\
    &\times\int d\kt\kt\,\biggl\{\Omega_+\biggl[\bigg(N_{\bq_1}+N_{\bq-\bq_1}+1\bigg)f(\ve_\kt)\bigg(1-f(\ve_\kt-\Omega_+)\bigg)+N_{\bq_1}N_{\bq-\bq_1}\bigg(f(\ve_\kt)-f(\ve_\kt-\Omega_+)\bigg)\biggr] \\ 
    &\quad\quad\quad+\Omega_-\bigg[N_{\bq_1}N_{\bq-\bq_1}\bigg(f(\ve_\kt)-f(\ve_\kt-\Omega_-)\bigg)-N_{\bq_1}f(\ve_\kt-\Omega_-)\bigg(1-f(\ve_\kt)\bigg)+N_{\bq-\bq_1}f(\ve_\kt)\bigg(1-f(\ve_\kt-\Omega_-)\bigg)\bigg]
    \biggr\}.
    \end{split} 
\end{flalign} 
Per usual, we may linearize the electron dispersion near the FS such that $\ve_{\kt}\approx\vfp (\kt-\kp)$, where $\vfp=\kp/m_\perp$ and set all instances of $\kt=\kp$. The integrals over $\ve_\kt$ may be performed just as in Eq.~\eqref{energy},
\begin{flalign}
\begin{split}
    \frac{\partial E_e^{\mathrm{2ph}}}{\partial t} &= -\frac{2 V}{(2\pi)^5}m_\perp\sqrt{m_\perp m_\parallel }\int_{-1}^1 dt_1\int_0^\infty dq_1q_1^2 \int_{-1}^1 dtA(\mu,t) \int_0^{2\kp A(\mu,t)} dqq\abs{g_{\bq_1,\bq_-\bq_1}}^2  \\
    &\times\biggl\{\Omega_+^2\biggl[\bigg(N_{\bq_1}+N_{\bq-\bq_1}+1\bigg)\tilde N(\Omega_+)-N_{\bq_1}N_{\bq-\bq_1}\biggr] +\Omega_-^2\bigg[-N_{\bq_1}N_{\bq-\bq_1}+N_{\bq_1}\tilde N(-\Omega_-)+N_{\bq-\bq_1}\tilde N(\Omega_-)\bigg)\bigg]
    \biggr\}.
    \end{split} 
\end{flalign} 
where $\tilde{N}$ specifies a Bose function at the electron temperature $T_e$. The Bose functions may be expanded in $\delta T/T$ which yields
\begin{flalign}
\begin{split}
    \frac{\partial E_e^{\mathrm{2ph}}}{\partial t}&= -\frac{\delta T}{T^2}\frac{2 V}{(2\pi)^5}m_\perp\sqrt{m_\perp m_\parallel }\int_{-1}^1 dt_1\int_0^\infty dq_1q_1^2 \int_{-1}^1 dtA(\mu,t) \int_0^{2\kp A(\mu,t)} dqq\abs{g_{\bq_1,\bq_-\bq_1}}^2  \\
    &\times\bigg\{\Omega_+^3  \frac{e^{\Omega_+/T}}{(e^{\omega_{\bq_1}/T}-1)(e^{\omega_{\bq-\bq_1}/T}-1)(e^{\Omega_+/T}-1)}+ \Omega_-^3
          \frac{e^{\omega_{\bq_1}/T}-e^{\omega_{\bq-\bq_1}/T}}{(e^{\omega_{\bq_1}/T}-1)(e^{\omega_{\bq-\bq_1}/T}-1)(e^{\Omega_-/T}-1)(e^{-\Omega_-/T}-1)}\bigg\}
    \biggr\}.
    \end{split} 
\end{flalign} 

\subsubsection{Energy relaxation rate} Finally, we use the relation in Eq.~\eqref{generic rate} to write the ERR,
\begin{flalign}
\begin{split}
   \frac{1}{\tau_E^{\mathrm{2ph}}} &= \frac{1}{T^3}\frac{E_\mathrm{F}m_\perp\sqrt{m_\perp m_\parallel }}{16\pi^7n}
   \int_{-1}^1 dt_1
   \int_0^\infty dq_1q_1^2 
   \int_{-1}^1 dtA(\mu,t) \int_0^{2\kp A(\mu,t)} dqq\abs{g_{\bq_1,\bq_-\bq_1}}^2  \\
    &\times\bigg\{\Omega_+^3  \frac{e^{\Omega_+/T}}{(e^{\omega_{\bq_1}/T}-1)(e^{\omega_{\bq-\bq_1}/T}-1)(e^{\Omega_+/T}-1)}+ \Omega_-^3
          \frac{e^{\omega_{\bq_1}/T}-e^{\omega_{\bq-\bq_1}/T}}{(e^{\omega_{\bq_1}/T}-1)(e^{\omega_{\bq-\bq_1}/T}-1)(e^{\Omega_-/T}-1)(e^{-\Omega_-/T}-1)}
    \biggr\}. \label{general 2to eer aniso}
    \end{split} 
\end{flalign} 
In the isotropic case, $m_\parallel=m_\perp=m$, $A(1,t)=1$, $\kp=\kf$, and we have a simple reduction to Eq.~\eqref{eer}.
Here we emphasize that the ERR's temperature scaling is indeed dependent on the specific form of the coupling matrix elements---i.e. there is no universal scaling law. We now insert the hydrostatic 2TO vertex and relabel $\Xi_h^{(2)}\equiv g_2$ (see Sec.~\ref{sec: final vertex}).
Integrating over dimensionless variables, $\bx=s\bq/T$ and $\by=s\bq_1/T$, and writing $u=\omega_0/T$, yields (see Eq. (17) in MT)
\begin{flalign}
\begin{split}
   \frac{1}{\tau_E^{\mathrm{2TO}}} &= \frac{T^3 }{ E_{\mathrm{2TO},\perp}\Tbgp}\frac{3}{2\pi^4}
    \int_{-1}^1 dt_\bx A(\mu,t_\bx) \int_0^{A(\mu,t_\bx)\Tbgp/T} dxx
   \int_{-1}^1 dt_{\bx,\by}\int_0^\infty dyy^2 
   \frac{1+t_{\bx,\by}^2}{\sqrt{u^2+y^2}\sqrt{u^2+|\bx-\by|^2}}\\
    &\times\left\{\left(\sqrt{u^2+y^2}+\sqrt{u^2+|\bx-\by|^2}\right)^3  
    \frac{e^{\sqrt{u^2+y^2}+\sqrt{u^2+|\bx-\by|^2}}}{\left(e^{\sqrt{u^2+y^2}}-1\right)\left(e^{\sqrt{u^2+|\bx-\by|^2}}-1\right)\left(e^{\sqrt{u^2+y^2}+\sqrt{u^2+|\bx-\by|^2}}-1\right)}\right.
    \\
    &\left.+\left(\sqrt{u^2+y^2}-\sqrt{u^2+|\bx-\by|^2}\right)^3
          \frac{e^{\sqrt{u^2+y^2}}-e^{\sqrt{u^2+|\bx-\by|^2}}}{\left(e^{\sqrt{u^2+y^2}}-1\right)\left(e^{\sqrt{u^2+|\bx-\by|^2}}-1\right)\left(e^{\sqrt{u^2+y^2}-\sqrt{u^2+|\bx-\by|^2}}-1\right)\left(e^{-\sqrt{u^2+y^2}+\sqrt{u^2+|\bx-\by|^2}}-1\right)}
    \right\},
    \\
    &=\frac{T^3 }{ E_{\mathrm{2TO},\perp}\Tbgp}
    \bar F_\mathrm{2TO}\left(\frac{T}{\Tbgp},\frac{\omega_0}{\Tbgp},\frac{m_\perp}{m_\parallel}\right),\label{gapped anisotropic eer to}
    \end{split} 
\end{flalign}  
where we used $n=E_\mathrm{F}\sqrt{m_\perp m_\parallel}\kp/(3\pi^2)$, $\bar F_\mathrm{2TO}$ denotes the anisotropic case and contains an additional integral over $t_\bx$ compared to $F_\mathrm{2TO}$, and $E_{\mathrm{2TO},\perp}=E_\mathrm{2TO}(m\rightarrow m_\perp)$.
To properly compare Eq.~\eqref{gapped anisotropic eer to} with Eq.~\eqref{isotropic 2to dimless variables} we use the definition of the dispersion in the isotropic and anisotropic cases along with the relation,
\begin{flalign}
    \kf^3=k_\mathrm{F\perp}^2\kl,
\end{flalign}
where we consider a scenario where the volume of the Fermi surface is preserved under deformation.
Combining this with the relation $\kl/\kp=\sqrt{m_\parallel/m_\perp}$ we have 
\begin{flalign}
    \frac{\kf}{\kp}=\frac{1}{\mu^{1/6}}.
\end{flalign}
With $T_\mathrm{BG\perp}^2/E_{\mathrm{2TO},\perp}=(k_\mathrm{F\perp}^2m_\perp/\kf^2m)(T_\mathrm{BG}^2/E_\mathrm{2TO})=\mu^{2/3}(T_\mathrm{BG}^2/E_\mathrm{2TO})$ (under the assumption $g_2$ is the same in both isotropic and anisotropic cases), we can write Eq.~\eqref{gapped anisotropic eer to} with the same variables and in the same units as Eq.~\eqref{isotropic 2to dimless variables},
\begin{flalign}
\begin{split}
   \frac{1}{\tau_E^{\mathrm{2TO}}} &= 
    \frac{T^3}{E_{\mathrm{2TO}}\Tbg}
   \mu^{1/6}\bar F_\mathrm{2TO}\left(\frac{T}{\Tbg\ \mu^{1/6}},\frac{\omega_0}{\Tbg\ \mu^{1/6}},\mu\right) .\label{anisotropic 2to dimless variables isotropic units}
    \end{split} 
\end{flalign}

\subsection{Asymptotic limits (anisotropic)} \label{asymptotic 2to}
Here we present the limiting forms of Eq.~\eqref{gapped anisotropic eer to}, which in any case, if $\omega_0>T$, is exponentially suppressed. We therefore consider only a very small gap: $\omega_0\ll \Tbgp$.
For the 2TO scaling function, 
\begin{flalign}
\begin{split}
    \bar F_\mathrm{2TO}(z,\eta\ll z,\mu)&=\frac{3}{2\pi^4}
    \int_{-1}^1 dt_\bx \frac{1}{\sqrt{1-(1-\mu)t_\bx^2}} \int_0^{1/z \sqrt{1-(1-\mu)t_\bx^2}} dxx
   \int_{-1}^1 dt_{\bx,\by}\int_0^\infty dyy
   \frac{1+t_{\bx,\by}^2}{|\bx-\by|}\\
    &\times\left\{\left(y+|\bx-\by|\right)^3  
    \frac{e^{y+|\bx-\by|}}{\left(e^{y}-1\right)\left(e^{|\bx-\by|}-1\right)\left(e^{y+|\bx-\by|}-1\right)}\right.
    \\
    &\left.+\left(y-|\bx-\by|\right)^3
          \frac{e^{y}-e^{|\bx-\by|}}{\left(e^{y}-1\right)\left(e^{|\bx-\by|}-1\right)\left(e^{y-|\bx-\by|}-1\right)\left(e^{-y+|\bx-\by|}-1\right)}
    \right\},
    \end{split}
\end{flalign}
there will be three relevant scaling regimes: i) $T\ll\Tbgp$, ii) $\Tbgp\ll T\ll\Tbgl$, and iii) $T\gg\Tbgl$, just as in the case of DA coupling. It is again useful to let
\[
w=\sqrt{1-(1-\mu)t_\bx^2},
\]
and given the integrand is even with respect to $t$, we have
\begin{flalign}
\begin{split}
    \bar F_\mathrm{2TO}(z,\eta\ll z,\mu)&=
    \frac{3}{\pi^4}\frac{1}{\sqrt{1-\mu}}
    \int_{\sqrt{\mu}}^1 \frac{dw}{\sqrt{1-w^2}} \bar J\left(\frac{1}{z w}\right)
    \end{split}
\end{flalign}
where 
\begin{flalign}
\begin{split}
\bar J(a)&\equiv\int_0^{a} dxx
   \int_{-1}^1 dt_{\bx,\by}\int_0^\infty dyy 
   \frac{1+t_{\bx,\by}^2}{|\bx-\by|}
   \left\{\left(y+|\bx-\by|\right)^3  
    \frac{e^{y+|\bx-\by|}}{\left(e^{y}-1\right)\left(e^{|\bx-\by|}-1\right)\left(e^{y+|\bx-\by|}-1\right)}\right.
    \\
    &\hspace{5.7cm}
    \left.+\left(y-|\bx-\by|\right)^3
          \frac{e^{y}-e^{|\bx-\by|}}{\left(e^{y}-1\right)\left(e^{|\bx-\by|}-1\right)\left(e^{y-|\bx-\by|}-1\right)\left(e^{-y+|\bx-\by|}-1\right)}
    \right\}.
\end{split}
\end{flalign}
We again consider the limit $\mu\ll1$ so that even in the case of $T\gg\Tbgp$, one cannot naively expand the Bose factors. There are two separate regimes for $T\gg\Tbgp$: a high temperature, equipartition regime where $T\gg\Tbgl$, and an intermediate regime where $\Tbgp\ll T\ll\Tbgl$.

\subsubsection*{Low temperature limit: $T\ll\Tbgp$} \label{subsec:low T da}
In the BG regime, where $T\ll\Tbgp$ ($\Tbgp$ is below the lesser of the two BG temperatures) the upper limit in the integral over $x$ can be taken to infinity for all $w$ and the asymptotic limit of the scaling function is 
\begin{flalign}
\begin{split}
    \bar F_\mathrm{2TO}(z\ll1,\eta\ll z,\mu)&=
    \frac{3}{\pi^4}\frac{1}{\sqrt{1-\mu}}
    \int_{\sqrt\mu}^1 \frac{dw}{\sqrt{1-w^2}}
    \bar J(\infty)\\
    &
    =\frac{3\alpha}{\pi^4}\bar M_\mathrm{L}(\mu),\qquad \bar M_\mathrm{L}(\mu)=M_\mathbf{L}(\mu)=\frac{\arccos(\sqrt\mu)}{\sqrt{1-\mu}},
    \end{split}
\end{flalign}
where we have considered a small gap $\omega_0\ll\Tbgp$, and $\alpha$ was defined above in the isotropic low temperature limit.
The ERR is therefore
\begin{flalign}
\begin{split}
    \boxed{\frac{1}{\tau^{\mathrm{2TO}}_E}
    =\frac{T^3}{E_{\mathrm{2TO},\perp}\Tbgp}\frac{3\alpha}{\pi^4}M_\mathrm{L}(\mu)}, \qquad T\ll\Tbgp,\label{low T 2to}
    \end{split}
\end{flalign}
which clearly yields the isotropic result in Eq.~\eqref{isotropic low T 2to} when $\mu=1$ ($M_\mathrm{L}(1)=1$). 
In the limit $\mu\ll1$, $M_\mathrm{L}(\mu)\approx\pi/2$.

In summary, in the BG regime we have
\begin{flalign}
    \frac{1}{\tau^{\mathrm{2TO}}_E}\propto T^3, \qquad T\ll\Tbgp.
\end{flalign}

\subsubsection*{High temperature limit: $T\gg\Tbgl$ }
In the equipartition regime, $1/z w\ll1$ for all $w$. We revert back to the discussion for the isotropic high temperature limit shown in Eq.~\eqref{high T 2to scaling} to get,
\begin{flalign}
\begin{split}
    F_\mathrm{2TO}(z\gg1,\eta\ll 1,\mu\gg1/z^2) &= \frac{3}{\pi^4}
    \frac{1}{\sqrt{1-\mu}}
    \int_{\sqrt{\mu}}^1 \frac{dw}{\sqrt{1-w^2}} \bar J\left(\frac{1}{z w}\ll1\right)\\&
    =\frac{8}{\pi^4}
    \frac{1}{\sqrt{1-\mu}}
    \int_{\sqrt{\mu}}^1 \frac{dw}{\sqrt{1-w^2}}
    \int_0^{1/z w} dxx
   \int_0^\infty dyy^3
   \frac{e^{2y}}{(e^{y}-1)^2(e^{2y}-1)}
   \\
   &=\frac{4\beta}{\pi^4z^2}\bar M_\mathrm{H}(\mu),
    \end{split}
\end{flalign}
where $\beta$ was defined in the isotropic high temperature limit, and 
\begin{flalign}
    \bar M_\mathrm{H}(\mu)=\frac{1}{\sqrt{1-\mu}}\int_{\sqrt{\mu}}^1\frac{dw}{w^2\sqrt{1-w^2}}=\frac{1}{\sqrt{\mu}}.
\end{flalign}
The ERR is 
\begin{flalign}
\begin{split}
\boxed{\frac{1}{\tau_E^\mathrm{2TO}}
=T\frac{T_\mathrm{BG\perp}}{E_\mathrm{2TO,\perp}}\frac{4\beta}{\pi^4}\bar M_\mathrm{H}(\mu)},\qquad T\gg\Tbgl. \label{high T 2to}
\end{split}
\end{flalign}
In the isotropic case, $\mu=1$ and $M_\mathrm{H}(1)=1$ and the results in Eq.~\eqref{high T 2to scaling} and Eq.~\eqref{high T err 2to isotropic} are recovered. 
In summary, we have
\begin{equation}
\frac{1}{\tau_E^\mathrm{2TO}} \propto T,\qquad T\gg\Tbgl.
\end{equation}

\subsubsection*{Intermediate temperature limit: $\Tbgp\ll T\ll\Tbgl$}
Now, $w=\sqrt\mu=\Tbgp/\Tbgl\ll1$ at the upper limit so that the interval includes a region $w\lesssim\Tbgp/T\ll1$ where the argument of $J$ and $J$ itself are $O(1)$. We may expand in small $w$ and $\mu$ so that $\sqrt{1-w^2}\approx1$ and $\sqrt{1-\mu}$. We let $r=z w$ and have for the scaling function,
\begin{flalign}
\begin{split}
    \bar F_\mathrm{2TO}(z\gg 1,\eta\ll z,\mu\ll 1/z^2)&=
    \frac{3}{\pi^4z}
    \int_{\sqrt\mu z}^{z} dw\bar J(1/w)\\
    &=\frac{3\gamma}{\pi^4 z},
    \end{split}
\end{flalign} 
where 
\begin{flalign}
    \gamma=\int_{0}^{\infty} dw\bar J(1/w)\approx 47.11.
\end{flalign}
The ERR is then
\begin{flalign}
\begin{split}
   \boxed{\frac{1}{\tau_E^{\mathrm{2TO}}} =\frac{T^2}{E_{\mathrm{2TO,\perp}}}
   \frac{3\gamma}{\pi^4}},\qquad\Tbgp\ll T\ll\Tbgl.\label{intermediate 2to eer}
    \end{split} 
\end{flalign}
In summary, we have quadratic temperature scaling in the intermediate regime,
\begin{flalign}
    \frac{1}{\tau_E^\mathrm{2TO}}\propto T^2,\qquad \Tbgp\ll T\ll\Tbgl.
\end{flalign}
This result can also be determined by taking the exact limit $\mu=0$, analogous to having a cylindrical Fermi surface, which is discussed below.

\subsubsection{Crossover temperatures} 
As in Sec.~\ref{crossover isotropic}, we now determine the crossover temperatures between each of the three scaling regimes by matching Eqs.~\eqref{low T 2to} and \eqref{intermediate 2to eer} and Eqs.~\eqref{intermediate 2to eer} and \eqref{high T 2to}. We find a theoretical crossover between 
the BG and intermediate temperature regimes to occur at 
\[
    T_\mathrm{cr,1}=\frac{\gamma}{\alpha}\frac{\sqrt{1-\mu}}{\arccos(\mu)}\Tbgp\approx 0.6\frac{\sqrt{1-\mu}}{\arccos(\mu)}\Tbgp.
\]
However, this is after we have already taken $\mu\rightarrow0$ in the intermediate regime. Doing the same for the BG regime we have, 
\begin{flalign}
T_\mathrm{cr,1}\approx0.6\frac{2}{\pi}\Tbgp\approx0.37\ \Tbgp,
\end{flalign}
to leading order in $\mu$. The theoretical crossover between the intermediate and equipartition temperature regimes occurs at 
\begin{flalign}
    T_\mathrm{cr,2}=\frac{4\beta}{3\gamma\sqrt{\mu}}\Tbgp \approx \frac{0.03}{\sqrt\mu}\Tbgp
\end{flalign}
A true intermediate crossover regime of finite width does not occur until $T_\mathrm{cr,2}> T_\mathrm{cr,1}$, which based on this result is only the case for $\mu\lesssim0.005$. 

\begin{figure}[!]
\centering
  \includegraphics[scale=0.7]{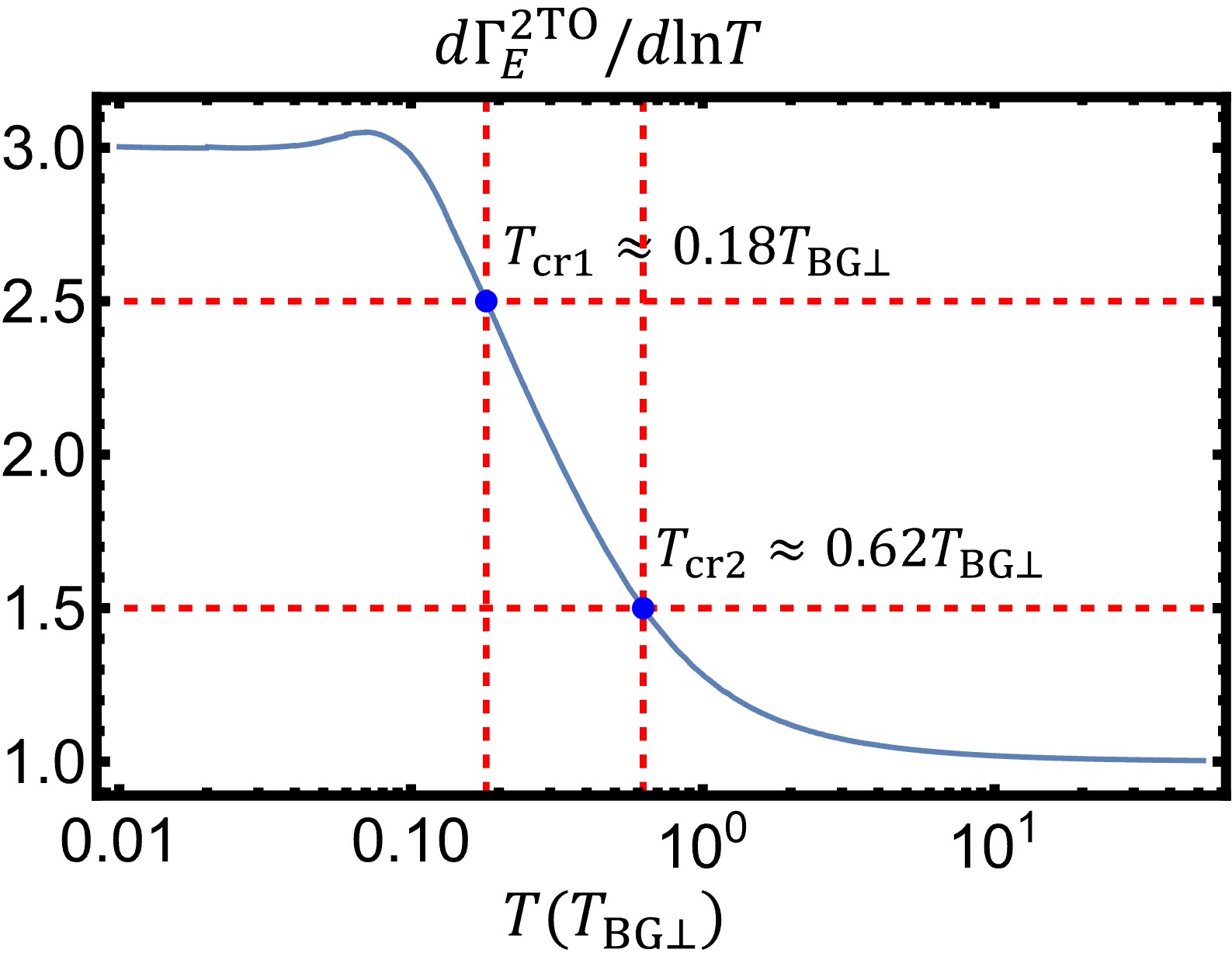}
  \caption{The effective exponent $n(T)$ is plotted for $\mu=0.8$. We now define two crossover temperatures by the relations $n(T_\mathrm{cr1})=2.5$ and $n(T_\mathrm{cr2})=1.5$, which gives $T_\mathrm{cr}\approx0.18\ \Tbgp$ and $T_\mathrm{cr}\approx0.62\ \Tbgp$. }  
  \label{fig:crossover anisotropic}
\end{figure}

One can also determine the crossover temperatures numerically for a given value of $\mu$. A recent experiment \cite{Wakabayashi2025Orbital-resolvedARPES} shows that the ellipsoidal Fermi surface of the conduction band in doped STO, at number density $n=3.58\times10^{20}\ \mathrm{cm}^{-3}$, has a transverse mass, $m_\perp=0.63m_e$, and longitudinal mass, $m_\parallel=8m_e$, where $m_e$ is the free electron mass. We now determine two crossover temperatures, $T_\mathrm{cr1}$ and $T_\mathrm{cr2}$,
by calculating the temperature at which the scaling exponent, 
$n(T)\equiv d \ln \Gamma^\mathrm{2TO}_E/d\ln T$,
equals the average of the BG and intermediate as well as the intermediate and equipartition $T$ limits, namely, when $n(T_\mathrm{cr})=2.5$ and 1.5, respectively. For $\mu=0.63/8\approx0.08$, this method yields $T_\mathrm{cr1} \approx0.18\ \Tbgp$ and $T_\mathrm{cr2} \approx0.62\ \Tbgp$---see Fig.~\ref{fig:crossover anisotropic}. 

It is straightforward to show that the number density for three degenerate ellipsoidal Fermi surfaces is given by
\begin{flalign}
    \begin{split}
        n=\frac{k^{3}_\mathrm{F\perp}}{\pi^2\sqrt\mu},
    \end{split}
\end{flalign}
where we have also accounted for spin degeneracy.
Given that $\kl=\kp/\sqrt\mu$, one may deduce the BG temperatures,
\begin{flalign}
    \begin{split}
        \Tbgp&=\frac{2\hbar s\kp}{\kb}\approx 101\mathrm{K}, \\
        \Tbgl&=\frac{2\hbar s\kl}{\kb}\approx 358 \mathrm{K},
    \end{split}
\end{flalign}
where we considered the case $n=3.58\times10^{20}\ \mathrm{cm}^{-3}$ and $\mu=0.08$, and used $s=6.6\times 10^5\ \mathrm{cm/s}$ \cite{Yamada1969NeutronSrTiO3}. Then the crossover temperatures are 
\begin{flalign}
    \begin{split}
        T_\mathrm{cr1}&\approx18\ \mathrm{K}, \\
        T_\mathrm{cr2}&\approx 63\ \mathrm{K}.
    \end{split}
\end{flalign}
The experimental work on energy relaxation in STO thin films \cite{Kumar2025AbsenceSpectroscopy}---which inspired this work---also considered number densities $\sim 10^{20}\ \mathrm{cm}^{-3}$. However, a temperature range of only $10-50$K. Therefore, we propose one must go to higher temperatures to reach the equipartition regime and see two phonon processes at play.

\subsubsection*{Cylindrical limit}
As in Sec.~\ref{subsec:intermed T da} one may also consider the limiting case $\mu=0$ so that the ellipsoidal dispersion becomes a cylinder which may be cut at finite $\pm\kl$ \cite{Kukkonen1978TBismuth,Kang2018Thermopower-conductivitySrTiO3}. Upon a transformation to cylindrical coordinates we restate the definition,
\[
A(\mu=0,\theta)=\frac{1}{\sin\theta}=\frac{q}{\qp}=\frac{x}{x_\perp}.
\]
Then the integral defined in Eq.~\eqref{gapped anisotropic eer to} becomes
\begin{flalign}
\begin{split}
\bar F_\mathrm{2TO}\left(z\gg1,\eta\ll z,\mu=0\right)
    &\rightarrow\frac{3}{2\pi^4}\int_{-1/\sqrt\mu z}^{1/\sqrt\mu z}d\xl
    \int_0^{1/z}d\xp
    \int_{-\infty}^{\infty}d\yl
    \int_0^{\infty}d\yp\yp
    \frac{y^2+\yl^2}{y^3|\bx-\by|}\\
    &\times\bigg\{(y+|\bx-\by|)^3  \frac{e^{y+|\bx-\by|}}{(e^{y}-1)(e^{|\bx-\by|}-1)(e^{y+|\bx-\by|}-1)}
    \\
    &+
    (y-|\bx-\by|)^3
          \frac{e^{y}-e^{|\bx-\by|}}{(e^{y}-1)(e^{|\bx-\by|}-1)(e^{y-|\bx-\by|}-1)(e^{-y+|\bx-\by|}-1)}
    \biggr\},\label{F intermed 2to}
    \end{split}
\end{flalign}
where 
\[x=\sqrt{\xl^2+\xp^2},\quad y=\sqrt{\yl^2+\yp^2},\quad |\bx-\by|=\sqrt{x^2+y^2-2x\,y\,t_{\bx,\by}},\quad t_{\bx}=\frac{\xl}{x},\quad t_{\bx,\by}=\frac{\yl}{y},
\]
In this temperature regime we take $\Tbgl/T\rightarrow\infty$ and expand in small $\xp$ so that each instance of $|\bx-\by|\approx\sqrt{(\xl-\yl)^2+\yp^2}$. Then in the intermediate regime we have
\begin{flalign}
    \bar F_\mathrm{2TO}\left(z\gg1,\eta\ll z,\mu=0\right)\approx \frac{3\xi}{\pi^4z}
\end{flalign}
and then for the ERR,
\begin{flalign}
\begin{split}
\frac{1}{\tau_E^{\mathrm{2TO}}}&=\frac{T^2}{E_{\mathrm{2TO},\perp}}\frac{3\xi}{\pi^4}
\end{split}
\end{flalign}
where we get a factor of $\Tbgp/T$ from the integral over $x_\perp$, and
\begin{flalign}
\begin{split}
\xi=&
    \frac{1}{2}\int_{-\infty}^{\infty}d\xl
    \int_{-\infty}^{\infty}d\yl
    \int_0^{\infty}d\yp\yp
    \frac{y^2+\yl^2}{y^3|\bx-\by|}\\
    &\times\bigg\{(y+|\bx-\by|)^3  \frac{e^{y+|\bx-\by|}}{(e^{y}-1)(e^{|\bx-\by|}-1)(e^{y+|\bx-\by|}-1)}
    \\
    &+
    (y-|\bx-\by|)^3
          \frac{e^{y}-e^{|\bx-\by|}}{(e^{y}-1)(e^{|\bx-\by|}-1)(e^{y-|\bx-\by|}-1)(e^{-y+|\bx-\by|}-1)}
    \biggr\}\Bigg|_{|\bx-\by|=\sqrt{(\xl-\yl)^2+\yp^2}}\\
    &\approx 48.18.
    \end{split}
\end{flalign}
which nearly matches that of Eq.\eqref{intermediate 2to eer}, differing by numerical precision.

Finally, we note that if we instead let $T/\Tbgl\gg1$ in Eq.~\eqref{F intermed 2to}, we can also expand the integrand in small $\xl$ and reverting back to finite integration limits we have
\begin{flalign}
\frac{1}{\tau_E^{\mathrm{2TO}}}=T\frac{\Tbgl}{E_{\mathrm{2TO},\perp}}\times\mathrm{const}.
\end{flalign}
This differs from Eq.~\eqref{high T 2to} only by a numerical prefactor, and the high $T$ limit is recovered. If we consider the BG regime, $T\ll\Tbgp,\Tbgl$, where we take both limits $\Tbgp/T\rightarrow\infty$ and $\Tbgl/T\rightarrow\infty$, the temperature dependence from the integral portion drops out entirely and the $T^3$ result in Eq.~\eqref{low T 2to} is recovered up to a numerical prefactor.

\bibliographystyle{plain}
\bibliography{jcovey}